\documentclass[fleqn,usenatbib]{mnras}

\usepackage{color}
\usepackage[T1]{fontenc}
\usepackage{xcolor}
\usepackage{ulem}
\usepackage{graphicx}	% Including figure files
\usepackage{amsmath}	% Advanced maths commands
\usepackage{cases}
\usepackage{adjustbox}

\usepackage{graphicx}	% Including figure files
\usepackage{amssymb}	% Extra maths symbols
\usepackage{newtxtext,newtxmath}
\usepackage{cases}
\usepackage{array,multirow}
\usepackage{upgreek}
\usepackage{textcomp}
\usepackage{rotating}
\usepackage{fontawesome}
\newcommand{\Msol}{\,{\rm M}_{\odot}}
\newcommand{\Rsol}{\,{\rm R}_{\odot}}
\newcommand{\gram}{\;\mathrm{g}}
\newcommand{\cm}{\;\mathrm{cm}}

\newcommand{\km}{\;\mathrm{km}}

\newcommand{\yr}{\;\mathrm{yr}}
\newcommand{\mstar}{\;M_{\star}}
\newcommand{\Mbh}{M_{\bullet}}

\usepackage{tikz}
\newcommand*\emptycirc[1][0.7ex]{\tikz\draw (0,0) circle (#1);} 

\newcommand*\fullcirc[1][1ex]{\tikz\fill (0,0) circle (#1);} 

\defcitealias{Ryu+2022}{Paper~1}
\defcitealias{Ryu+2023}{Paper~2}

\title[Hydrodynamics of three-body black hole star encounters]{Close encounters of black hole - star binaries with stellar-mass black holes}

\author[T. Ryu et al.]{%
Taeho Ryu$^{1,2}$,\thanks{E-mail: tryu@mpa-garching.mpg.de}
Ruggero Valli$^{1}$,
R\"udiger Pakmor$^{1}$,
Rosalba Perna$^{3,4}$,
Selma E. de Mink$^{1,5}$,
\newauthor Volker Springel$^{1}$\vspace*{0.1cm}\\%
$^{1}$ Max Planck Institute for Astrophysics, Karl-Schwarzschild-Str.~1, 85748 Garching, Germany\\%
$^{2}$ Physics and Astronomy Department, Johns Hopkins University, Baltimore, MD 21218, USA\\%
$^{3}$ Department of Physics and Astronomy, Stony Brook
  University, Stony Brook, NY 11794-3800, USA\\%
$^{4}$ Center for Computational Astrophysics, Flatiron Institute, New York, NY 10010, USA\\%
$^{5}$ Anton Pannekoek Institute for Astronomy, University of Amsterdam, Science Park 904, 1098XH Amsterdam, The Netherlands
}

\date{Accepted XXX. Received YYY; in original form ZZZ}

\pubyear{2022}

\begin{document}
\label{firstpage}
\pagerange{\pageref{firstpage}--\pageref{lastpage}}
\maketitle

\begin{abstract}
Dynamical interactions involving binaries play a crucial role in the evolution of star clusters and galaxies. We continue our investigation of the hydrodynamics of three-body encounters, focusing on binary black hole (BBH) formation, stellar disruption, and electromagnetic (EM) emission in dynamical interactions between a BH-star binary and a stellar-mass BH, using the moving-mesh hydrodynamics code {\small AREPO}. This type of encounters can be divided into two classes depending on whether the ﬁnal outcome includes BBHs. This outcome is primarily determined by which two objects meet at the ﬁrst closest approach. BBHs are more likely to form when the star and the incoming BH encounter ﬁrst with an impact parameter smaller than the binary’s semimajor axis. In this case, the star is frequently disrupted. On the other hand, when the two BHs encounter ﬁrst, frequent consequences are an orbit perturbation of the original binary or a binary member exchange. For the parameters chosen in this study, BBH formation, accompanied by stellar disruption, happens in roughly 1 out of 4 encounters. The close correlation between BBH formation and stellar disruption has possible implications for EM counterparts at the binary’s merger. The BH that disrupts the star is promptly surrounded by an optically and geometrically thick disk with accretion rates exceeding the Eddington limit. If the debris disk cools fast enough to become long-lived, EM counterparts can be produced at the time of the BBH merger.
\end{abstract}


%%%%%%%%%%%%%%%%%%%%%%%%%%%%%%%

\section{Introduction}\label{sec:intro}

Dynamical interactions between stars and the compact objects they leave behind play an important role in dense environments, such as globular and nuclear star clusters and disks of Active Galactic Nuclei (AGNs). On global, large scales they can influence cluster thermodynamics \citep{Hut1992}, while on local, small-scales close interactions can alter the original birth composition of isolated stars and binaries.

Dynamical formation of binary black holes (BBHs)  is one of the leading pathways  \citep[e.g.,][]{Downing+2010,Portegieszwart2000,Samsing2014,Rodriguez2015,Antonini2016,Askar+2017,Banerjee2018,Perna2019,Fragione2019,DiCarlo+2019,Rodriguez+2019,ArcaSedda+2020,Mapelli2021} to forming the binaries which have been observed via gravitational wave (GW) emission at their merger by the LIGO and Virgo observatories \citep{LIGO2021}.
Subsequent dynamical encounters between BBHs and tertiary BHs can further influence the orbital parameters of the binaries, and hence their timescale to merger by GW radiation \citep[e.g.,][]{Trani+2019,Samsing+2020,Wang2021,ArcaSedda+2021}.

Despite the relatively high fractions of stars and compact objects in binaries, hydrodynamic simulations of close encounters involving binaries have begun only recently.  \citet{Lopez2019} and \citet[][\citetalias{Ryu+2022} in the following]{Ryu+2022} studied close encounters between BBHs and single stars. They found that, in addition to altering the spin of the accreting BHs, tidal disruption events (TDEs) can have a significant impact on the  binary BH's orbit, in ways which can be quantitatively different than the case of pure scattering. The EM signatures produced by these close encounters can also differ significantly from those of TDEs by isolated BHs: depending on the geometry of the encounter, the accretion rate can display periodic modulations with the orbital period. Detections of such events can provide constraints on the formation of BBH mergers \citep{Samsing2019}.

More recently, \citet[][\citetalias{Ryu+2023} in the following]{Ryu+2023} performed the first investigation of close encounters between binary stars and single BHs. Their hydrodynamic simulations showed  a variety of possible outcomes, from full disruptions of both stars, to a full disruption of one star and a partial disruption of the other, to dissociation into bound and unbound single stars. Among these cases of dissociation, interesting outcomes include the formation of a runaway star, and of a fast-moving BH that accretes the tidally disrupted debris of the other star. In other outcomes, the binary stars are dissociated, and one of the stars is exchanged with the intruding BH, resulting in the formation of an X-ray binary.

Here we extend the line of investigation begun in \citetalias{Ryu+2022} and \citetalias{Ryu+2023} by performing a suite of hydrodynamic simulations of nearly parabolic close encounters between BH-star binaries and single BHs. Similarly to \citetalias{Ryu+2023}, we use the moving-mesh code {\small AREPO} \citep{Arepo,ArepoHydro,Arepo2}, whose quasi-Lagrangian approach to hydrodynamics is well adapted to the problem. Our study aims to elucidate how such encounters can lead to a variety of outcomes, including EM transients due to the disruption of the star and the formation (via member exchange) of tight BBHs surrounded by debris material, potentially leading to a situation in which the BBH merger could be accompanied by an EM counterpart.

Our paper is organized as follows: \S~\ref{sec:rate} presents the estimate for the rate of this type of encounters in globular clusters. \S~\ref{sec:method} describes the details of the numerical simulations and the initial conditions. We present our simulation results in \S~\ref{sec:results}, with particular emphasis on a classification of the outcomes and its dependence on encounter parameters. We discuss these results in the context of their possible EM counterparts in \S~\ref{sec:discussion}, and we finally summarize our work and conclude in \S~\ref{sec:conclusion}.

\section{Encounter rate in globular clusters}\label{sec:rate}

We begin by estimating the encounter rate of three-body interactions between BH-star binaries and single stars in globular clusters. Following \citetalias{Ryu+2023}, we first calculate the differential rate of a single BH encountering a BH-star binary as ${\rm d}\mathcal{R}/{\rm d} N_{\bullet}\simeq n\Sigma v_{\rm rel}$. Here, $n$ is the binary number density in the vicinity of the BH, $n \simeq f_{\rm b}n_{\rm s}$, where $f_{\rm b}$ is the non-interacting star - BH binary fraction, $f_{\rm b}\simeq 10^{-4}-10^{-5}$ \citep{Morscher+2015,Kremer+2018}, and $n_{\rm s}$ is the number density of stellar-mass objects near the center of the clusters. The variable $v_{\rm rel}$ represents the relative velocity between the binary and the BH while $\Sigma$ is the encounter cross-section. For $v_{\rm p}=\sqrt{{2G (\Mbh+ \mstar)}/{r_{\rm p}}}\gg \sigma$, we can write $\Sigma \simeq  \uppi G (\Mbh+ \mstar) r_{\rm p}/\sigma^{2}$ where $\sigma$ is the velocity dispersion. Adopting our results that strong encounters occur when $r_{\rm p}< a$, and assuming that this relation applies to binaries of any size and mass ratio, we can approximate $\Sigma \simeq \uppi  G (\Mbh+ \mstar) a\sigma^{-2}$. Then, we find that ${\rm d}\mathcal{R}/{\rm d}N_{\bullet}$ can be expressed as
\begin{align}\label{eq:rate}
   \frac{{\rm d}\mathcal{R}}{{\rm d} N_{\bullet}} & \simeq \frac{n  \uppi G (\Mbh+ \mstar) a}{\sigma},\nonumber\\
              &\simeq 4 \times 10^{-13} \yr^{-1} \left(\frac{f_{\rm b}}{10^{-4}}\right) \left(\frac{n_{\rm s}}{10^{5}{\rm pc}^{-3}}\right)\left(\frac{\Mbh+ \mstar}{20\Msol}\right)\nonumber\\
              &\times\left(\frac{a}{100\Rsol}\right) \left(\frac{\sigma}{15\km\sec}\right)^{-1}.
\end{align}
Assuming more than a few tens of single stellar-mass black holes exist in dense clusters at the present day \citep{Morscher+2015,Askar+2018,Kremer+2018}, and $\simeq$150 globular clusters in the Milky Way \citep{Harris+2010}, the rate of strong three-body encounters per Milky Way-like galaxy is, 
\begin{align}\label{eq:rate1}
   \mathcal{R}& \simeq 6\times 10^{-9} \yr^{-1} \left(\frac{N_{\bullet}}{15000}\right)\left(\frac{f_{\rm b}}{10^{-4}}\right) \left(\frac{n_{\rm s}}{10^{5}{\rm pc}^{-3}}\right)\left(\frac{\Mbh+ \mstar}{20\Msol}\right)\nonumber\\
              &\times\left(\frac{a}{100\Rsol}\right) \left(\frac{\sigma}{15\km\sec}\right)^{-1}.
\end{align}
Two outcomes that can be produced in this type of encounters are EM transients due to disruption of the star and formation of BBHs. In particular, because BBHs would likely form in $\simeq 25\%$ of all encounters (see \S~\ref{subsec:dependence}), the rate $\mathcal{R}$ of \textit{BBH-forming} events would be $O(10^{-9})\yr^{-1}$. The rate for encounters involving massive stars would be relatively high, compared to that for low-mass stars before all the massive stars turn into compact objects. However, as discussed in \S\ref{subsec:variety}, the total number of this type of encounters over the full cluster lifetime would be higher for less massive stars because of their longer lifetime and higher abundance. Note that $f_{\rm b}$ depends on cluster parameters such as the initial binary fraction and the cluster age \citep{Morscher+2015}, and calculating $\mathcal{R}$ requires a detailed modeling of cluster evolution, as well as the star formation history. Thus, for a more precise estimate of $\mathcal{R}$ a more careful consideration of cluster evolution history is required.

\section{Simulation details}\label{sec:method}

\subsection{Numerical methods}

Our numerical methods and setup are essentially the same as in \citetalias{Ryu+2023}. We perform a suite of 3D hydrodynamic simulations of close encounters using the massively parallel gravity and magnetohydrodynamics moving-mesh code {\small AREPO} \citep{Arepo,ArepoHydro,Arepo2}, which combines advantages of the two conventional hydrodynamical schemes, the Eulerian finite-volume method and the Lagrangian smoothed particle method, such as shock capturing without introducing an artificial viscosity, low advection errors, an efficient tracking of supersonic flow, and an automatically adaptive adjustment of spatial resolution. We use the {\small HELMHOLTZ} equation of state \citep{HelmholtzEOS} which accounts for radiation pressure, assuming local thermodynamic equilibrium. We include $8$ isotopes ($\mathrm{n}$, $\mathrm{p}$, $^{4}\mathrm{He}$, $^{12}\mathrm{C}$, $^{14}\mathrm{N}$, $^{16}\mathrm{O}$, $^{20}\mathrm{Ne}$, $^{24}\mathrm{Mg}$, \citealt{Pakmor+2012}). We follow the advection of the elements which are then used for the update of the thermodynamics quantities (e.g., pressure). We do not follow the nuclear reactions, which should be fine given the short duration of the simulations and the reaction rates expected for the temperatures and densities that occur in our simulations.

\begin{figure}
	\centering
	\includegraphics[width=8.6cm]{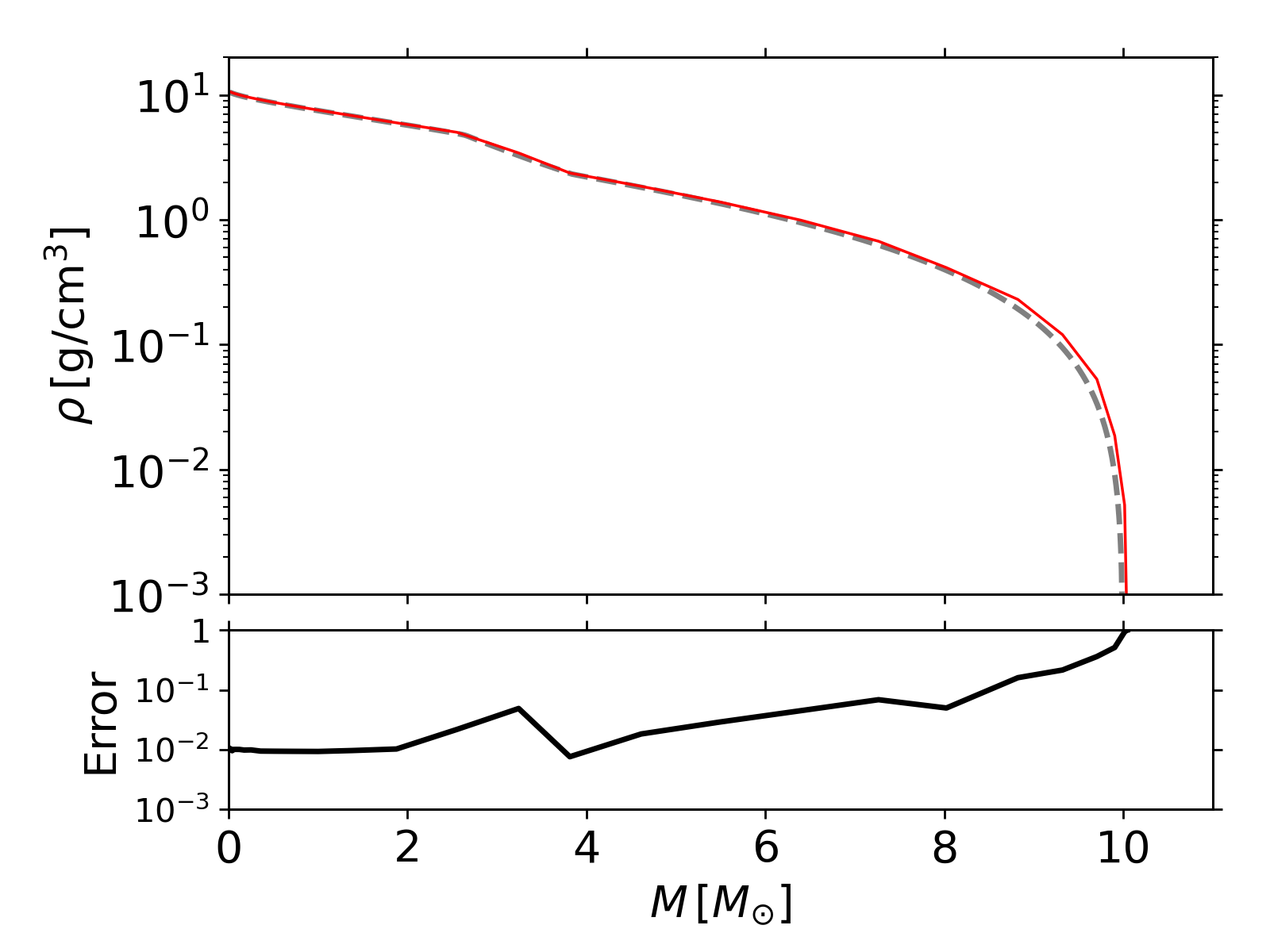}
\caption{\textit{Top} panel: The radial density profile of the main-sequence stars with $M_{\star}=10\Msol$ (red) relaxed for five stellar dynamical times. \textit{Bottom} panel: The relative error with respect to the MESA model, as a function of mass. The dashed grey line in the \textit{top} panel indicates the profiles for the {\sc MESA} model, which are just sitting below the solid line.}
	\label{fig:density}
\end{figure}

\subsection{Stellar model}

The initial state of the star was taken as an evolved main-sequence (MS) star computed using the stellar evolution code {\sc MESA} (version r22.05.1) \citep{Paxton+2011,paxton:13,paxton:15,MESArelaxation,paxton:19,jermyn22}. The star has an initial mass of $10\, {\rm M}_\odot$ and a metallicity $Z=0.006$, { which is lower than solar and {consistent with what is found for} globular clusters \citep[e.g., ][]{vandenberg_ages_2013}, whose high stellar density facilitates dynamical interactions.}\footnote{{Note that the exact value of the metallicity does not affect our main results because the two most often cases in our simulations are fly-bys of stars around the BHs, or near-collisional disruptions of stars, and the outcomes of these two cases are not significantly affected by a slight change in the stellar internal structure due to a different metallicity.}}. Convection is modeled according to the mixing length theory with a mixing length parameter of $1.5$. We use the \citet{ledoux_stellar_1947} criterion to determine the boundary of the convective regions and include an exponential overshoot prescription \citep{herwig_evolution_2000} with parameters $f = 0.014$ and $f_0 = 0.004$. We treat semiconvection as in \citet{langer_semiconvective_1983, langer_evolution_1985} assuming it is fully efficient. Wind mass loss is modeled with the prescription from \citet{vink_mass-loss_2001}. { We evolve the star until about halfway through the main sequence, which we choose as the time when the central hydrogen mass fraction drops to 0.3, at which point the star has developed a $2.6\Msol$ convective helium core with a radius of $0.8\Rsol$}. The stellar radius is $R_{\star}=5.4\Rsol$, and the central density is $\simeq 11\,{\rm  g\,cm^{-3}}$. Since we evolve the stellar model as a single star, we neglect possible past interactions that could have affected the structure. For example, if the black hole binary system formed through binary evolution, the star may have accreted mass \citep[e.g., ][]{renzo_evolution_2021}.  If the system formed through  dynamical capture, the star may have lost mass.  We expect such effects to be small and not affect our results significantly. 

We use the {\sc MESA} stellar model as initial state for the {\small AREPO} simulation. After mapping the 1D {\sc MESA} model into a 3D {\small AREPO} grid with $N\simeq 5\times10^{5}$ cells, the 3D single star is first relaxed. It usually takes up to five stellar dynamical time  until it is fully relaxed. The stellar dynamical time is defined as $\sqrt{R_{\star}^{3}/GM_{\star}}$ where $R_{\star}$ and $M_{\star}$ are the radius and mass of the star, respectively.  Note that we increase the resolution for each single star by almost a factor of 2 compared to that in \citetalias{Ryu+2023}, showing the results were converged with $N\gtrsim 2.5\times10^{5}$. This is to more conservatively guarantee the convergence of our results, and to better resolve stars that may be partially disrupted during encounters. In addition to that, the resolution becomes finer over time by adopting refinement near the BHs (\S~\ref{subsec:refinement}): some simulations with violent interactions have $10^{7}$ cells at the end of the simulations. The density profiles of the relaxed stars considered in our simulations are depicted in Figure~\ref{fig:density}. {As shown in the figure, the relative {difference} of our 3D star with respect to the {\sc MESA} model is $1\%$ for the inner region up to $2\Msol$.} The match is better than  $10\%$ throughout most of the star, except for the surface. We expect that this is sufficient for the aims of this study.

\begin{figure*}
	\centering
	\includegraphics[width=8.6cm]{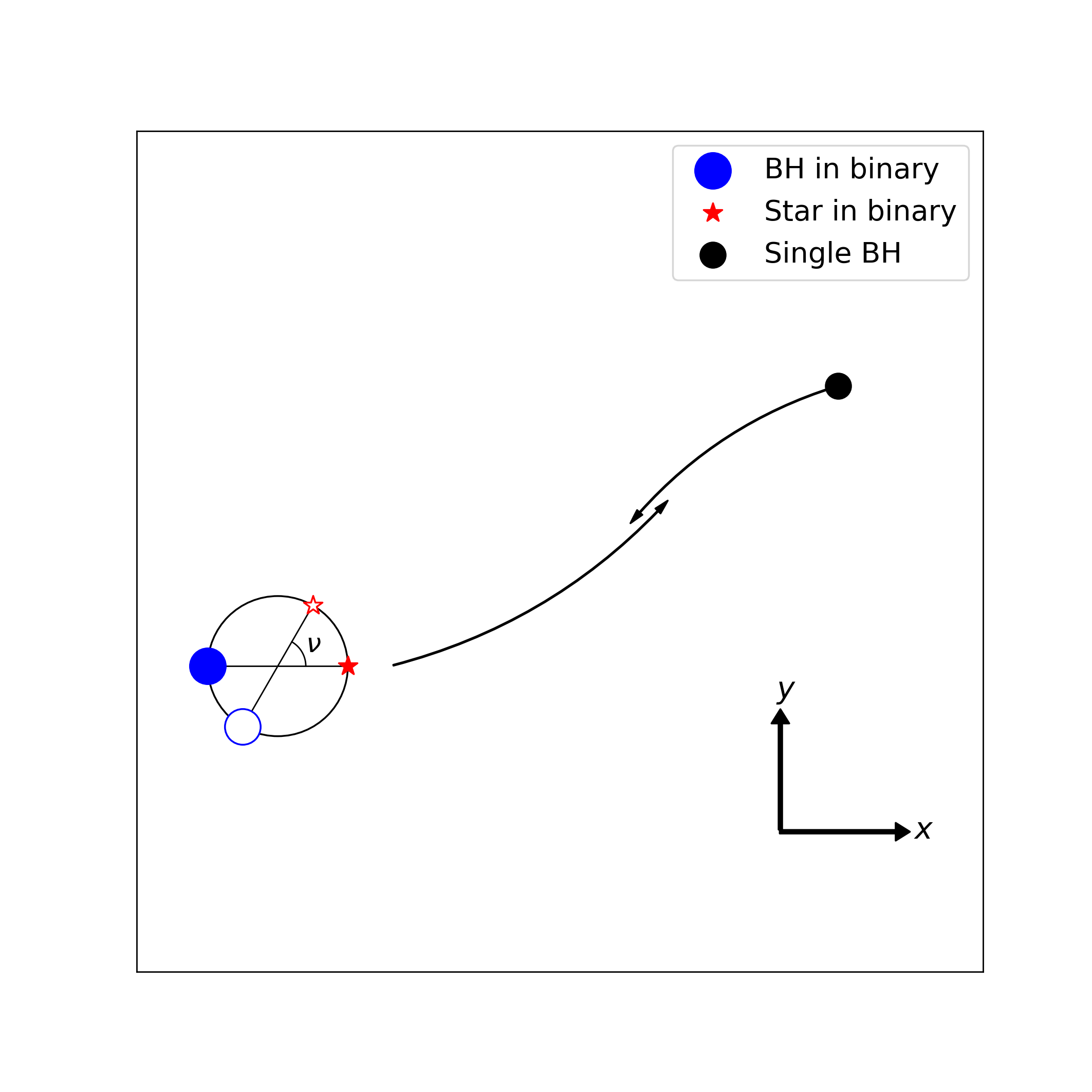}
		\includegraphics[width=8.6cm]{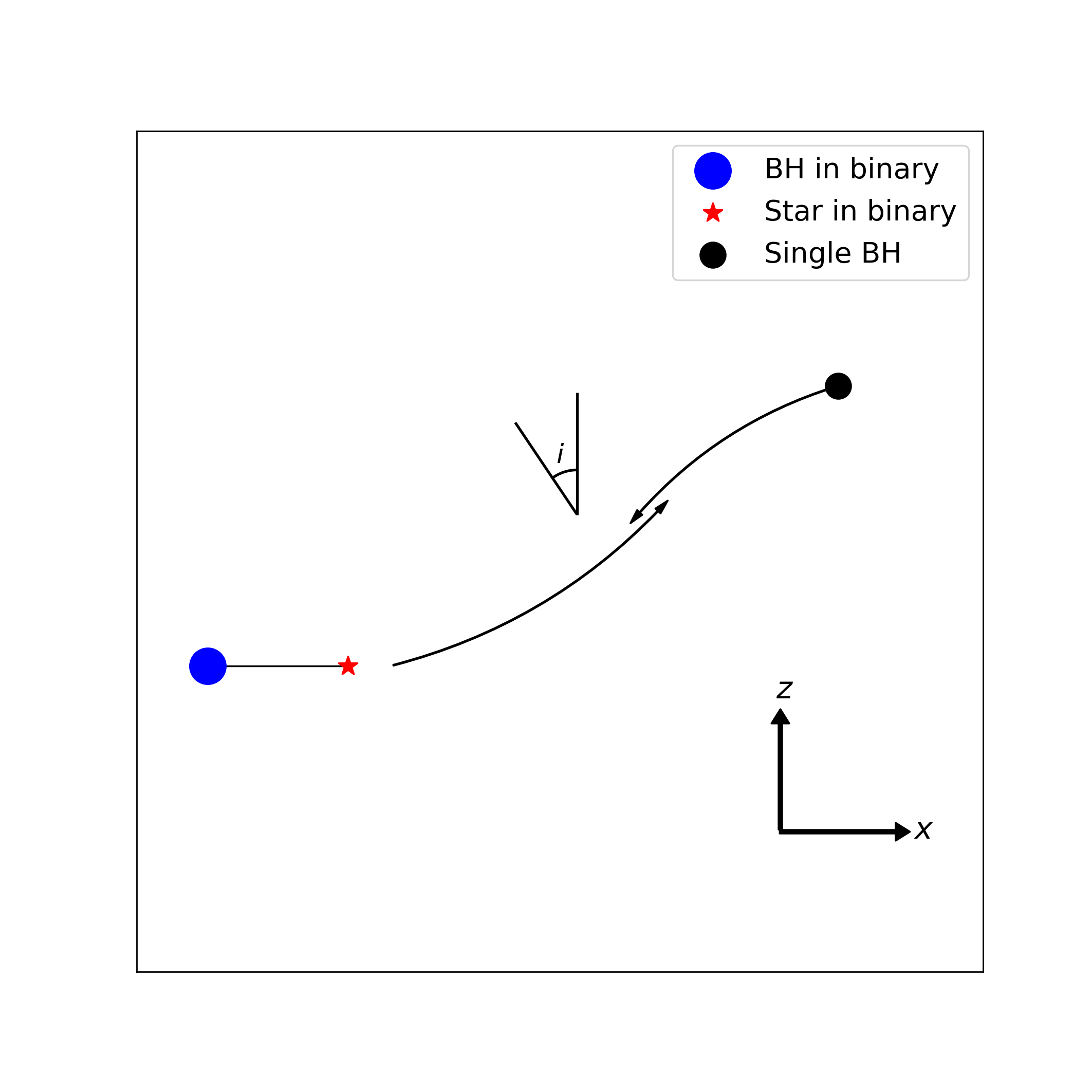}
\caption{Schematic diagrams for the initial configuration of the BH-star binary (blue solid circle and red solid star) and single BH (black circle) for a prograde case with an inclination angle $i<90^{\circ}$ and  phase angle $\phi=0^{\circ}$, projected onto the $x-y$ plane (\textit{left}) and the $x-z$ plane (\textit{right}). The arrows indicate the instantaneous direction of motion. The open symbols, on the same circle with the solid symbols, indicate the case with $\phi >0^{\circ}$.}
	\label{fig:init}
\end{figure*}

\subsection{Black holes}

As in \citetalias{Ryu+2023}, we model the BH using {a sink particle assuming it is not rotating initially}. It only interacts gravitationally with gas and grows in mass via accretion of gas. We set the gravitational softening length of the BH ($\simeq 0.01\Rsol$) to be ten times the minimum softening length of the cells of the stars. 

We follow the same procedure for accretion described in \citetalias{Ryu+2023}. However, we significantly improve the resolution near the BH using refinement (see \S~\ref{subsec:refinement}), which leads to more accurate estimates of the accretion rate with stricter conditions for accretion than in \citetalias{Ryu+2023}. We search for cells bound to the BH (i.e.~negative orbital energy relative to the BH) within $10^{3\,}r_{\rm g}$ (c.f., $1.5\times10^{4}\, r_{\rm g}$ in \citetalias{Ryu+2023}) where $r_{\rm g} = G M_{\bullet}/c^{2}$ is the gravitational radius of the BH and $M_{\bullet}$ denotes the mass of the BH. We still apply the same inverse-distance kernel \citep{MonaghanLattanzio1985} to put more weight onto closer cells. Although {the change in the momentum and the mass of the BH due to accretion is taken into account}, our simulations do not include potential radiative feedback produced by accretion.

\subsection{Mesh refinement}\label{subsec:refinement}
The simulation code can adjust the local mesh resolution by adaptively splitting or merging cells if certain prescribed refinement criteria are satisfied (for more details, see \S~6 in \citealt{Arepo}). We apply the refinement technique to cells in the vicinity of each BH to better resolve the stream structure there. At every time step, the code refines cells near a BH if all of the following conditions are met:
\begin{enumerate}
    \item the distance from the BH fulfils $r< 5000~r_{\rm g}$,
    \item the cell density is $\rho>2\times 10^{-4}\gram\cm^{-3}$,
    \item the cell mass exceeds $>6\times10^{22}\gram$,
    \item and $\Delta d/r > 0.26$ for $500~r_{\rm g} < r < 5000~r_{\rm g}$ and $\Delta d > 500\, r_{\rm g}$ for $r< 500\,r_{\rm g}$, where $\Delta d$ is the cell size.
\end{enumerate}
{The refinement radius in condition (i) is chosen to be larger than the accretion radius ({$1000~r_{\rm g}$ in this work}) to ensure that gas streams inside the accretion radius are well resolved. }
Condition (ii) is designed to apply the refinement only to the cells that represent ``real'' gas, not vacuum regions. Criterion (iii) avoids a runaway creation of low-mass cells. Finally, the resolution limit imposed through condition (iv) guarantees that there are at least $O(10^{2})$ cells within the accretion radius. On the other hand, at every time step, the code can also derefine cells  within $r< 5000\,r_{\rm g}$ around each BH if the cell mass is $<1.5\times10^{22}\gram$, meaning the mass of cells within $r< 5000\,r_{\rm g}$ never becomes smaller than this mass resolution limit.

We ran a few simulations with five different resolution limits within the range $0.05\leq \Delta d \leq 0.4$. This confirmed that the global evolution of the systems, such as their final interaction outcomes, is not affected by the refinement. The accretion rate has converged when the cell size fulfils $\Delta d/r <0.3$. Note that the number of cells within a volume at distance $r$ increases approximately by a factor of 8 when $\Delta d$ decreases by a factor of 2.

\subsection{Star -- black hole binary}\label{subsec:binary}
Before we carry out our encounter experiments, we relax binaries consisting of a fully relaxed star and a BH for $10$ stellar dynamical times. We parameterize the binary's semimajor axis $a$ using an approximate analytic estimate of the Roche lobe radius \citep{Eggleton1983}, 
\begin{align}
   \frac{r_{\rm RL}}{a}= \frac{0.49\, q^{2/3}}{0.6\,q^{2/3}+\ln(1+q^{1/3})},
\end{align}
where $r_{\rm RL}$ is {the volume averaged Roche lobe radius of the star}, $q=M_{\star}/M_{\bullet}$ is the mass ratio,  and $a$ is the orbital separation. We define $a_{\rm RL} \equiv a(R_{\rm RL}=R_{\star})$ as the separation at which the star fills its Roche lobe. For $q=0.5$ and $r_{\rm RL} = R_{\star}$,  $a_{\rm RL} \simeq 3.12\,R_{\star}\simeq 16.9\Rsol$. 

We have performed this binary relaxation process for every binary with different orbital parameters (3~different binaries in total). The semimajor axis and the eccentricity of the relaxed binaries differ by less than 1\% from their initial values.

\begin{table*}
\begin{tabular}{ c c c c c c c  c c c c c c c} 
\hline
Model number & Model name & \multicolumn{2}{c}{$a$}  &\multicolumn{2}{c}{$b$} & $\phi~[^{\circ}]$  &  $i$  & $t_{\rm p}$ & $P$ & $v_{\rm orb}$\\
\hline
\multicolumn{2}{c}{Unit} & $a_{\rm RL}$ & $\Rsol$ & - & $\Rsol$ & $^{\circ}$ & $^{\circ}$ & hours & days & ${\rm km\,s^{-1}}$ \\
\hline
1 & $a4b2\phi0i30$   & 4 & 67.5 & 2 & 67.5 & 0  & 30 & 50 & 12 & 291 \\
2 & $a4b1\phi0i30$   & 4 & 67.5 & 1  & 33.8 & 0 & 30  & 18  & 12 & 291 \\
3 & $a4b1/2\phi0i30$   & 4 & 67.5 & 1/2  & 16.9 & 0  & 30  & 6.3 & 12 & 291 \\
4 & $a4b1/4\phi0i30$  & 4 & 67.5 & 1/4  & 8.45 & 0  & 30 & 2.3& 12 & 291 \\
5 & $a4b2\phi180i30$   & 4 & 67.5 & 2 & 67.5  & 180 & 30 & 50& 12 & 291 \\
6 & $a4b1\phi180i30$   & 4 & 67.5 & 1  & 33.8 & 180 & 30  & 18& 12 & 291 \\
7 & $a4b1/2\phi180i30$   & 4 & 67.5 & 1/2  & 16.9 & 180 & 30 & 6.3 & 12 & 291 \\
8 & $a4b1/4\phi180i30$  & 4 & 67.5 & 1/4  & 8.45 & 180 & 30   & 2.3& 12 & 291 \\
\hline
9 & $a4b2\phi0i150$   & 4 & 67.5 & 2  & 67.5   & 0 & 150 & 50& 12 & 291 \\
10 & $a4b1\phi0i150$   & 4 & 67.5 & 1 & 33.8   & 0  & 150  & 18 & 12 & 291 \\
11 & $a4b1/2\phi0i150$   & 4  & 67.5 & 1/2 & 16.9 & 0  & 150  & 6.3 & 12 & 291 \\
12 & $a4b1/4\phi0i150$   & 4 & 67.5  & 1/4  & 8.45  & 0  & 150 & 2.3 & 12 & 291 \\
13 & $a4b2\phi180i150$   & 4 & 67.5  & 2  & 67.5 & 180  & 150   & 50& 12 & 291 \\
14 & $a4b1\phi180i150$   & 4 & 67.5  & 1 & 33.8  & 180  & 150 & 18 & 12 & 291 \\
15 & $a4b1/2\phi180i150$  & 4 & 67.5  & 1/2 & 16.9  & 180  & 150   & 6.3 & 12 & 291 \\
16 & $a4b1/4\phi180i150$   & 4 & 67.5  & 1/4 & 8.45  & 180  & 150  & 2.3 & 12 & 291 \\
\hline
17 & $a2b1/2\phi0i30$  & 2  & 33.7  & 1/2  & 8.44 & 0 & 30  & 1.2  & 4.1 & 412\\
18 & $a2b1/2\phi180i30$  & 2  & 33.7 & 1/2  & 8.44 & 180 & 30   & 1.2 & 4.1 & 412\\
19 & $a2b1/2\phi0i150$   & 2 & 33.7 & 1/2   & 8.44 & 0  & 150 &1.2  & 4.1 & 412\\
20 & $a2b1/2\phi180i150$   & 2 & 33.7 & 1/2   & 8.44 & 180  & 150 &1.2 & 4.1 & 412 \\
21 & $a6b1/2\phi0i30$   & 6 & 101 & 1/2  & 25.3 & 0 & 30   & 12  & 22 & 238\\
22 & $a6b1/2\phi180i30$   & 6 & 101 & 1/2  & 25.3 & 180 & 30 & 12  & 22& 238\\
23 & $a6b1/2\phi0i150$   & 6  & 101 & 1/2  & 25.3  & 0 & 150  & 12  & 22& 238\\
24 & $a6b1/2\phi180i150$   & 6 & 101 & 1/2  & 25.3 & 180 & 150  & 12  & 22& 238\\
\hline
25 & $a4b1/2\phi0i0$  & 4 & 67.5  & 1/2 & 16.9   & 0 & 0  & 6.3 & 12 & 291 \\
26 & $a4b1/2\phi0i60$   & 4 & 67.5 & 1/2 & 16.9  & 0 & 60 & 6.3 & 12 & 291 \\
27 & $a4b1/2\phi0i120$   & 4 & 67.5 & 1/2  & 16.9  & 0 & 120 & 6.3 & 12 & 291 \\
28 & $a4b1/2\phi180i0$  & 4 & 67.5  & 1/2  & 16.9  & 180 & 0 & 6.3 & 12 & 291 \\
29 & $a4b1/2\phi180i60$   & 4 & 67.5  & 1/2 & 16.9  & 180 & 60 & 6.3 & 12 & 291 \\
30 & $a4b1/2\phi180i120$   & 4 & 67.5 & 1/2 & 16.9  & 180 & 120 & 6.3 & 12 & 291 \\
\hline
31 & $a4b1/2\phi45i30$   & 4 & 67.5  & 1/2 & 16.9  & 45  & 30   & 6.3 & 12 & 291 \\
32 & $a4b1/2\phi90i30$   & 4 & 67.5  & 1/2 & 16.9  & 90 & 30  & 6.3 & 12 & 291 \\
33 & $a4b1/2\phi135i30$  & 4 & 67.5  & 1/2 & 16.9  & 135 & 30  & 6.3 & 12 & 291 \\
34 & $a4b1/2\phi225i30$   & 4 & 67.5   & 1/2 & 16.9  & 225  & 30   & 6.3 & 12 & 291 \\
35 & $a4b1/2\phi270i30$   & 4 & 67.5  & 1/2 & 16.9  & 270 & 30  & 6.3 & 12 & 291 \\
36 & $a4b1/2\phi315i30$  & 4 & 67.5  & 1/2 & 16.9  & 315 & 30  & 6.3 & 12 & 291 \\
\hline
\end{tabular}
\caption{The initial model parameters for encounters between a circular binary ($q=0.5$) with total mass of $30\Msol$ and a $10\Msol$ BH. The model name (second column) contains the information of key initial parameters: for the model names with the format $a(1)b(2)\phi(3)i(4)$, the numerical values encode (1) the initial semimajor axis of the binary $a/a_{\rm RL}$, (2) the impact parameter $b$, (3) the initial phase angle $\phi$ in degrees, and (4) the initial inclination angle $i$ in degrees.  Here, $a_{\rm RL} \simeq 3.12\,R_{\star}\simeq 16.9\Rsol$ is the separation when the star in the binary fills its Roche lobe. The last three columns show the dynamical time $t_{\rm p}$ at pericenter (see its definition in the text), in units of hours, the orbital period $P$ of the binary, and the relative velocity $v_{\rm orb}$ of the binary members. }\label{tab:initialparameter}
\end{table*}

\subsection{Initial conditions}\label{subsec:initial}

Following the same terminology as in \citetalias{Ryu+2023}, we refer to quantities with a subscript containing $b-\bullet$ as those relating to the orbit between a binary and a single BH. We assume a parabolic encounter with eccentricity $1-e_{b-\bullet}=10^{-5}$ between a single $10\Msol$ BH with a binary consisting of a $20\Msol$ BH and a $10\Msol$ star. The exact choices of the system parameters are somewhat arbitrary, but BHs with such masses have been found in X-ray binaries \citep[e.g.]{binder_wolfrayet_2021}. Encounters between objects of comparable masses are expected in the dense centers of young mass-segregated star clusters. We later discuss potential effects of different masses and orbits of encountering objects in \S~\ref{subsec:variety}, based on our simulation results.  We consider three semi-major axes for the initial binary systems: $a/a_{\rm RL}=2$, $4$ and $6$, corresponding to an orbital period of 4, 12, and 22 days, respectively.  We assume the binaries are circular at the start of our simulations. This is primarily to simplify the initial conditions, but this may not be unreasonable given that close binaries are often found to be circular \citep{Almeida+2017}.

The distance between the binary's center of mass and the BH at the first closest approach $r_{\rm p, b-\bullet}$ is parameterized using the impact parameter $b$, i.e., $r_{\rm p,b-\bullet}=0.5\, b a$ where $a$ is the binary semimajor axis. We consider $b = 1/4$, $1/2$, $1$, and $2$ for $a/a_{\rm RL}=4$, and $1/2$ for $a/a_{\rm RL}=2$ and $6$. The binary's angular momentum direction is always along the $z$-axis in our simulations. We illustrate the initial configuration of the stellar binary and the BH in Figure~\ref{fig:init}. 

We investigate the dependence of encounter outcomes on key encounter parameters, that is inclination angle $i=0$, $30^{\circ}$, $60^{\circ}$, $120^{\circ}$ and $180^{\circ}$, $b = 1/4$, 1/2, 1 and 2, and the phase angle $\phi=0^{\circ}-180^{\circ}$ with $\Delta\phi = 45^{\circ}$. We define $\phi$ as the initial angle between the line connecting the two members in the binary and the coordinate $x-$axis (see Figure~\ref{fig:init}). We start by studying the dependence on the two phase angles of the binary ($\phi=0^{\circ}$ and $180^{\circ}$) by fixing all the other parameters. To achieve this, we initially rotate the binary while the initial separation between the center of mass of the binary and the BH is fixed at $r=5\,a$. This allows us to examine the outcomes from the first contact of the single BH with a different member of the binary. However, given the relatively high computational costs, instead of simulating encounters with every combination of $i$ and $b$, we perform simulations for the encounters of the intermediate-size binaries ($a/a_{\rm RL}=4$) with different combinations of $b=1/4$, 1/2, 1 and 2, and $i=30^{\circ}$, $150^{\circ}$, and $\phi = 0^{\circ}$ and $180^{\circ}$. For the smallest and largest binaries ($a/a_{\rm RL}=2$ and 6), we only consider $i=30^{\circ}$ and $150^{\circ}$ while $b=1/2$. In addition, we further examine the dependence of $i$ on the outcome properties by considering $i=0$, $60^{\circ}$, $120^{\circ}$ and $180^{\circ}$ (for $b=1/2$). Last, we also study the impact of the phase angle $\phi$ on the encounter outcomes by simulating encounters with six additional phase angles ($\phi = 45^{\circ}$, $90^{\circ}$, $135^{\circ}$, $225^{\circ}$, $270^{\circ}$, and $315^{\circ}$).

In Table~\ref{tab:initialparameter}, we summarize the initial parameters considered in our simulations. Each of the models is integrated in time up to a few   $100\,t_{\mathrm p}$ as needed to identify the final outcomes. Here, $t_{\rm p}= \sqrt{r_{\rm p}^3/GM}$ is the dynamical time at $r=r_{\rm p}$, where $M$ is the total mass of the three objects ($40\Msol$). The value of $t_{\rm p}$ for each model is given in Table~\ref{tab:initialparameter}.

The total computational cost for each run varies, mainly depending on how long the interactions last until the final outcomes are produced. Using 200 - 300 CPU-cores of the Intel Xeon CascadeLake-AP processor (Xeon Platinum 9242), the total compute time per run has been around  70000 - 100000 core hours.

\begin{figure*}
	\centering
	\includegraphics[width=4.3cm]{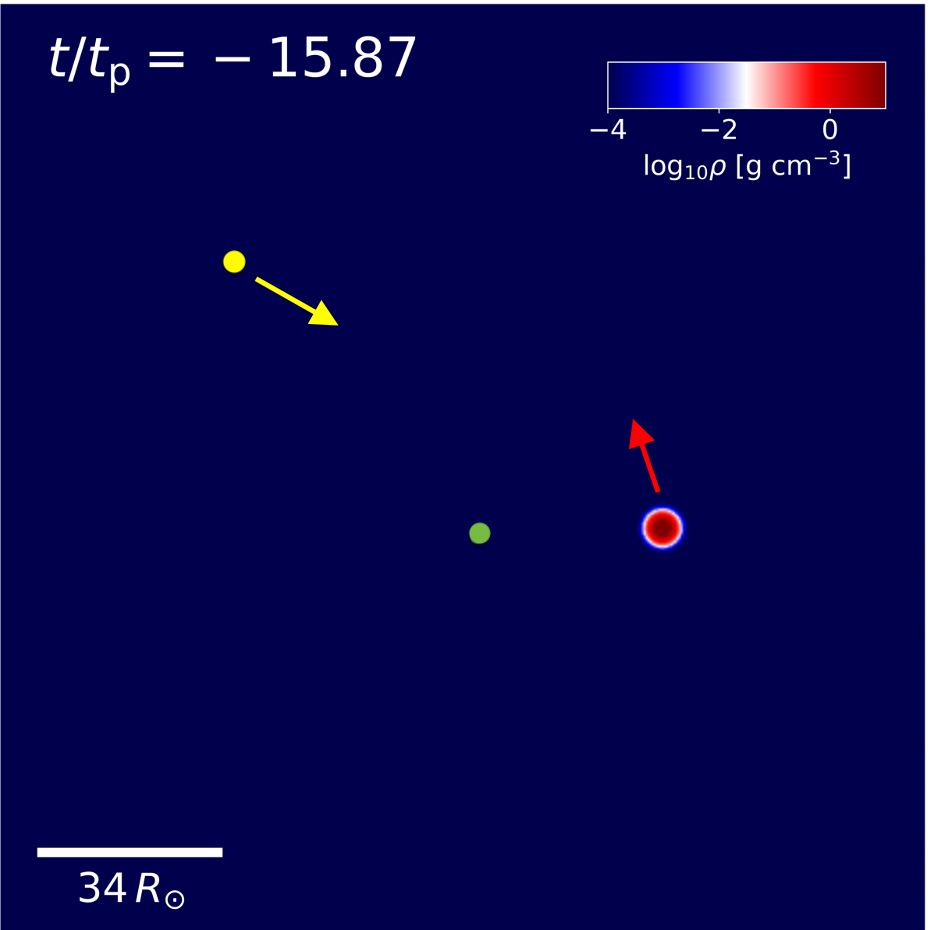}
	\includegraphics[width=4.3cm]{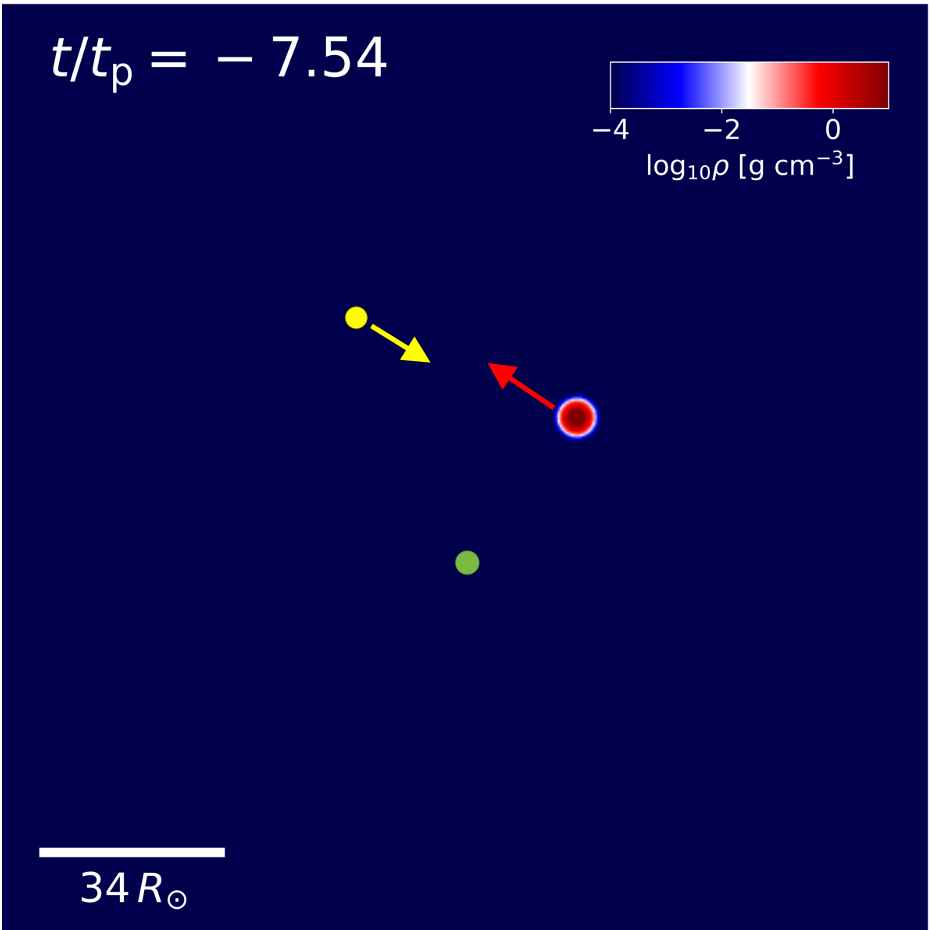}
	\includegraphics[width=4.3cm]{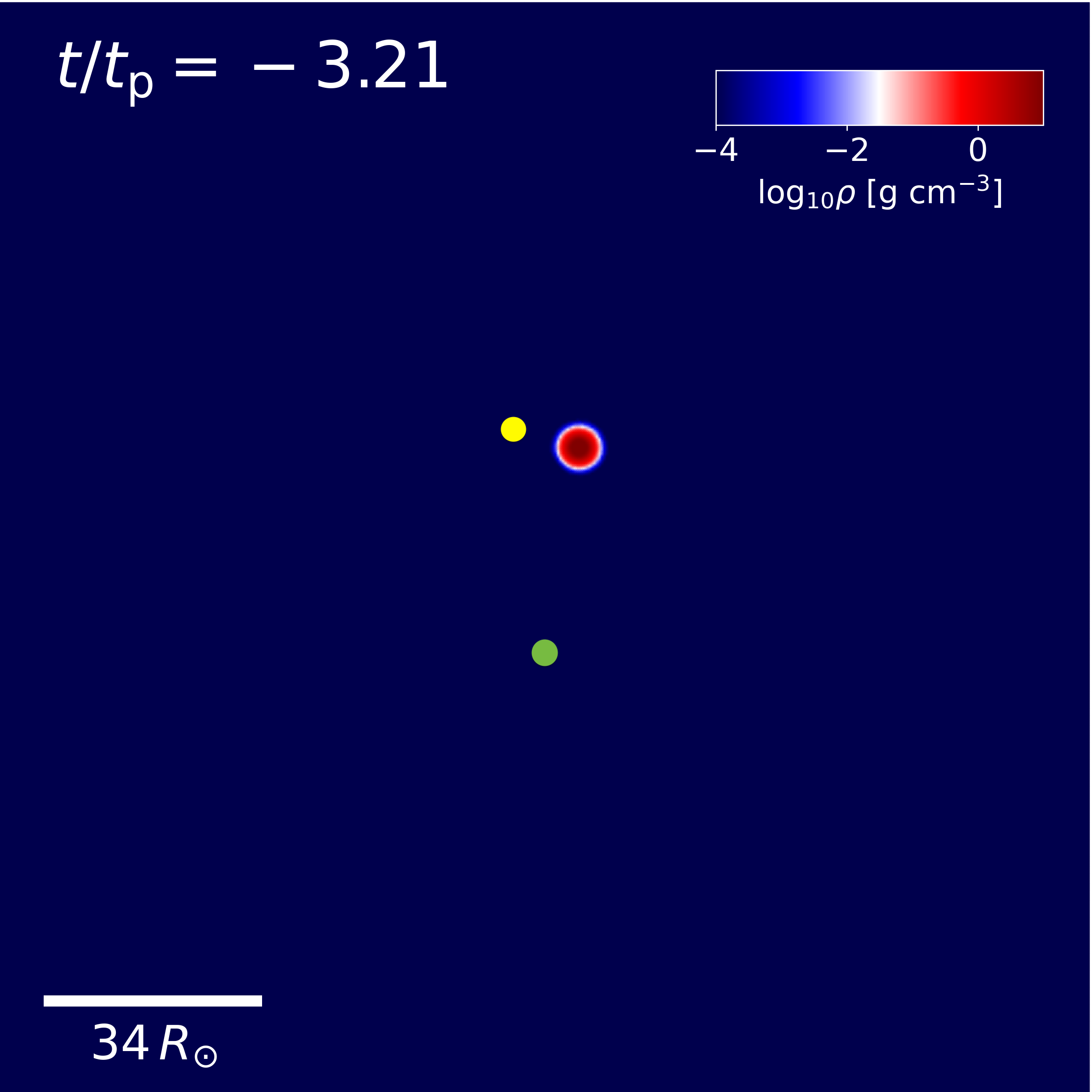}
	\includegraphics[width=4.3cm]{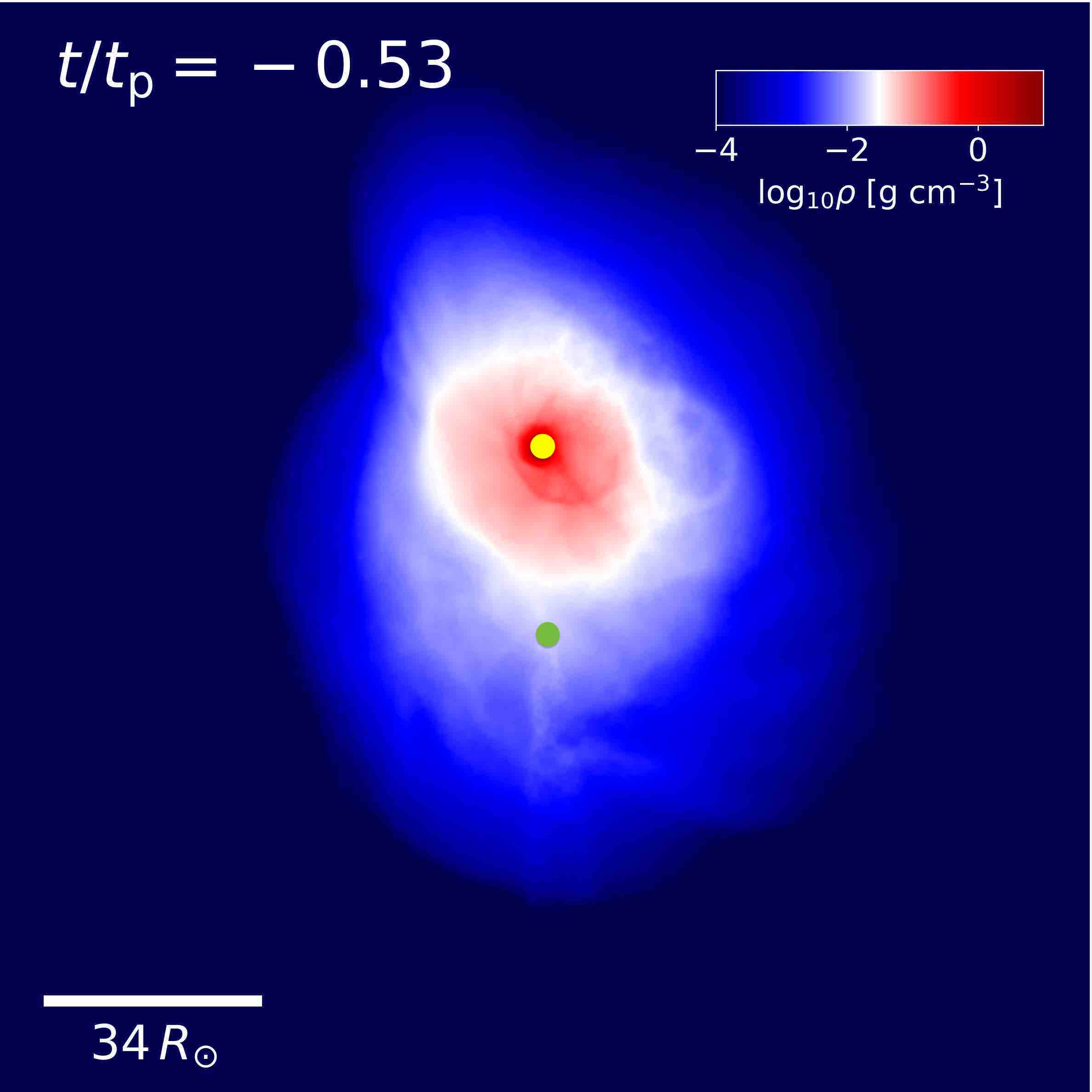}\\
	\includegraphics[width=4.3cm]{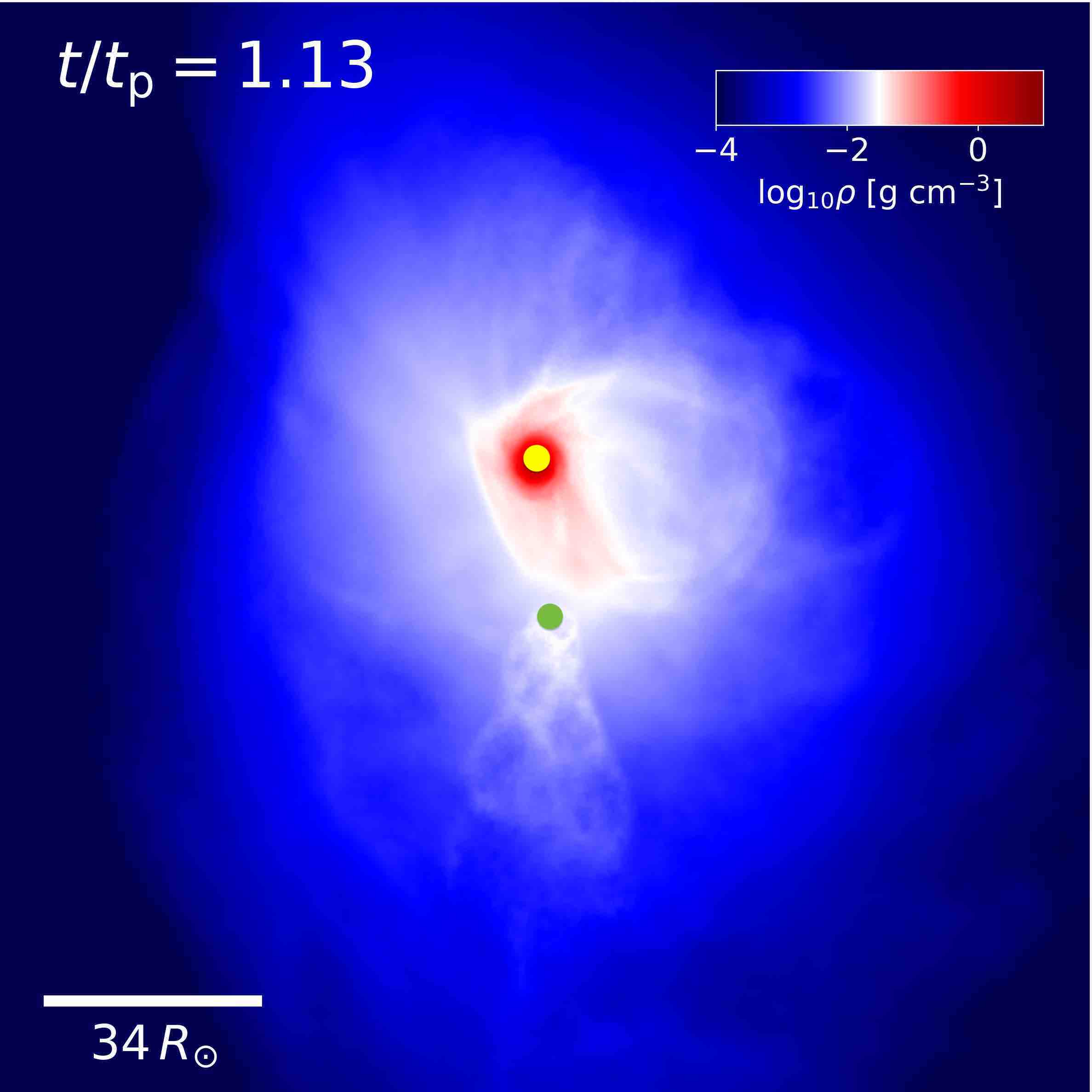}
	\includegraphics[width=4.3cm]{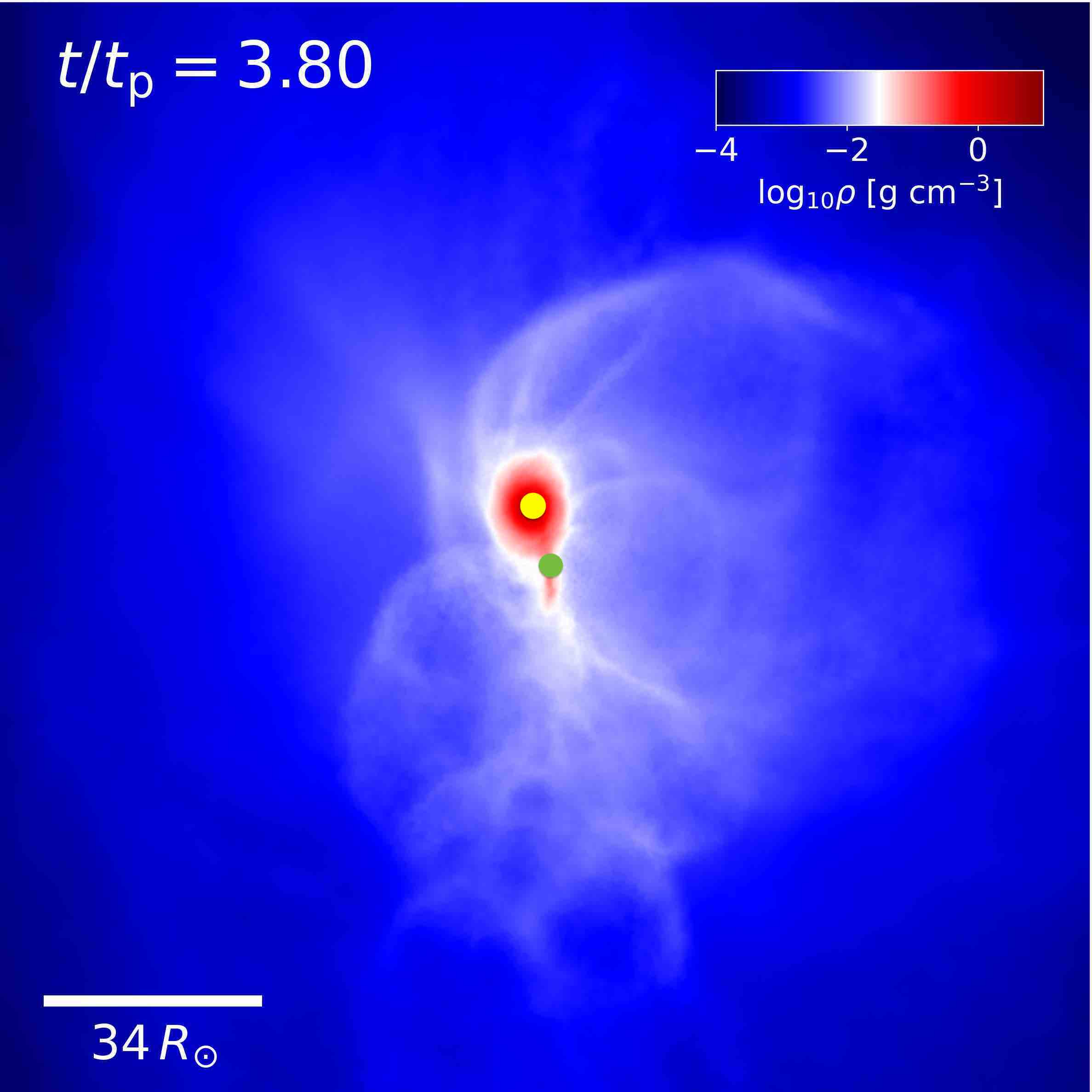}
	\includegraphics[width=4.3cm]{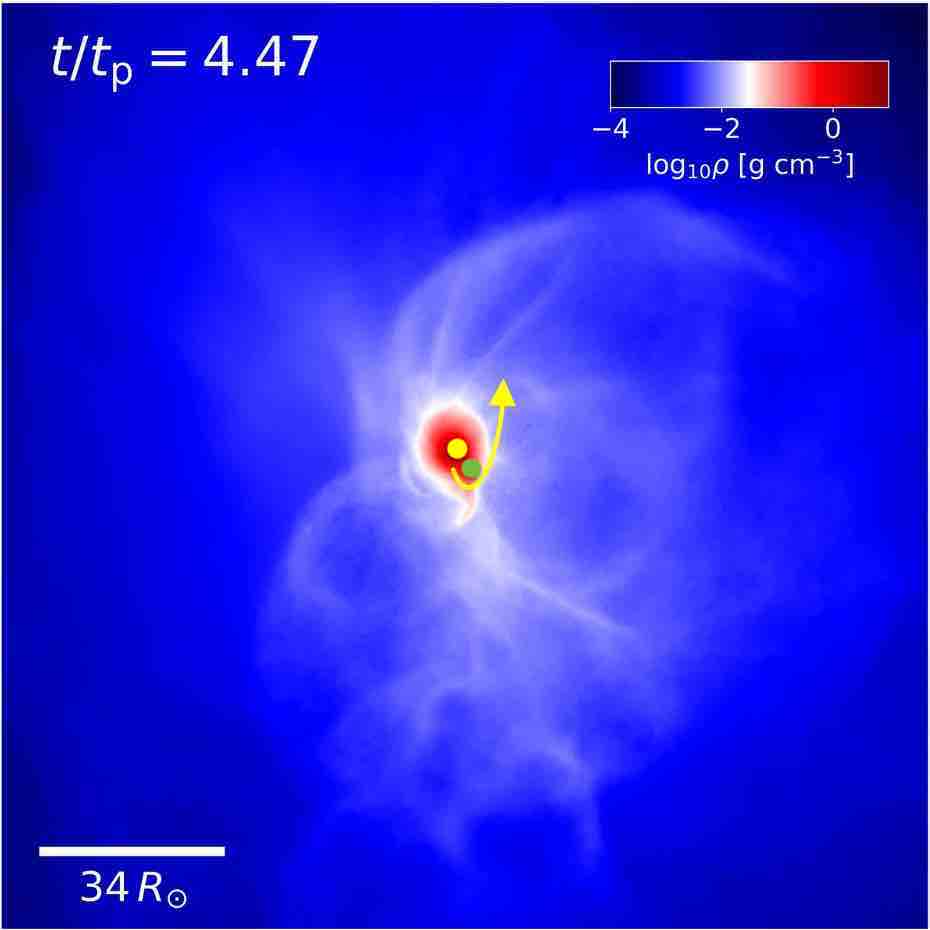}
    \includegraphics[width=4.3cm]{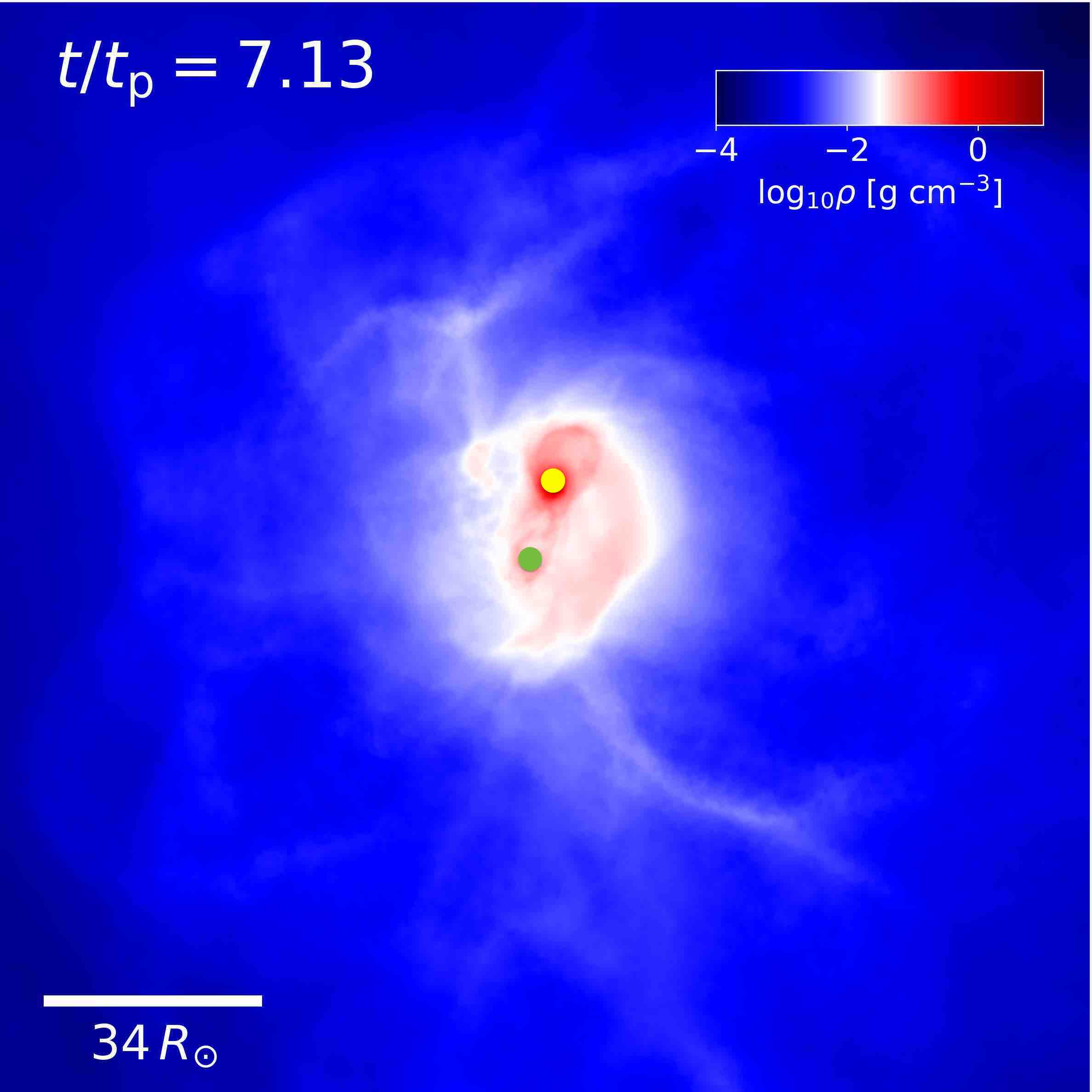}	
\caption{{An example of a \textit{BBH-forming} encounter, Model 20.$a2b1/2\phi180i150$, showing} the density distribution in the binary orbital plane at a few different times in units of $t_{\rm p}$. The color bar gives the logarithmic density in ${\rm g}\,{\rm cm}^{-3}$. The time is measured since the expected pericenter passage between the binary's center of mass and the single BH. At $t/t_{\rm p}\simeq-16$ (top-$1^{\rm st}$), the binary (star - green dot) and the single BH (yellow dot) approach each other. At $t/t_{\rm p}=-3.21$ (top-$3^{\rm rd}$), the incoming BH strongly encounters with the star in the binary, followed by the collision of the star (top-$4^{\rm th}$). The BH that disrupted the star is gravitationally captured by the other BH, forming a merging binary with $a\simeq 6.70\Rsol$ and $e\simeq 0.943$, corresponding to $t_{\rm GW}\simeq 10^{4}\yr$ (bottom panels). }
	\label{fig:example1}
\end{figure*}

\begin{table*}
\resizebox{\textwidth}{!}{
\begin{tabular}{ c c c c | c c c c c | c c } 
\hline
Model number & Model name & BBH?  & STAR? & Binary type & $a$  & $e$ & $\log_{10}t_{\rm GW}$ & $v$ & Single type & $v$\\
\hline
- & -  & - & - & -  & $\Rsol$ & - &  yr & km/s & - & km/s \\
\hline
1 & $a4b2\phi0i30^{\star}$      &  -   &   -    &  $\emptycirc(10)-\star(10)$  &  76.2   &  0.796   &   -   &   -  &  $\fullcirc$  &  - \\%factor of 3
2 & $a4b1\phi0i30^{\star}$     &  -  &  -   &  $\emptycirc(10)-\star(10)$   &  55.3  &  0.347   &  -   &   -    & $\fullcirc$  &   \\%factor of  6- 7
3 & $a4b1/2\phi0i30^{\star}$   &  -    &  -  &  $\emptycirc(10)-\star(10)$  &  110   &  0.352   &  -   &   -   &  $\fullcirc$  &  - \\%factor of 2
4 & $a4b1/4\phi0i30^{\star\star}$    &  -  &  -  &  -  &   -  &   -  &  -  &   -    &  -  &  - \\
5 & $a4b2\phi180i30$     &  Yes  &  Survived   &  $\fullcirc(20)-\emptycirc(10)$  &  63.3  &  0.1   &  12.2  &  56.0  &  $\star$  &  $167$ \\
6 & $a4b1\phi180i30$    &  Yes  &  Destroyed   &  $\fullcirc(20)\longrightarrow\emptycirc(10)$  &   257  &  0.685  &  13.3  &  61.0  &  -  &  - \\
7 & $a4b1/2\phi180i30$    &  Yes   &  Destroyed   &  $\fullcirc(20)\longrightarrow\emptycirc(10)$  &  114  &  0.518  &   12.5  &  31.3  &  -  & -  \\
8 & $a4b1/4\phi180i30$   &  Yes   &  Survived  &  $\fullcirc(20)-\emptycirc(10)$  &  41.9  &  0.727  &  9.91  &  60.5  &  $\star$  &  $183$ \\
\hline
9 & $a4b2\phi0i150$    &  No  &  Survived   &  $\fullcirc(20)-\star(10)$   &   62.0  &   0.156  &  -  &  57  &  $\emptycirc(10)$  &  172  \\
10 &  $a4b1\phi0i150$    &  No  &  Survived    &  $\fullcirc(20)-\star(10)$  &   42.2  &  0.724  &  -  &  67.8  &  $ \emptycirc(10)$   &  203 \\ %check
11 &  $a4b1/2\phi0i150$     &  No   &  Survived   &  $\fullcirc(20)-\star(10)$  &  50.3  &  0.722  &   -   &  57.1  &  $\emptycirc(10)$  &  170 \\ %check
12 &  $a4b1/4\phi0i150$    &  No    &  Survived   &  $\fullcirc(20)-\star(10)$   &  50  &  0.731  &  -  &  57.8  &  $ \emptycirc(10)$   &  172 \\ %check
13 &  $a4b2\phi180i150$    &  No    &  Survived   &  $\fullcirc(20)-\star(10)$  &   63.3  &   0.161  & - & 52.4  &   $\emptycirc(10)$  &  157 \\
14 &  $a4b1\phi180i150^{\star}$  &  -     &  -  &  $\fullcirc(20)-\star(10)$   &  66.1  &  0.857  &  -  & -  &  $\emptycirc(10)$  &  - \\
15 &  $a4b1/2\phi180i150$    &  Yes   &  Destroyed  &  $\fullcirc(20)\longleftrightarrow\emptycirc(10)$   &  12  &   0.661   &   8.031  &  38.6   & -  &  - \\
16 &  $a4b1/4\phi180i150^{\star}$     &  -   &  -  &  $\emptycirc(10)-\star(10)$  &  109  &   0.310  &  - &  -  &  $\fullcirc(20)$  &  - \\
\hline
17 &  $a2b1/2\phi0i30^{\star}$   &  -    &  -  &   $\emptycirc(10)-\star(10)$  &  18.0  &  0.396    &   -  &  -  &  $\fullcirc(20)$  & - \\
18 &  $a2b1/2\phi180i30$   &  Yes    &  Destroyed  &   $\fullcirc(20)\longleftrightarrow\emptycirc(10)$  &  61.0  &  0.406  &  11.7  &  64.0  &  -  &  - \\
19 &  $a2b1/2\phi0i150$     &  No   &  Destroyed &  -  &  -   &  -   &  -  &  -  &  $\fullcirc(20),\emptycirc(10)$  &  60.8, 234 \\%done
20 &  $a2b1/2\phi180i150$     &  Yes  &  Destroyed   &   $\fullcirc(20)\longleftrightarrow\emptycirc(10)$   &  6.70  &  0.943  &  4.36  &  47.6  & -  &  - \\
21 &  $a6b1/2\phi0i30^{\star\star}$     &  -  &  -  &  -  &  -  &   -  &  -  &  -  & - \\ %need to confirm
22 &  $a6b1/2\phi180i30$    &  Yes   &  Destroyed  &   $\fullcirc(20)\longleftrightarrow\emptycirc(10)$   &   354  &  0.787   &  13.2  &  69.1  &  -  &  -\\
23 &  $a6b1/2\phi0i150$      &  No &  Survived   &  $\fullcirc(20)-\star(10)$  &   76.0  &  0.728  &  -  &  65.7   &  $\emptycirc(10)$  &   199 \\
24 &  $a6b1/2\phi180i150$     &  Yes   &  Destroyed &   $\fullcirc(20)\longleftrightarrow\emptycirc(10)$  &  149  &  0.972  &  8.64  &  79.2  &  -  &  -\\
\hline
25 &  $a4b1/2\phi0i0^{\star\star}$    &  -  &  -  &   - &   -  &   -  &  -   &  - \\
26 &  $a4b1/2\phi0i60^{\star}$    &  -   &  -  &  $\fullcirc(20)-\emptycirc(10)$  &  101   &  0.725   &   11.5  &  -  & $\star$  &  - \\
27 &  $a4b1/2\phi0i120$     &  No  &  Survived  &   $\fullcirc(20)-\star(10)$  &   68  &  0.488   &  -  &  42.6  &  $\emptycirc(10)$  &  126 \\
28 &  $a4b1/2\phi180i0$     &  Yes &  Destroyed  &  $\fullcirc(20)-\emptycirc(10)$   &   370 &   0.831  &  13    &    38.6 &    - & - \\
29 &  $a4b1/2\phi180i60$     &  Yes  &  Destroyed   &  $\fullcirc(20)-\emptycirc(10)$  &  155  &  0.713   &  12.3  &  38.0 &   -  & - \\
30 &  $a4b1/2\phi180i120$   &  Yes   &  Destroyed   & $\fullcirc(20)-\emptycirc(10)$    &  94.6   &  0.774   &  11.0   &    94.5 &  -   & - \\
\hline
31 &  $a4b1/2\phi45i30^{\star\star}$     &  -  &  -  &  -  &   -  &  -   &   -  & - \\
32 &  $a4b1/2\phi90i30$     &  No  &  Survived  &  $\fullcirc(20)-\star(10)$  &    37  &  0.301   &  -  &  75.6  &  $\emptycirc(10)$  &  227 \\
33 &  $a4b1/2\phi135i30$   &  Yes   &  Survived  &  $\fullcirc(20)-\emptycirc(10)$    &  48.0  &   0.816  &   9.556  &  60.6  &  $\star$  &  181 \\
34 &  $a4b1/2\phi225i30^{\star}$   &  -    &  - &  $\emptycirc(10)-\star(10)$   &  58.4  &   0.555  &   - &  - &  $\fullcirc(20)$  &  - \\
35 &  $a4b1/2\phi270i30$    &  No  &  Survived  &  $\fullcirc(20)-\star(10)$   &  64.5 &   0.460  &   - &  69.6 &  $\emptycirc(10)$  &  232 \\
36 &  $a4b1/2\phi315i30$   &  No  &  Destroyed   &  - &  - &   - &   - &  - &  $\fullcirc(20),\emptycirc(10)$  &  93.1, 209 \\
\hline
\end{tabular}
}
\caption{The outcomes of each model: the model number (\textit{first} column), the model name (\textit{second} column), whether a BBH forms (\textit{third} column), and whether the star survives or is destroyed (\textit{fourth} column). The next five columns show the type of final binary product and its properties (semimajor axis, eccentricity, GW-driven merger time scale only for BBHs, and ejection velocity). The last two columns indicate the type of singles as a final product and their ejection velocity. The star symbol ($^{\star}$) at the end of some model names indicates the case where the three objects form a quasi-stable triple (see its definition in the main text) and the final outcomes are not determined until the end of simulations. For this case, the properties of the inner binary are provided. The double star symbols ($^{\star\star}$) indicate cases where the three objects do not form any quasi-stable object and the final outcomes are not determined. The symbol $\fullcirc(20)$ indicates the $20\Msol$ BH initially in the binary, $\emptycirc(10)$ marks the $10\Msol$ single incoming BH, and $\star(10)$ stands for the star initially in the binary. Arrows indicate the active BH in the BBHs; double-headed arrows signify that both BHs are active at the end of the corresponding simulation. }\label{tab:outcome}
\end{table*}

\begin{figure*}
	\centering
	\includegraphics[width=8.6cm]{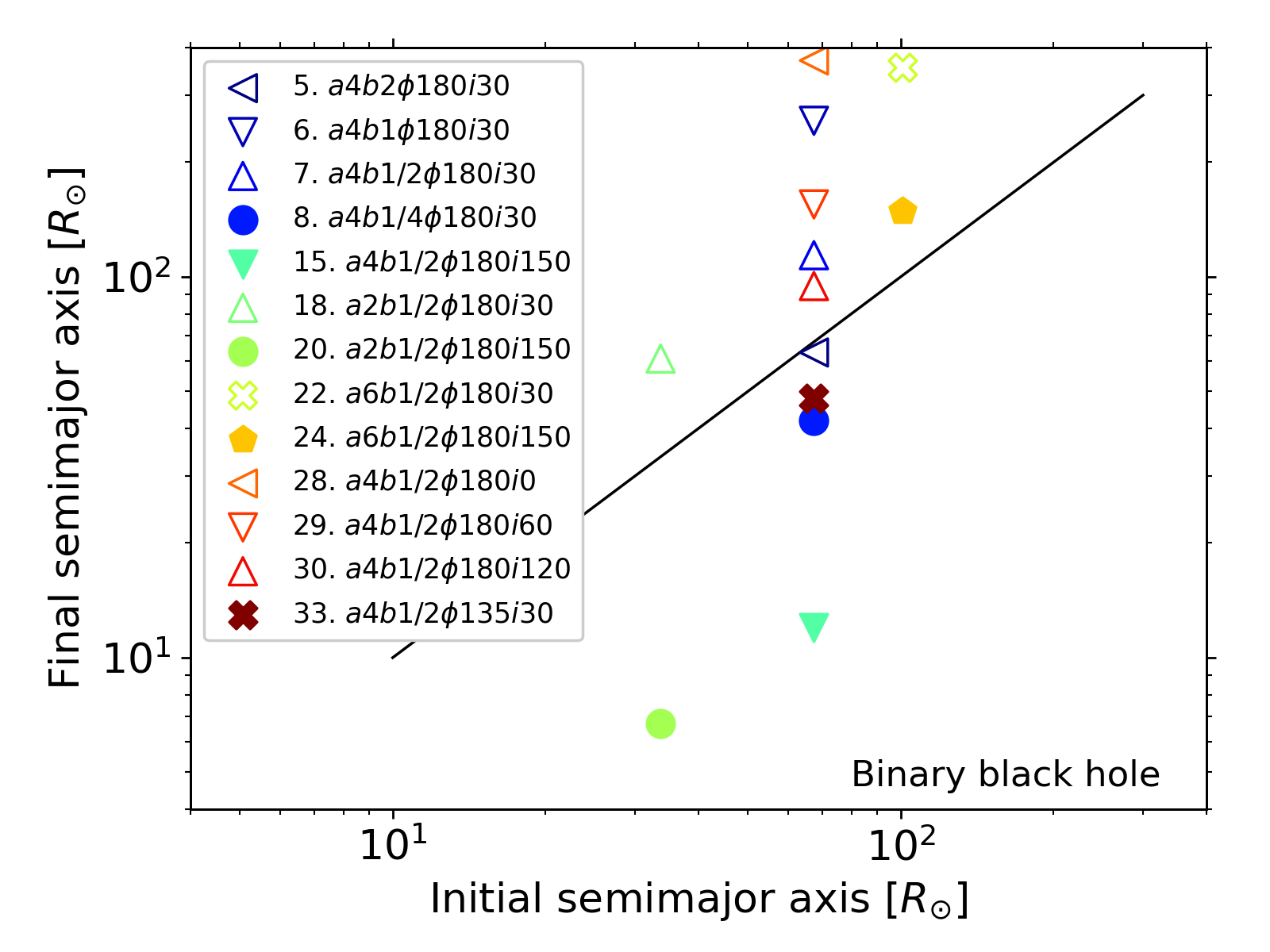}
	\includegraphics[width=8.6cm]{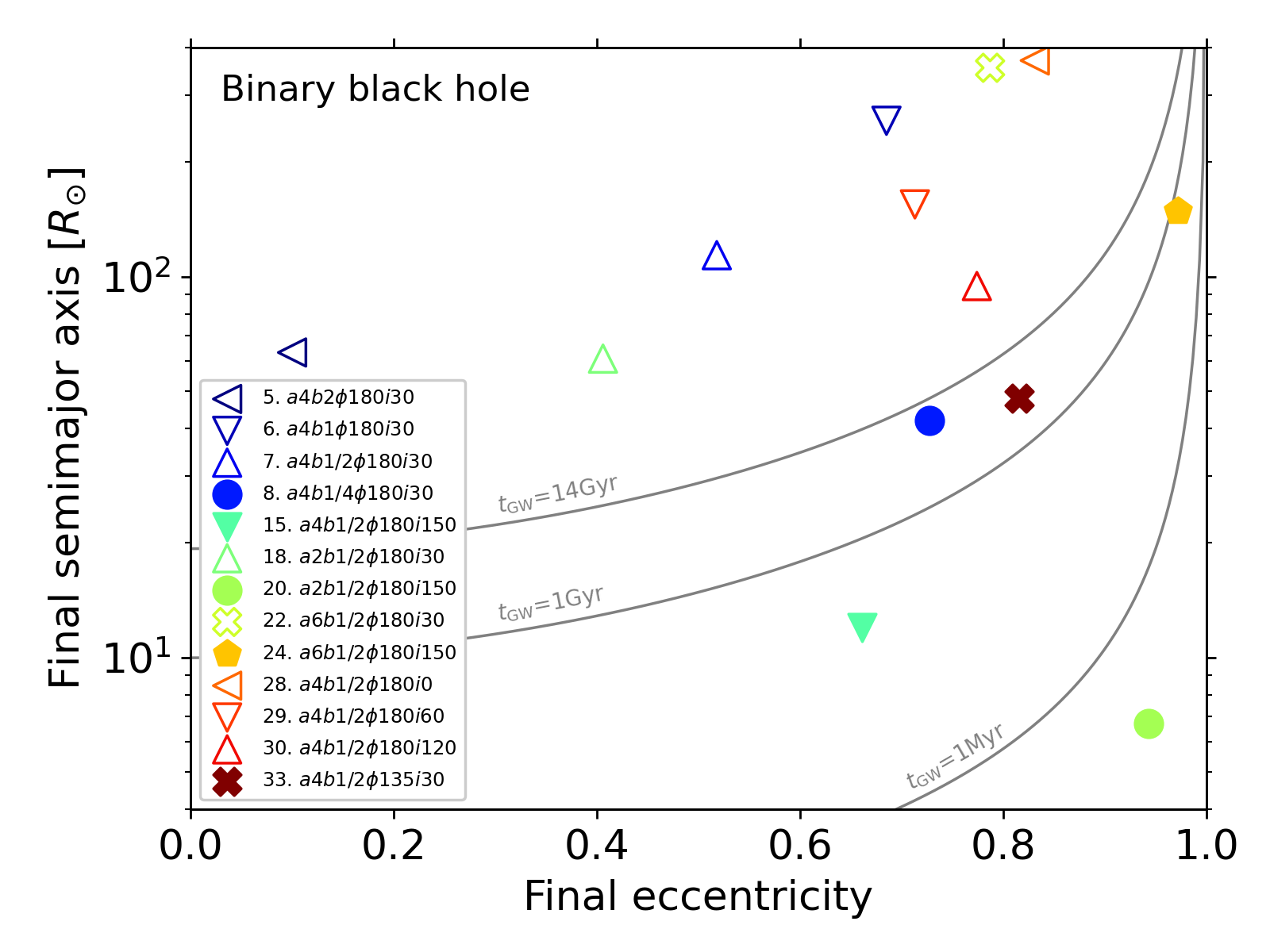}\\
	\includegraphics[width=8.6cm]{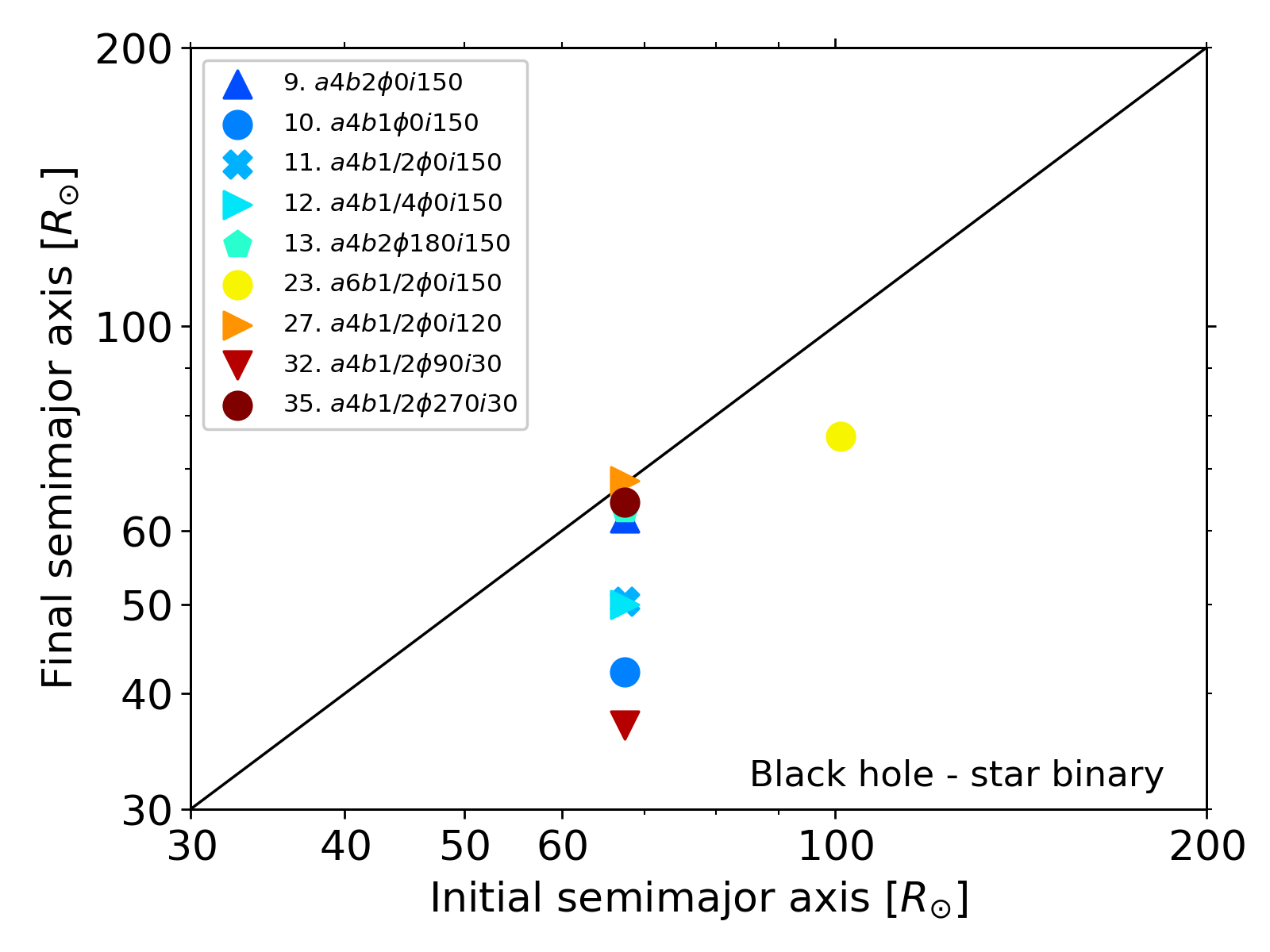}
	\includegraphics[width=8.6cm]{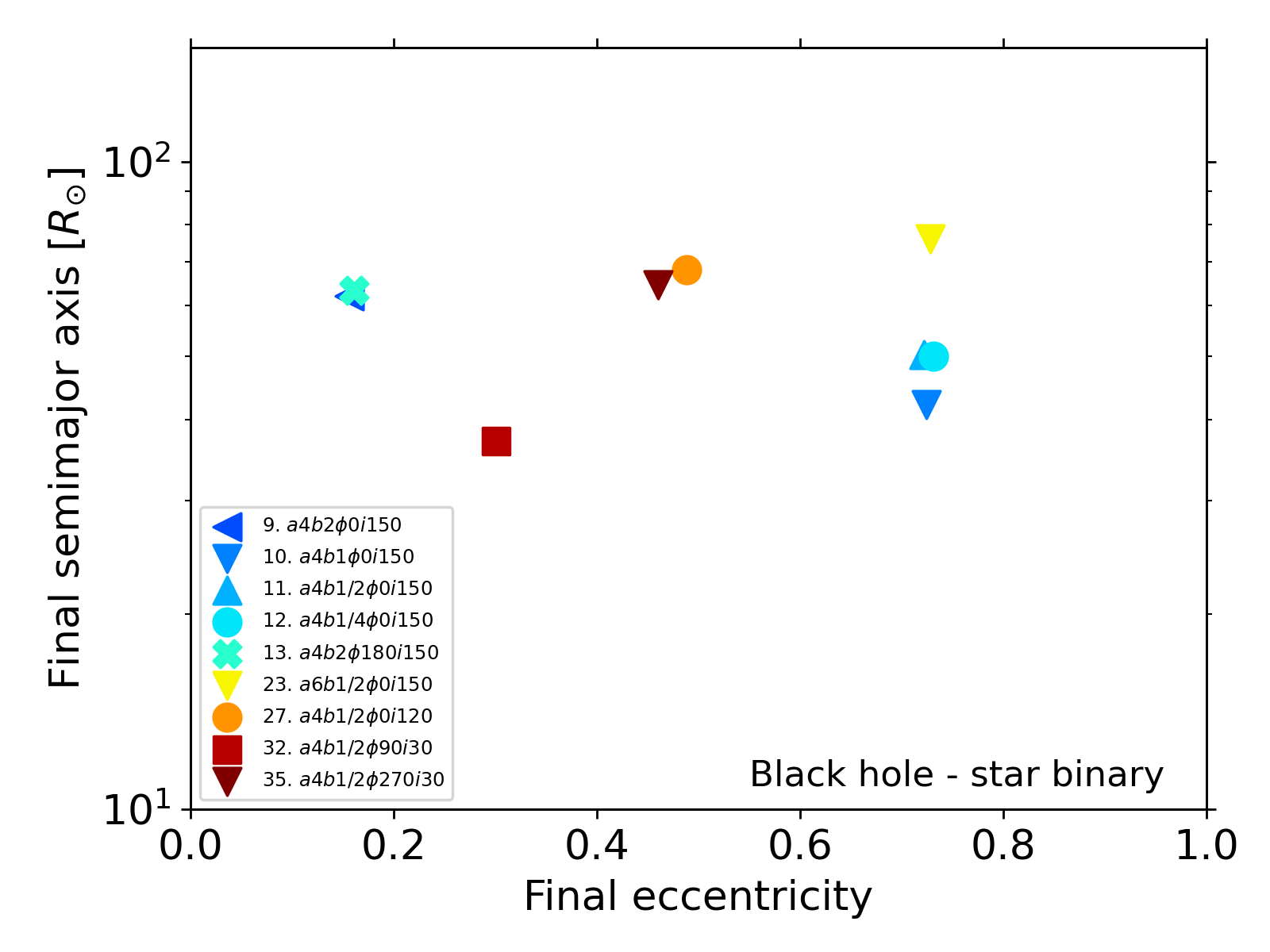}	
\caption{The orbital properties of the final binaries, for binary black holes (\textit{top} panels) and black hole -- star binaries (\textit{bottom} panels). The \textit{left} panels compare the initial semimajor axis with the final one, and the \textit{right} panels show the semimajor axis and the eccentricity of the final binaries. The black diagonal lines in the \textit{left} panels depict the cases where the size of the initial binaries and final binaries are identical. The grey curves in the \textit{top-right} panels indicate the gravitational wave-driven merger time scales of 14~Gyr, 1~Gyr and 1~Myr, respectively, for a binary consisting of $10\Msol$ and $20\Msol$ black holes. In the \textit{top} panels, the solid (hollow) markers indicate BBHs that would (not) merge in a Hubble time. On the other hand, the hollow markers in the \textit{bottom} panels indicate the models where the final outcome is an unstable triple and, for this case, the orbital properties of the inner binary are presented. }
	\label{fig:binary_orbit}
\end{figure*}

\begin{figure*}
	\centering
	\includegraphics[width=4.3cm]{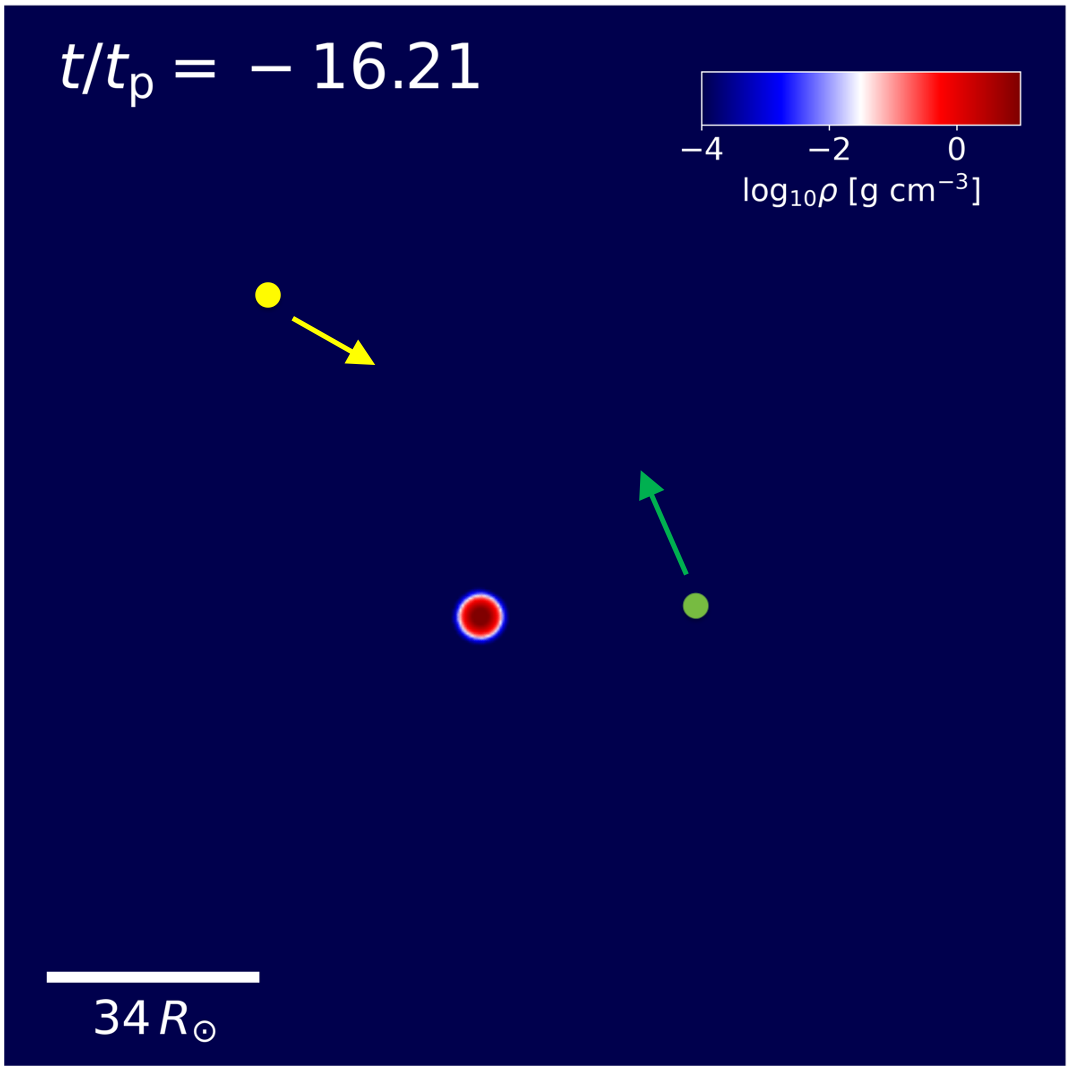}
	\includegraphics[width=4.3cm]{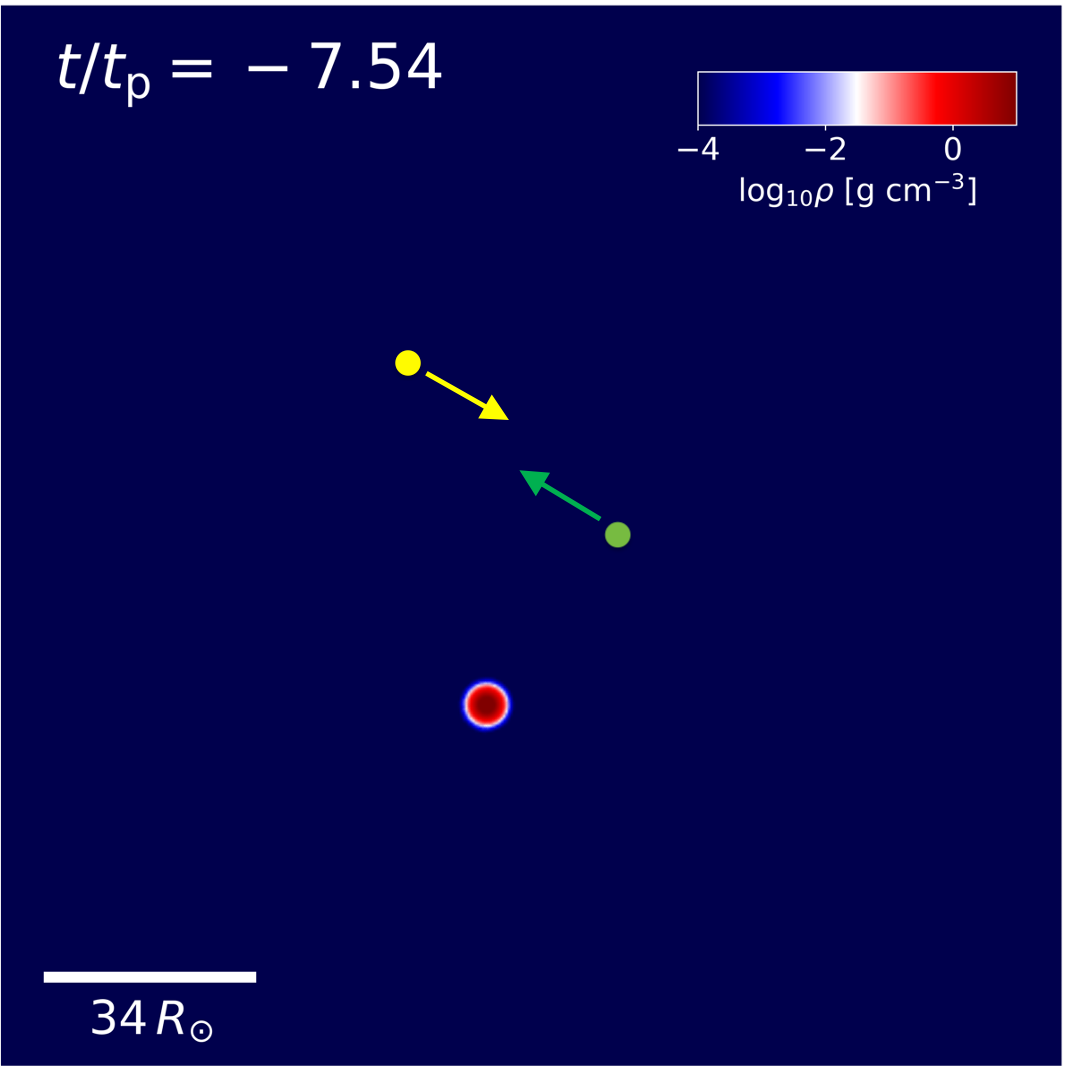}
	\includegraphics[width=4.3cm]{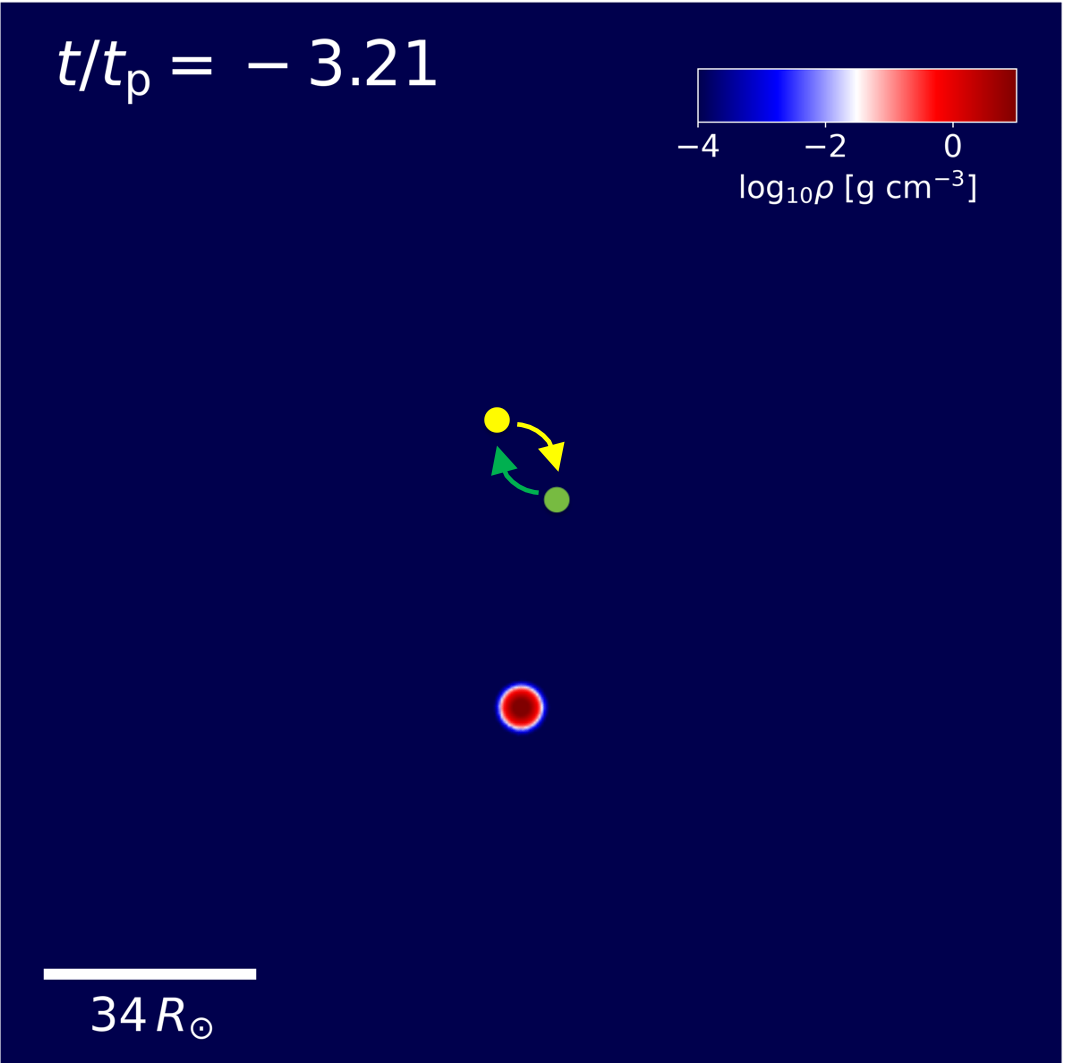}
	\includegraphics[width=4.3cm]{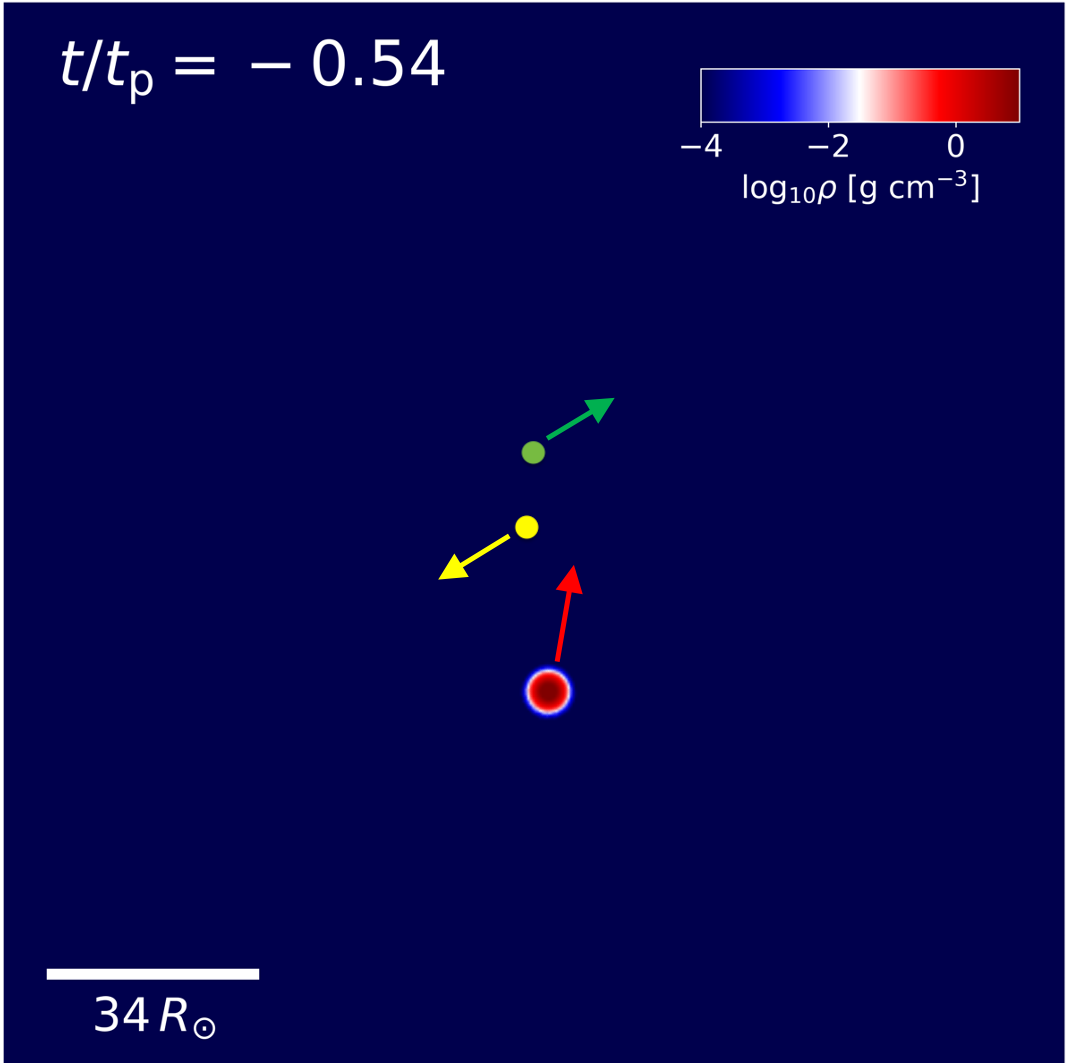}\\
	\includegraphics[width=4.3cm]{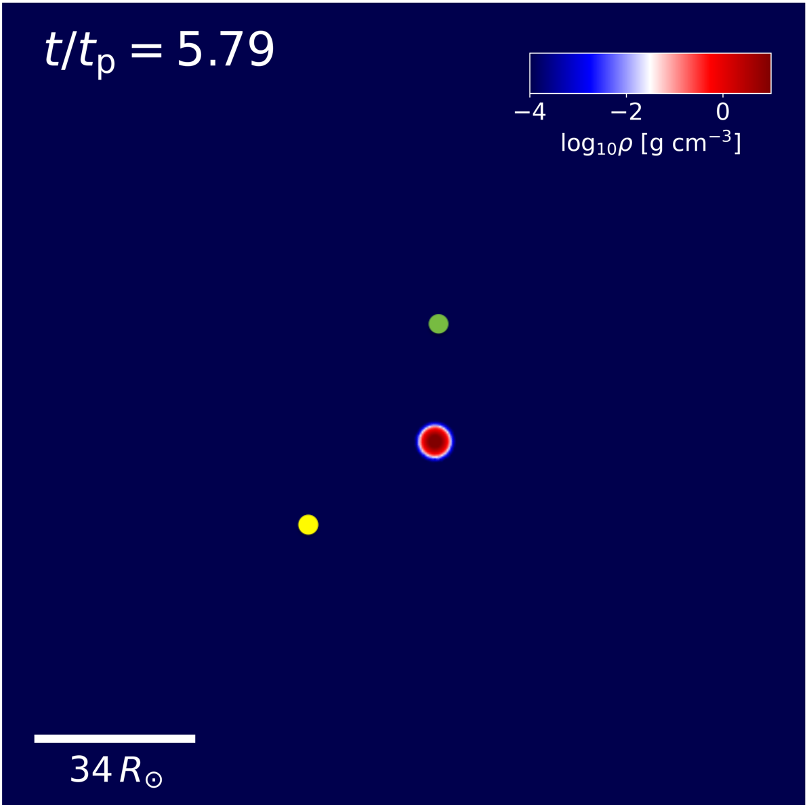}
	\includegraphics[width=4.3cm]{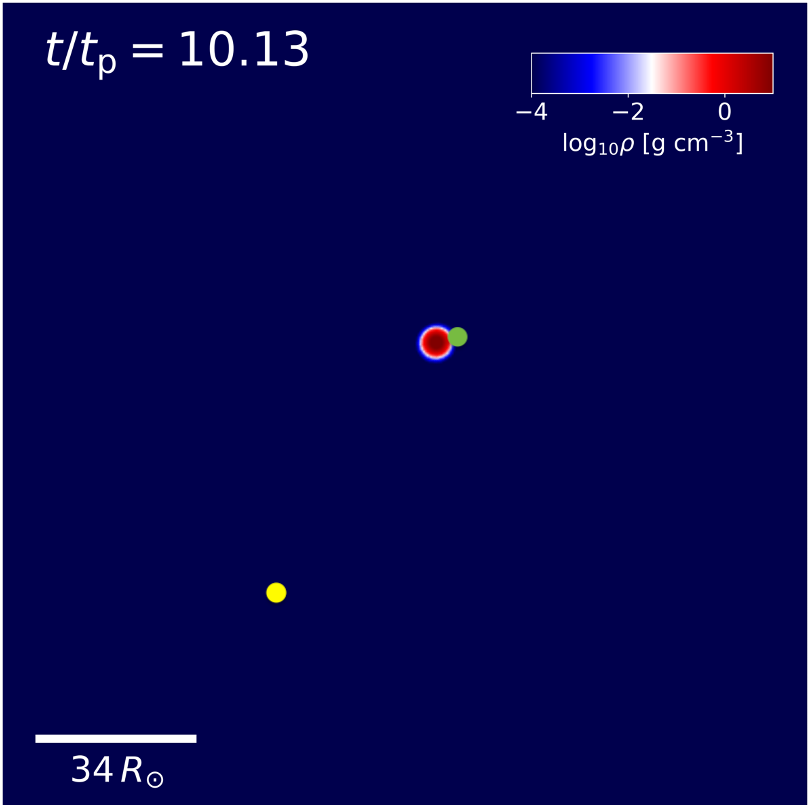}
	\includegraphics[width=4.3cm]{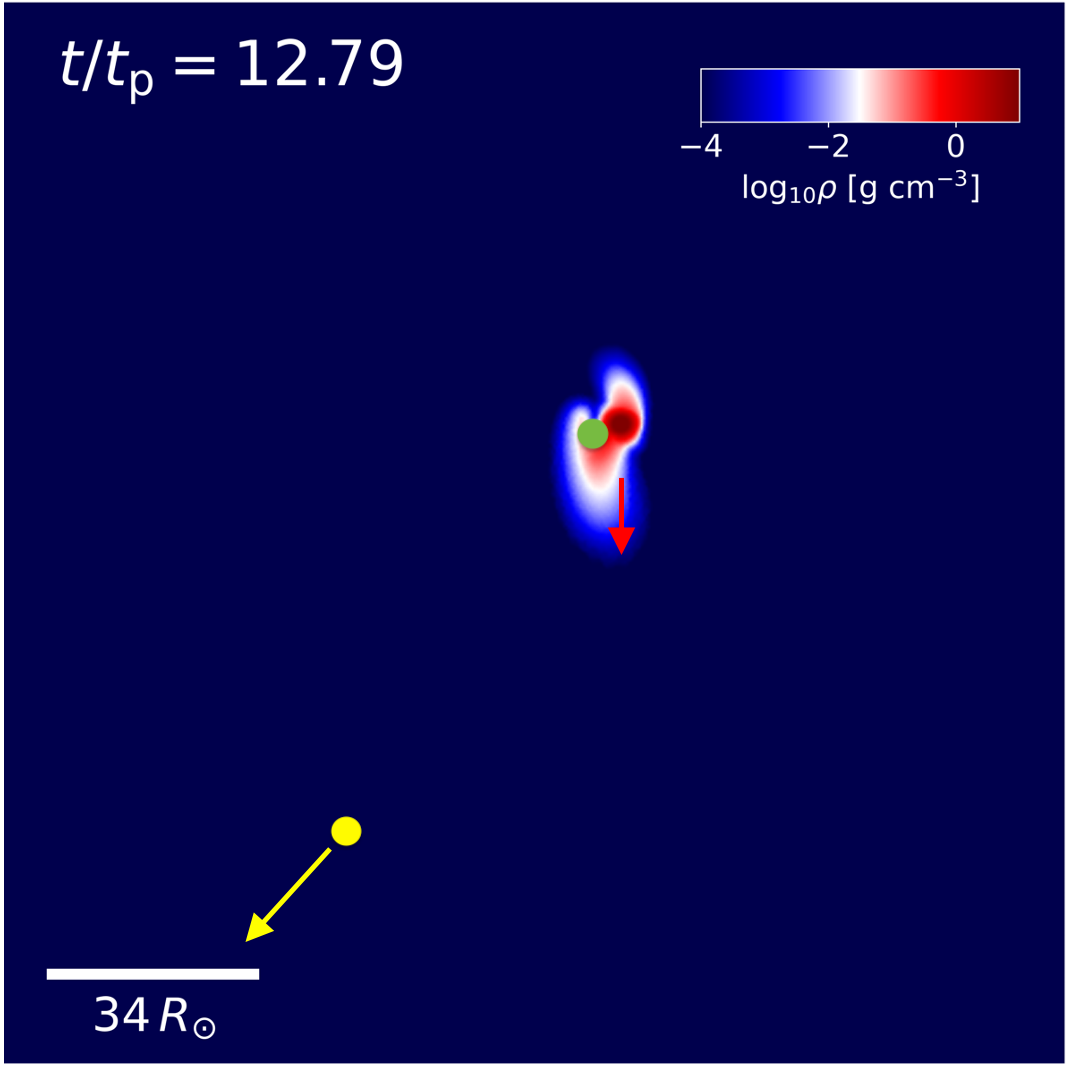}
	\includegraphics[width=4.3cm]{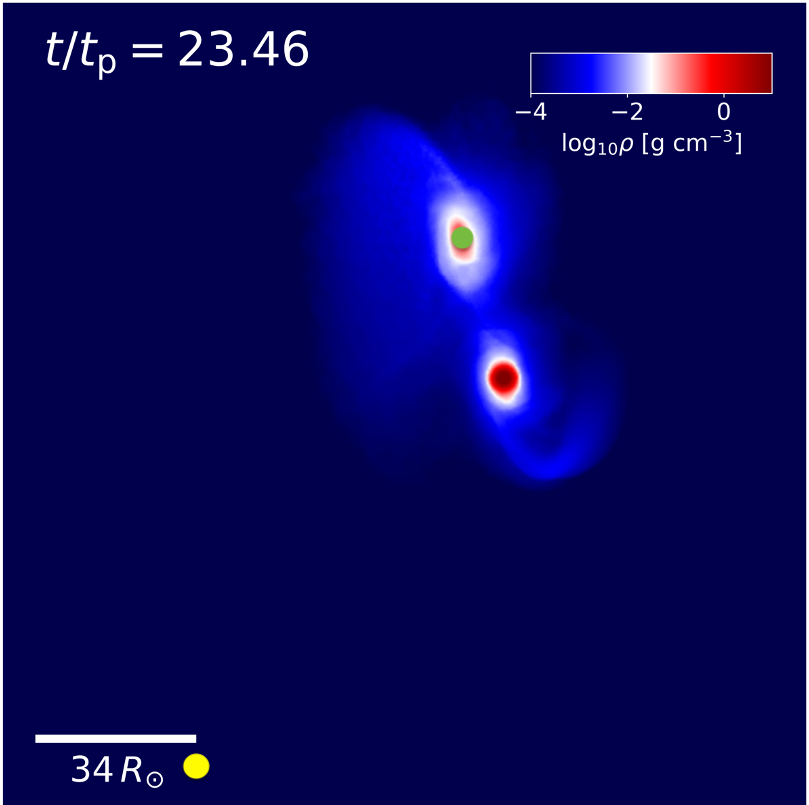}
\caption{{An example of a \textit{non-BBH-forming} encounter, Model 17.$a2b1/2\phi0i30$}. We depict the density distribution in the binary orbital plane  at a few different times in units of $t_{\rm p}$. At $t/t_{\rm p}\simeq-16$ (top-$1^{\rm st}$), the binary (star - green dot) and the single BH (yellow dot) approach each other. At $t/t_{\rm p}=-3.21$ (top-$3^{\rm rd}$), the two BHs encounter, resulting in the ejection of the initially single BH, while the initial binary  orbit is significantly perturbed to become an interacting binary  (bottom panels). Because of periodic interactions at pericenter, the binary orbit continues to evolve until the end of the simulation. The semimajor axis and eccentricity measured at the end of the simulation are $\simeq 18\Rsol$ and $\simeq0.4$, respectively. }
	\label{fig:example2}
\end{figure*}

\section{Results}\label{sec:results}

\subsection{Classification of outcomes}

The outcomes of three-body encounters between BH-star binaries and single BHs can be divided into three classes, depending on the final products.  

\begin{enumerate}
    \item \textit{BBH-forming encounters}: This class refers to encounters in which a BBH emerges. In this case, the impact parameter is mostly $\lesssim 1/2-1$. The incoming single BH frequently interacts first with the star by the time the binary's center of mass and the single BH arrive at pericenter (models with ``Yes'' in the fifth column in Table~\ref{tab:outcome}). In this situation, the star in the binary \textit{nearly collides} with the incoming single BH. We show one example for this type of encounter in Figure~\ref{fig:example1}. The incoming single BH loses a significant amount of its kinetic energy and is gravitationally captured by the other BH initially in the binary. Because of the member exchange due to a violent star-removing encounter, the size of the final binary is not necessarily correlated with the size of the initial binary. To illustrate this, we compare in the \textit{top-left} panel of Figure~\ref{fig:binary_orbit} the final $a$ of the BBHs with the semimajor axis of the initial BH-star binaries. The final $a$ covers over a wide range of values and is not necessarily comparable to $a$ of the initial binary.  These violent interactions can lead to the formation of \textit{merging} BBHs, as illustrated in the \textit{top-right} panels: the GW-driven merger time scale of 5 out of 14 final BBHs is less than a Hubble time. Note that the absolute magnitude of the binding energy of the merging BBHs is much larger than the typical kinetic energy of stars in both globular and nuclear clusters ($\simeq \sigma^{2}$ where $\sigma$ is the velocity dispersion). This suggests that  subsequent interactions with other stars would not dissociate these ``hard'' binaries, but rather make them more compact \citep{Heggie1975} and more eccentric \citep{Valtonen+2006}, which would facilitate their mergers.
    In addition, the disruption of the star prior to the BBH formation means that at least one member of the BBH is frequently surrounded by gas upon binary formation. When the BBH is compact, both BHs accrete gas. 
    
    \item \textit{Non-BBH-forming encounters}: In this class, the outcomes are member exchanges between the two BHs or perturbations of the initial binary's orbit (models with ``No'' in the fifth column of Table~\ref{tab:outcome}). This mostly occurs when the two BHs interact at the first contact between the binary and the single BH. We show one example for this type of encounter in Figure~\ref{fig:example2}, resulting in an orbit perturbation. The impact of the encounters is relatively weak compared to the \textit{BBH-forming encounters}. As shown in the \textit{bottom-left} panel of Figure~\ref{fig:binary_orbit}, the final value of $a$ scatters within less than a factor of 2 around the initial $a$. The eccentricity of the final BH-star binary is widely distributed between 0.1 and 0.9 (\textit{bottom-right} panel), similarly to those of final BBHs (\textit{top-right} panel). In this type of encounters, EM transient phenomena, such as tidal disruption events, collisions, or interacting binaries, can be created (e.g., Model 17. $a2b1/2\phi0i30$). In addition, the single BHs are ejected at $\gtrsim 60\, {\rm km\,s^{-1}}$, comparable to or higher than the escape velocity of globular clusters (i.e., tens of ${\rm km\,s^{-1}}$; \citealt{Gnedin+2002,Antonini+2016}). 
    
    \item \textit{Undetermined}: This class refers to cases where final outcomes are not determined (12 models in total, Models with ``-'' in the fifth column in Table~\ref{tab:outcome} and with superscript $\star$ or $\star\star$). Among these 12 models, there are eight encounters (models designated with superscript $\star$) in which the three objects form an unstable hierarchical triple, which we define as a triple where the outer binary is on a very large eccentric orbit so that the pericenter distance of the outer binary is smaller than the semimajor axis of the inner binary. In the table, we provide the orbital parameters of the inner binary. In the rest (models with superscript $\star\star$), interactions become extremely prolonged so that a final outcome has not (yet) emerged. 
\end{enumerate}

From now on, we will focus on the first two classes, i.e., \textit{BBH-forming} and \textit{Non-BBH-forming} encounters. These types of final outcomes and their properties are summarized in Table~\ref{tab:outcome}.

\begin{figure*}
	\centering
	\includegraphics[width=8.6cm]{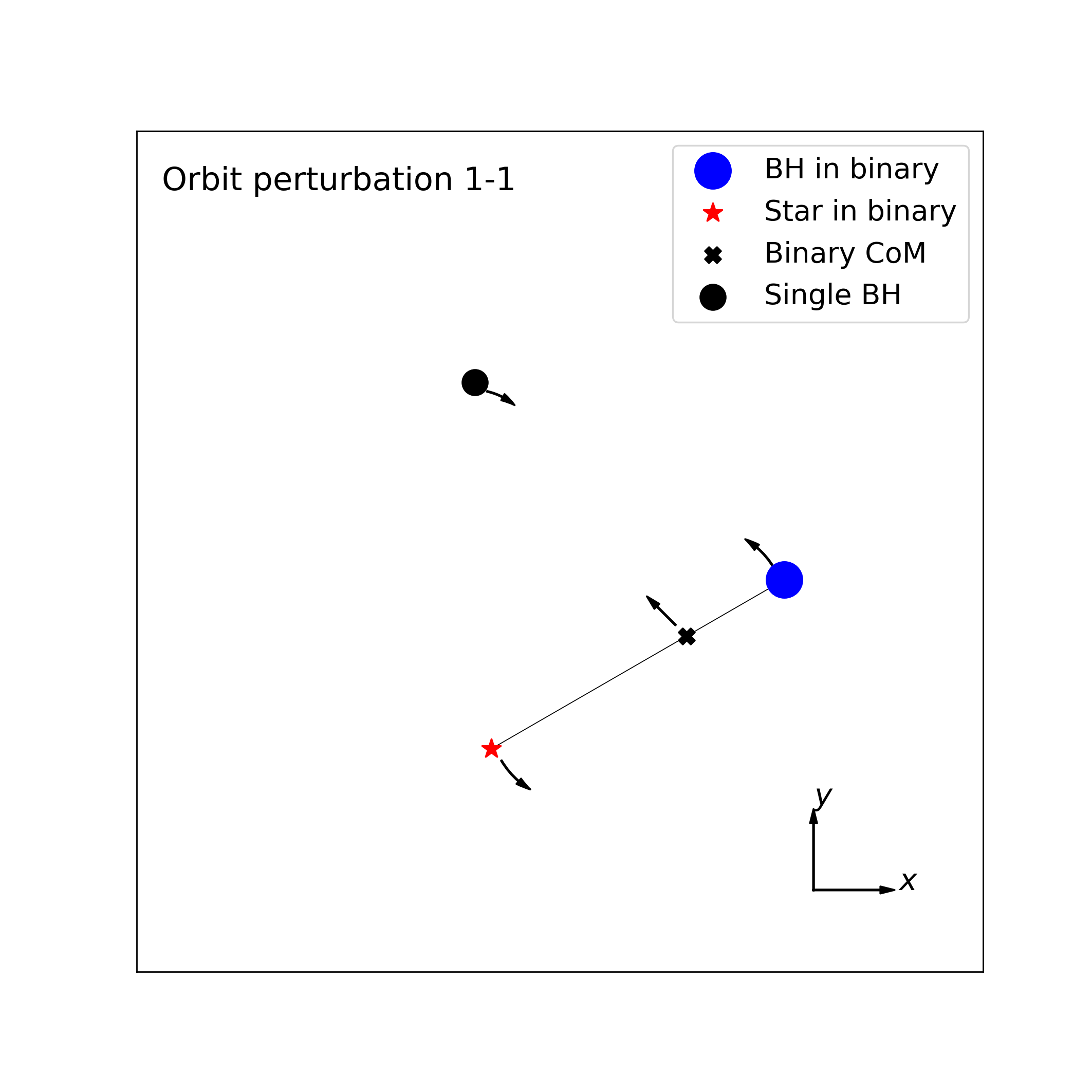}
	\includegraphics[width=8.6cm]{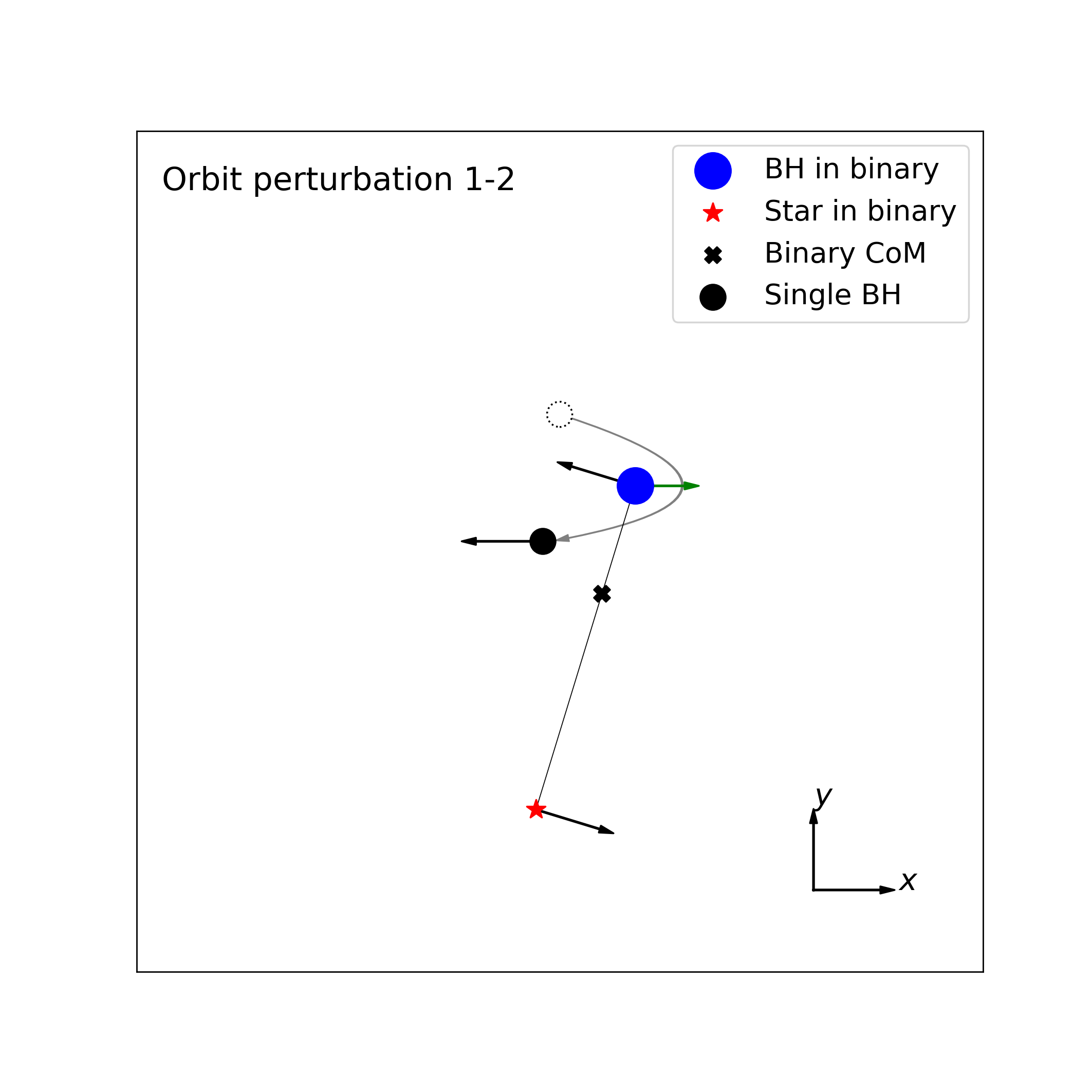}\\
	\includegraphics[width=8.6cm]{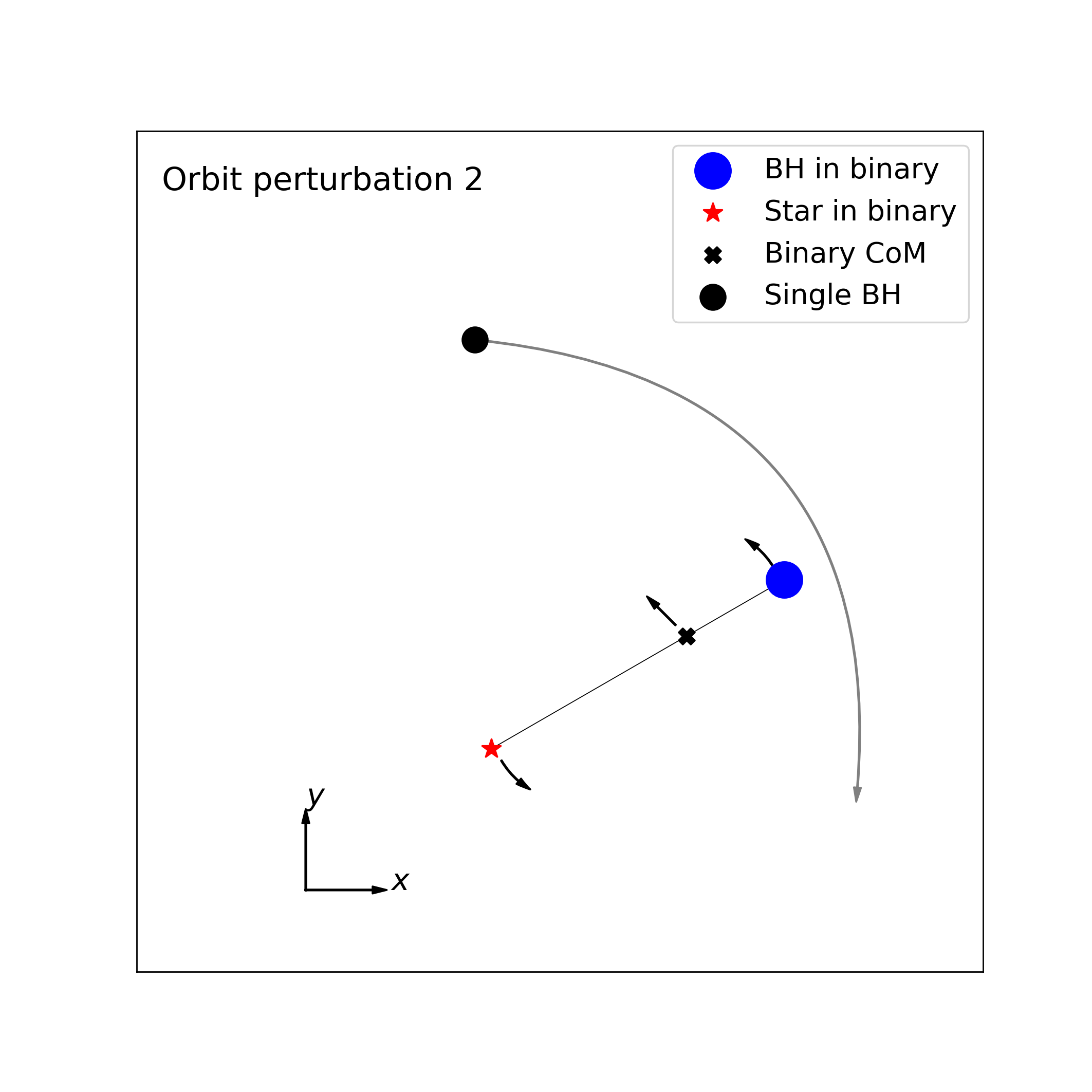}
	\includegraphics[width=8.6cm]{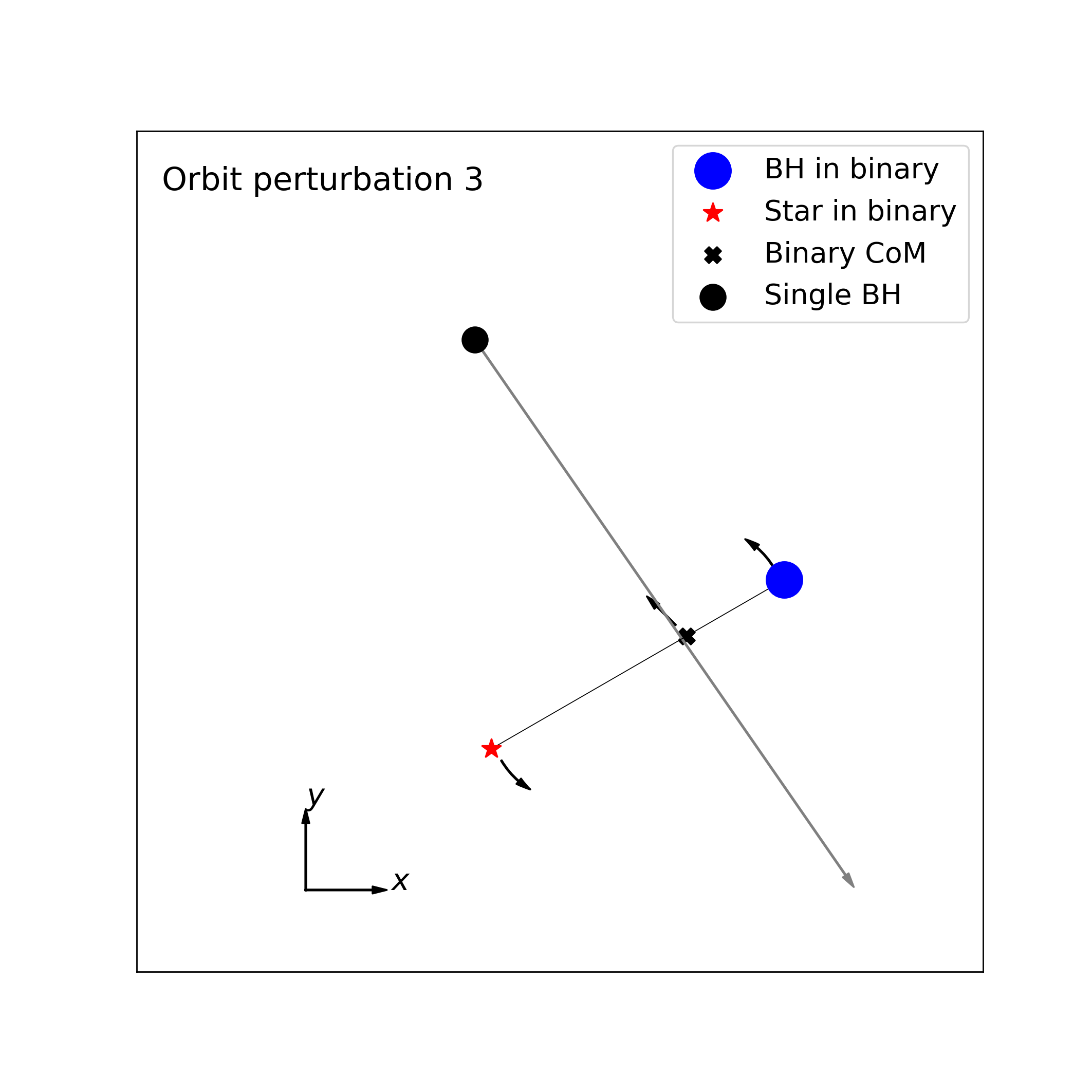}	
\caption{Schematic diagram showing three dominant configurations resulting in the perturbation of the original binary's orbit in our simulations. The black arrows indicate the direction of motion of objects, and long grey arrows the trajectory of the incoming BH (black circle).  In the first configuration (\textit{top} panels), for the prograde encounter with $\phi<90^{\circ}$ and $b \lesssim1$, the incoming BH undergoes a strong encounter with the BH (blue circle) initially in the binary (small closest approach distance compared to the binary semimajor axis), quickly turns around and advances to the left (\textit{top-right} panel). This quick turn-around motion gives a momentum kick (green arrow) to the blue circle to the right with respect to the star (red circle). The orbit of the initial binary is perturbed. The second configuration (\textit{bottom-left} panel) is a distant fly-by where the incoming BH does not significantly interact with any of the binary members, and this happens when $b \gtrsim1$. The last configuration (\textit{bottom-right} panel) shows the case where the incoming BH passes through the binary without strong interactions with any of the binary members (e.g., $b\simeq 1/4$ and $a/a_{\rm RL}=4$).}
	\label{fig:orbitperturb}
\end{figure*}

\begin{figure*}
	\centering
	\includegraphics[width=8.6cm]{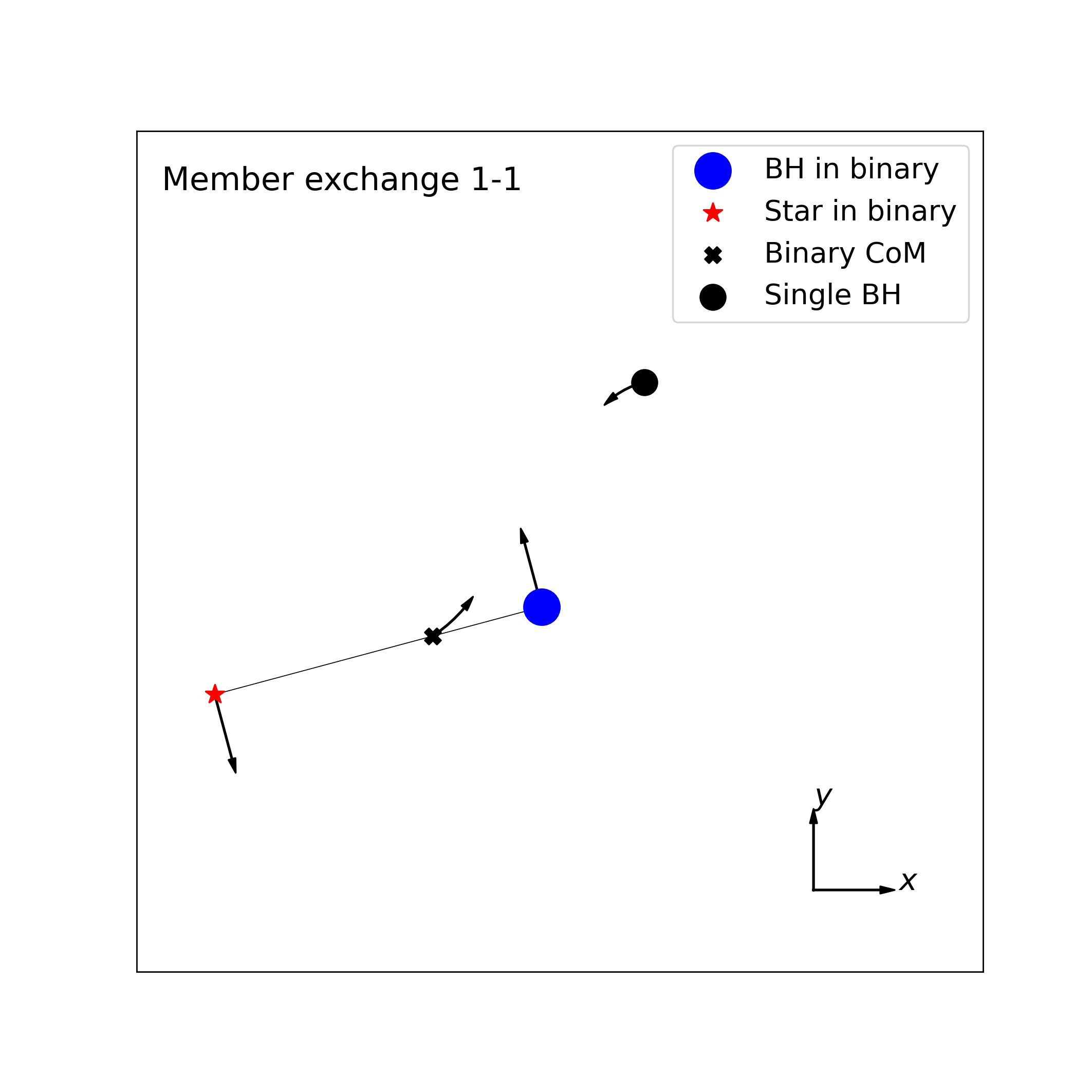}
	\includegraphics[width=8.6cm]{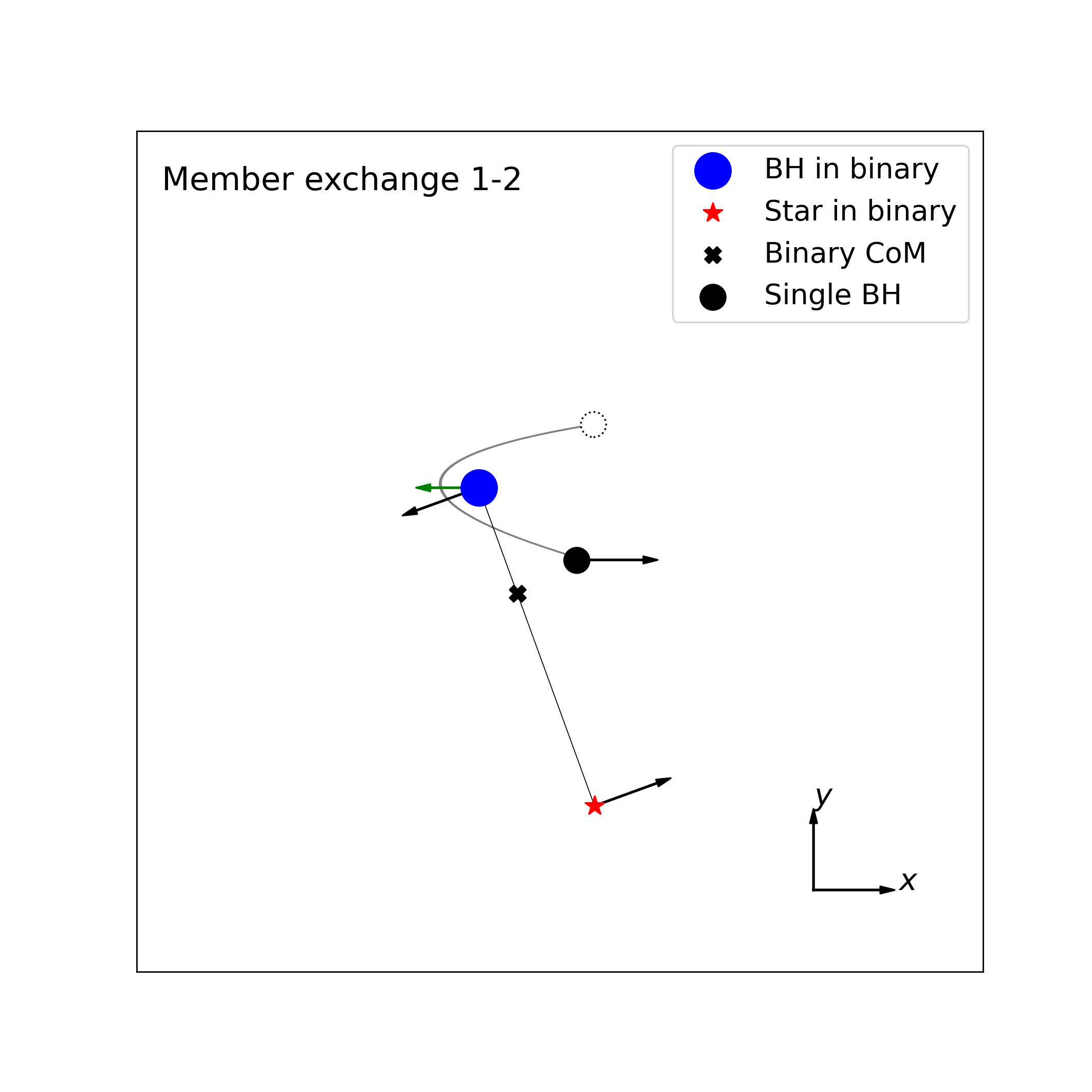}
\caption{Schematic diagram showing the dominant configuration resulting in an exchange of binary members. The black arrows indicate the direction of motion of objects, and long grey arrows (\textit{right} panel) the trajectory of the incoming BH (black circle).  In retrograde encounters with $\phi<90^{\circ}$, like configuration~1 for the orbit perturbation (Figure~\ref{fig:orbitperturb}), the incoming BH strongly interacts with the BH (blue circle) initially in the binary, quickly turns around and advances to the right. This motion results in a momentum kick (green arrow) to the blue circle to the left with respect to the star (red circle). The initially single BH gravitationally captures the star and forms a binary.  }
	\label{fig:memberexchange}
\end{figure*}

\subsection{Dynamical processes}\label{subsec:dynamics}

The two most crucial factors to determine the outcomes for the parameter space considered in our simulations are 1) the types of objects that meet at the closest encounter (BH - BH or BH - star), and 2) the net direction of the momentum kick relative to the bystander object (i.e., the object in the binary that does not interact with the incoming BH at the first closest approach) imparted by the interaction between the two meeting objects. As explained in the previous section, the first aspect substantially affects the chances of the survival of the star. The second aspect determines which objects end up in the final binary (i.e., member exchange, binary perturbation) and how the final binary's orbit looks like. 

For the \textit{non-BBH-forming} encounters with $b\simeq 1/2 - 1$, the most frequent outcomes are either a member exchange or a perturbation of the original binary. The latter happens in retrograde encounters. This case can be categorized into three configurations, which are illustrated in Figure~\ref{fig:orbitperturb}. In the first configuration (\textit{top} panels), the incoming BH strongly interacts with the other BH and turns around at a small pericenter distance compared to the binary semimajor axis. The initially single BH is rapidly ejected from the system in the direction roughly opposite to the incoming direction. This interaction imparts a momentum kick to the BH that perturbs the binary orbit. In the other two configurations, the incoming BH either moves around or passes through the binary, without significant interactions with any of the binary members (\textit{bottom} panels).

The dominant channel for member exchange is depicted in Figure~\ref{fig:memberexchange}. For the \textit{non-BBH-forming} encounters in a prograde orbit, the two BHs meet first and pass through their points of closest approach.
 Like the first configuration for orbit perturbation, their relative motion gives a momentum kick to the motion of the BH originally in the binary, relative to the star. The momentum kick gives an additional acceleration in the BH's receding motion from the star. The initially single BH, after turning around the other BH, moves in a similar direction with the star and gravitationally captures it. 

For the \textit{BBH-forming} encounters, the star and the initially single BH undergo close encounters, naturally resulting in a tidal disruption event or stellar collision. Both events can also impart a momentum kick to the disrupting BH. In our simulations, the momentum kick is not large enough to prevent the two BHs from forming a bound pair. For example, if the star and the incoming BH undergo a head-on collision, the incoming BH dramatically slows down and forms a merging BBH with the other BH (e.g., Model 20. $a2b1/2\phi180i150$). We caution that the head-on collision between the two equal-mass objects in our simulations is an extreme case yielding a dramatic drop in the BH's kinetic energy. The net effect of such star-removing events on the motion of the disrupting BH and the subsequent formation of a BBH depends on the mass ratio, relative velocity, and the direction of the momentum kick.

\subsection{Binary black hole formation}

Typical semimajor axes of BBHs formed in the \textit{BBH-forming} encounters range within $10 - 400\Rsol$ while eccentricities vary within $0.1 - 0.97$. Correspondingly, the GW-driven merger time scales of those merging binaries are in the range $10^{4}-10^{13}\yr$. Five of our models among these encounters are merging BBHs with $a\simeq 7-150\Rsol$ and $e\simeq 0.6-0.97$. As explained in \S~\ref{subsec:dynamics}, the dominant formation channel is the gravitational capture of the incoming BH by the BH originally in the binary after strong interactions between the incoming BH and the star. Naturally, a disruption event or a collision precedes the BBH formation. As a result, either the disrupting BH or both BHs accrete matter by the time they form a stable binary. 

\subsection{Dependence of outcomes on parameters}\label{subsec:dependence}

We examine the dependence of outcomes on a few key encounter parameters, phase angle $\phi$, impact parameter $b$, inclination angle $i$, and semimajor axis $a$, by varying one parameter at a time, keeping the rest of them fixed. Our simulations suggest that the two most important parameters that affect the formation of BBHs in this scenario of three-body encounters are the impact parameter and the phase angle. 
\begin{enumerate}
    \item \textit{Phase angle $\phi$}: this is found to be one of the key parameters that separates \textit{BBH-forming} encounters from \textit{non-BBH-forming} encounters. For the former, very likely outcomes are BBHs, frequently accompanied by a disruption of the star. On the other hand, for the latter, frequent outcomes are eccentric BH-star binaries produced via member exchange or weak tidal perturbations of the initial stellar orbit. In addition, even for the \textit{BBH-forming} encounters, the direction of the encounter between the initially isolated BH and the star at the first closest approach relative to the other BH determines the size of the semimajor axis of the BBH: if the momentum kick imparted on the encountering BH is given such that it adds to the encountering BH's momentum, a large binary forms (e.g., Model 6. $a4b1\phi180i30$ and Model 22. $a6b1/2\phi180i30$). Although our study is not appropriate for rate estimates, the dependence of outcomes on the phase angle may indicate that roughly $\simeq25\%$ of these three-body encounters between objects of similar mass with $b \lesssim 1$ may possibly lead to BBH formation with a high chance of creating EM transients. 

    \item \textit{Impact parameter $b$}: in general, the initial binary and the single BH can interact significantly (member exchange or stellar collisions) at the first closest approach when $r_{\rm p}\lesssim a $, which is also found in \citet{Ryu+2022, Ryu+2023}. Fly-by only occurs at $r_{\rm p}>a $ (Models 1. $a4b2\phi0i30$, and Model 9. $a4b2\phi0i150$). For this case, the initial binary orbit is weakly perturbed, resulting in a 10 - 20\% change in the semimajor axis. Relatively weak interactions also take place when the impact parameter is too small compared to the size of the binary, i.e., $r_{\rm p}<a/8$ (e.g., Model 4. $a4b1/4\phi0i30$, and Model 16. $a4b1/4\phi180i150$), as the single BH penetrates through the binary without interacting strongly with any of the binary members (see the \textit{bottom-right} panel of Figure~\ref{fig:orbitperturb}).

    \item \textit{Inclination angle $i$}: prograde encounters tend to result in strong interactions between the first two encounter objects, frequently leading to outcomes that involve member exchange (e.g., models with $\phi=0^{\circ}$ and $i=30^{\circ}$) or stellar collisions (e.g., models with $\phi=180^{\circ}$ and $i=150^{\circ}$). This is because the relative velocity between the two encountering objects is smaller, implying a larger gravitational focusing cross section ($\propto v^{2}/\sigma^{2}$ where $\sigma$ is a typical relative velocity at infinity). A typical configuration for member exchange in prograde encounters is drawn in Figure~\ref{fig:memberexchange}. On the other hand, the first interactions in retrograde orbits are relatively weak due to the large relative speed between the two encountering objects. As a result, frequent outcomes are perturbations of the initial binary orbit, as depicted in Figure~\ref{fig:orbitperturb}.

    \item \textit{Semimajor axis $a$}: given the same pericenter distance relative to $a$ ($r_{\rm p}\simeq 0.25a$) for the simulations with varying $a$, the type of the final outcomes does not show a strong dependence on $a$. However, the size of the final binary is closely correlated with that of the initial binary, e.g., $a\gtrsim 76\Rsol$ of final binaries in models with $a/a_{\rm RL}=6$ (or $a=101\Rsol$) and $a\lesssim 61 \Rsol$ in models with $a/a_{\rm RL}=2$ (or $a=32\Rsol$). 

\end{enumerate}

\begin{figure*}
	\centering
	\includegraphics[width=5.6cm]{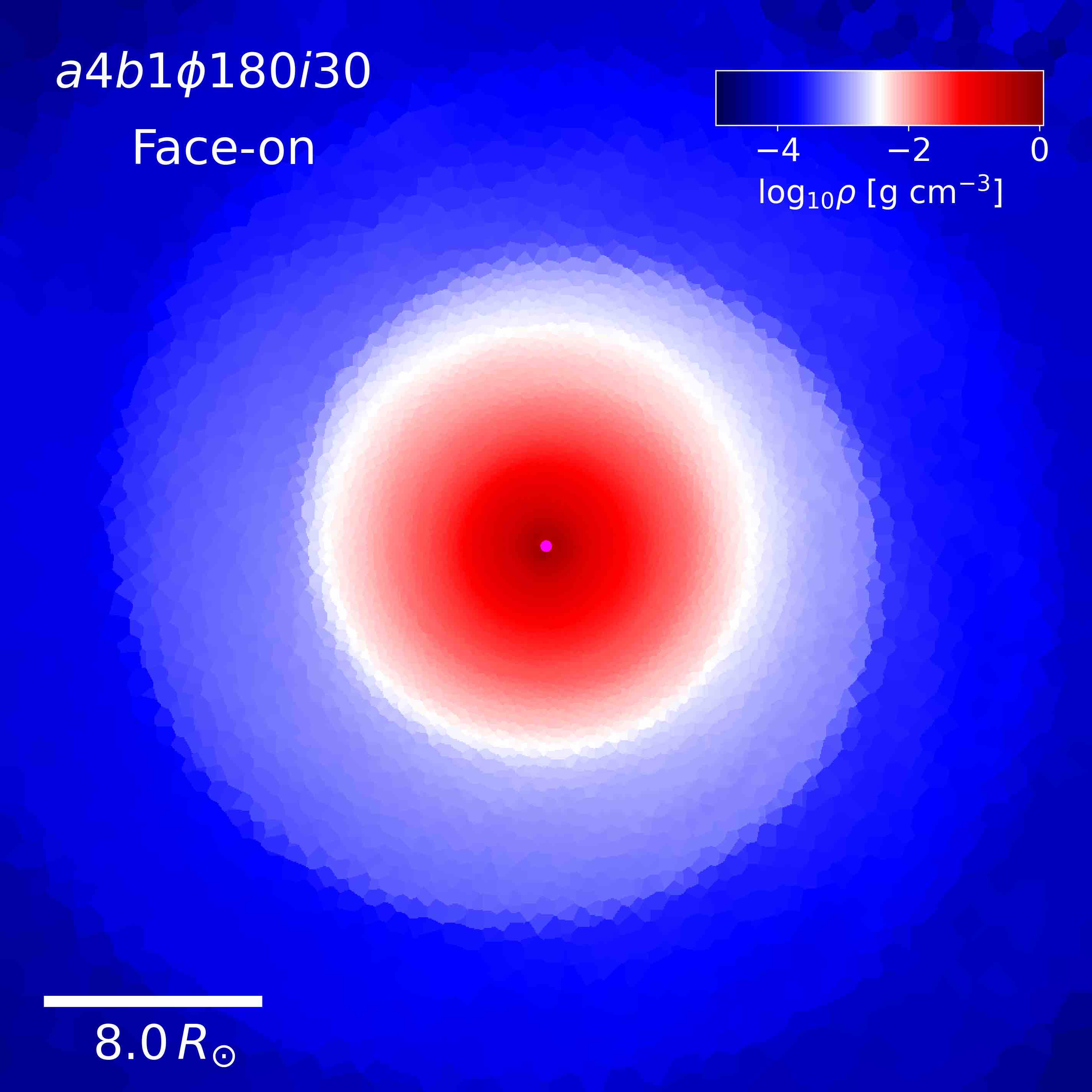}
	\includegraphics[width=5.6cm]{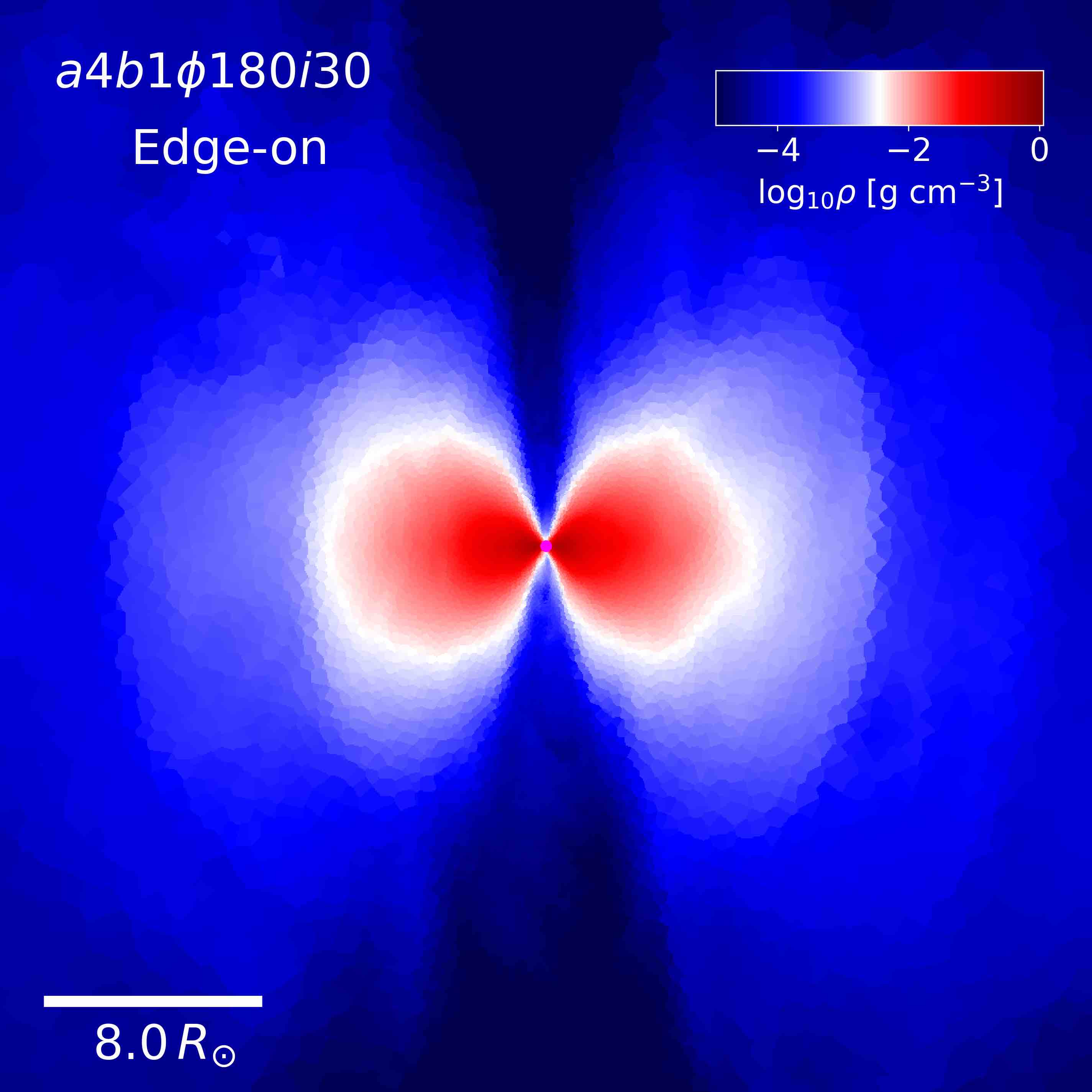}\\
    \includegraphics[width=5.6cm]{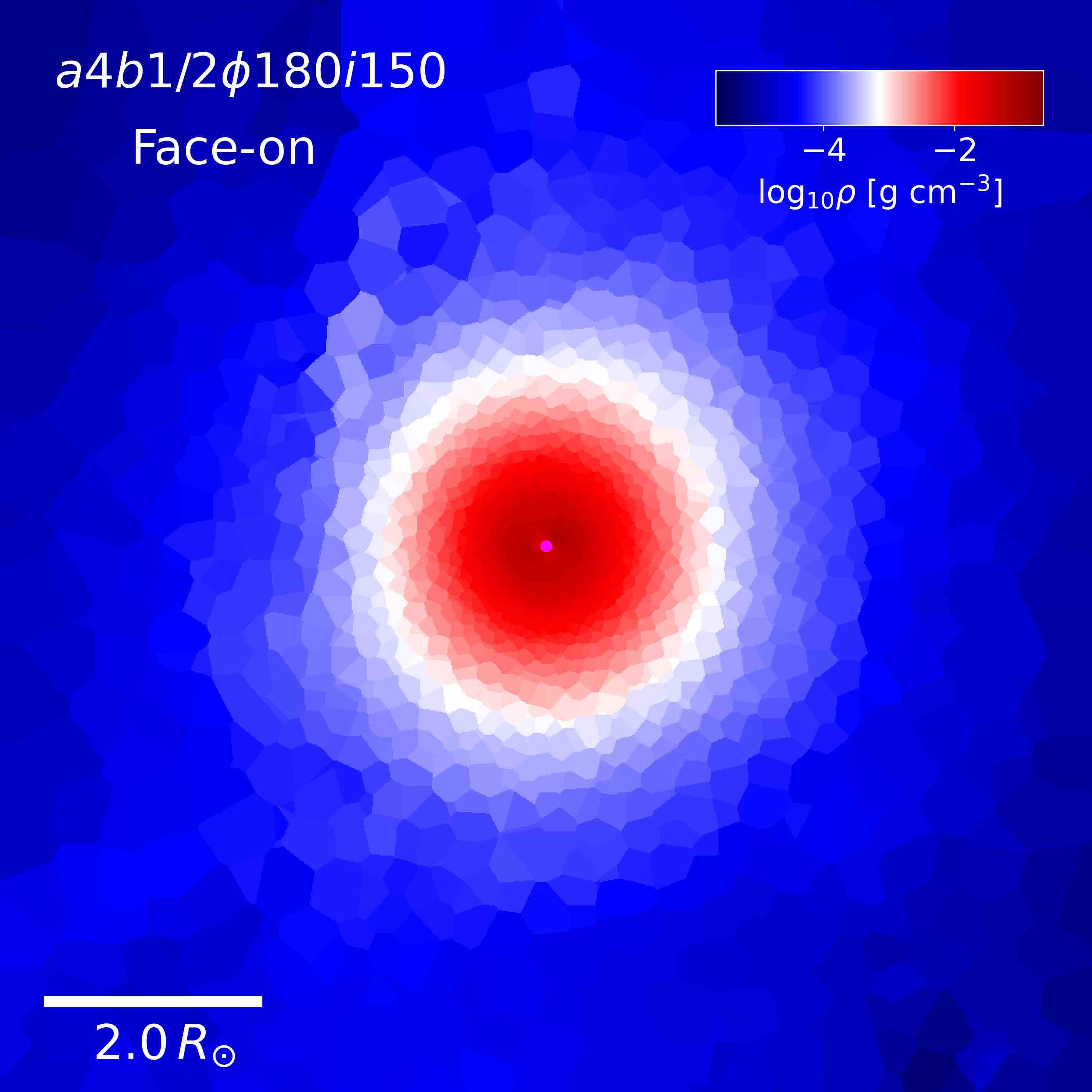}		
    \includegraphics[width=5.6cm]{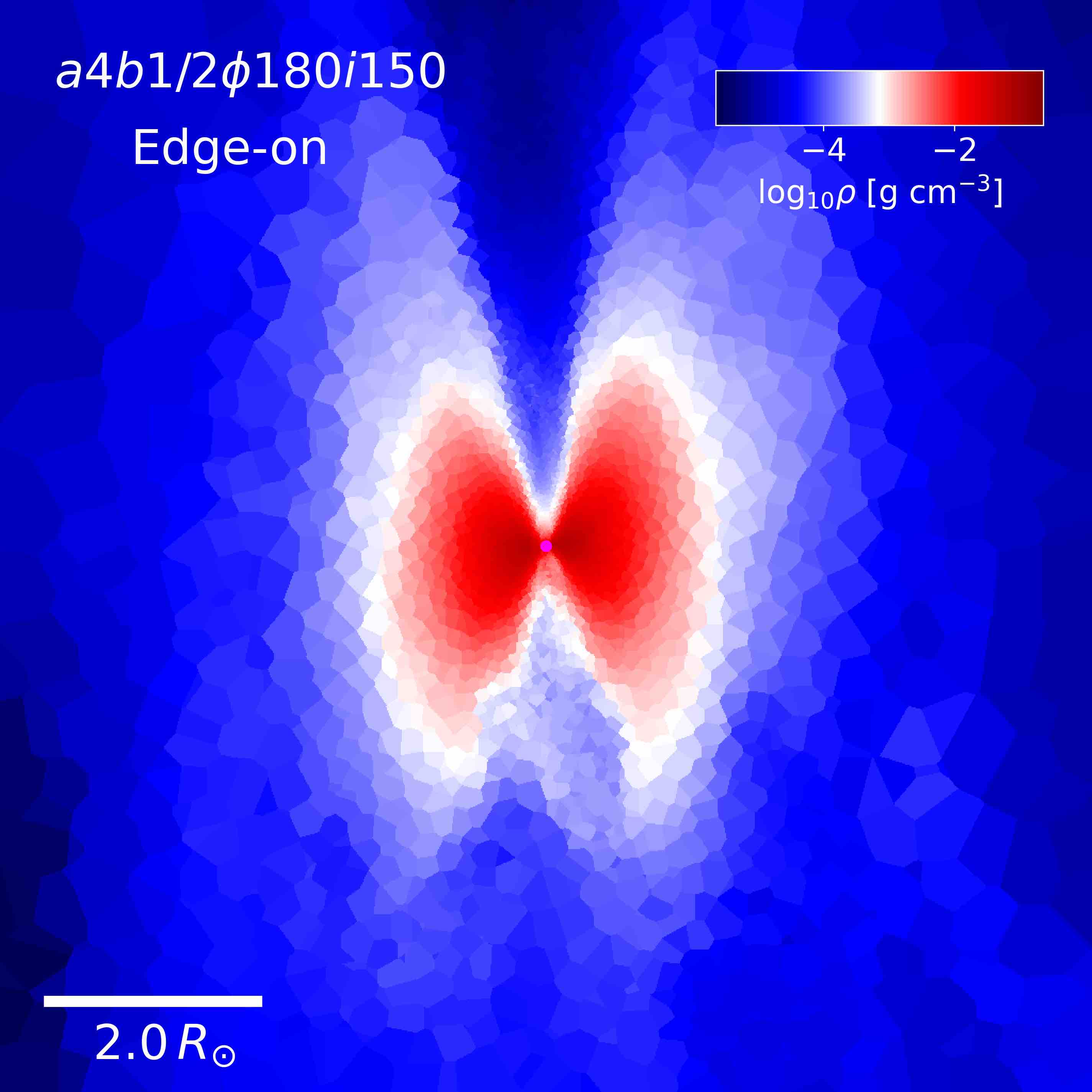}\\	
	\includegraphics[width=5.6cm]{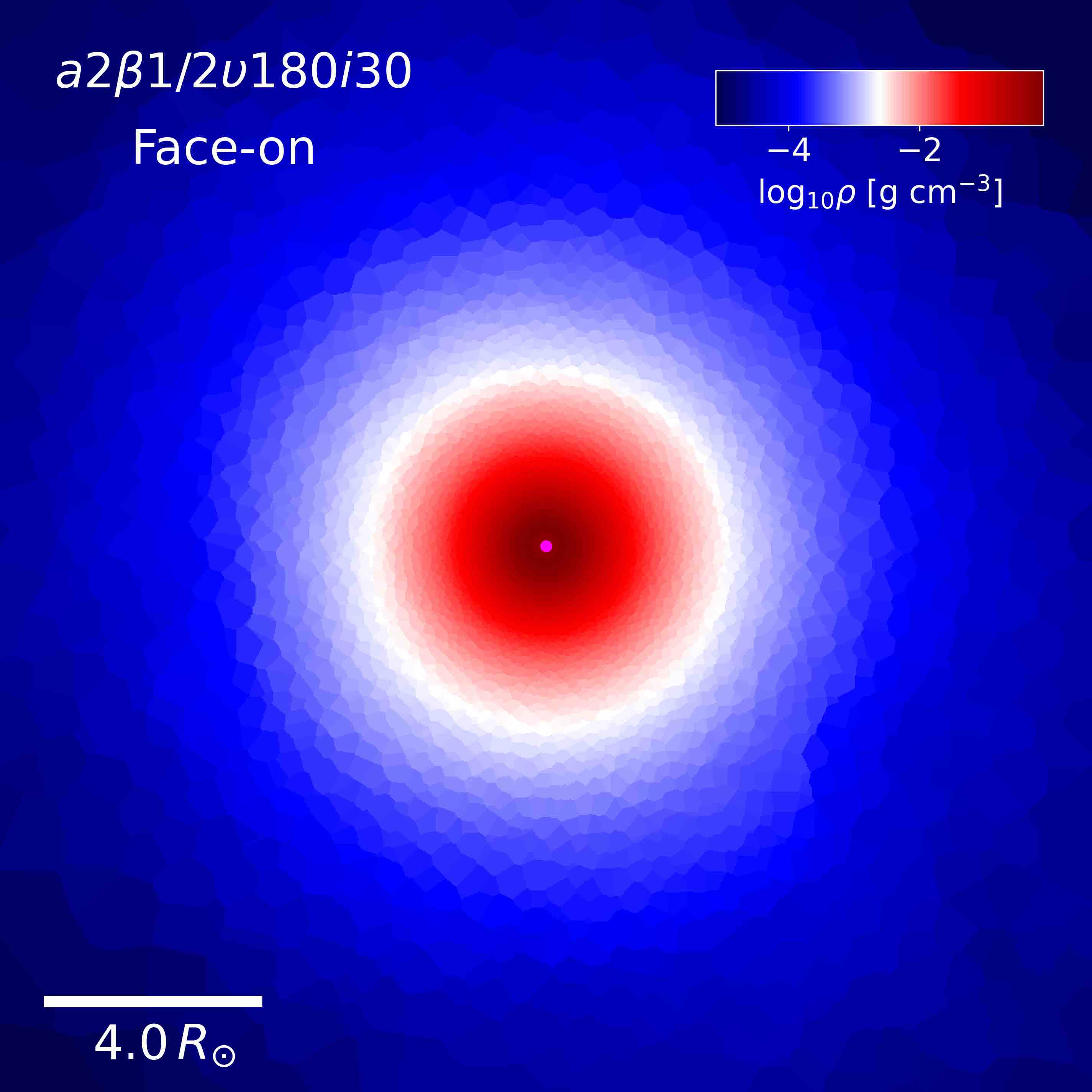}
	\includegraphics[width=5.6cm]{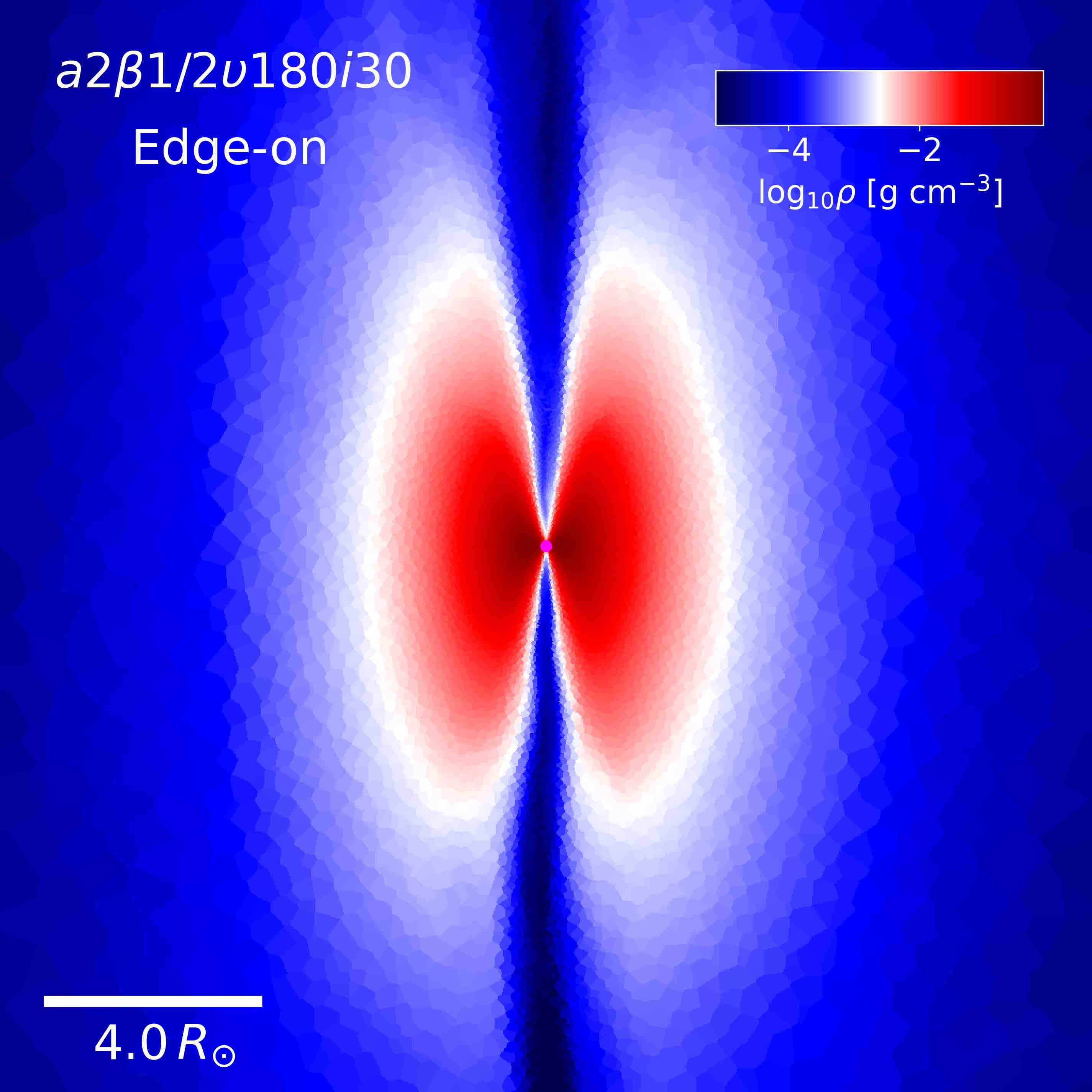}\\
	\includegraphics[width=5.6cm]{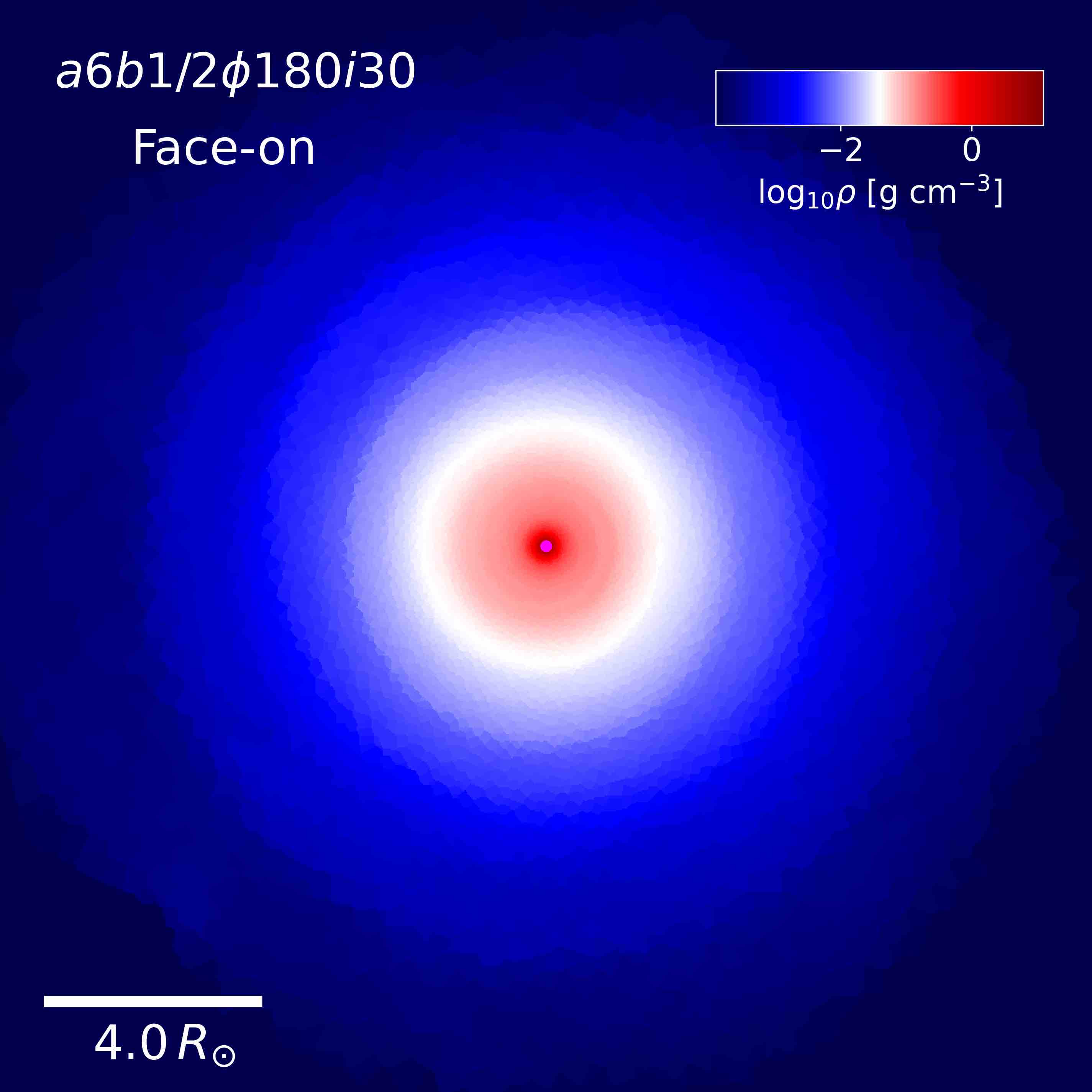}
	\includegraphics[width=5.6cm]{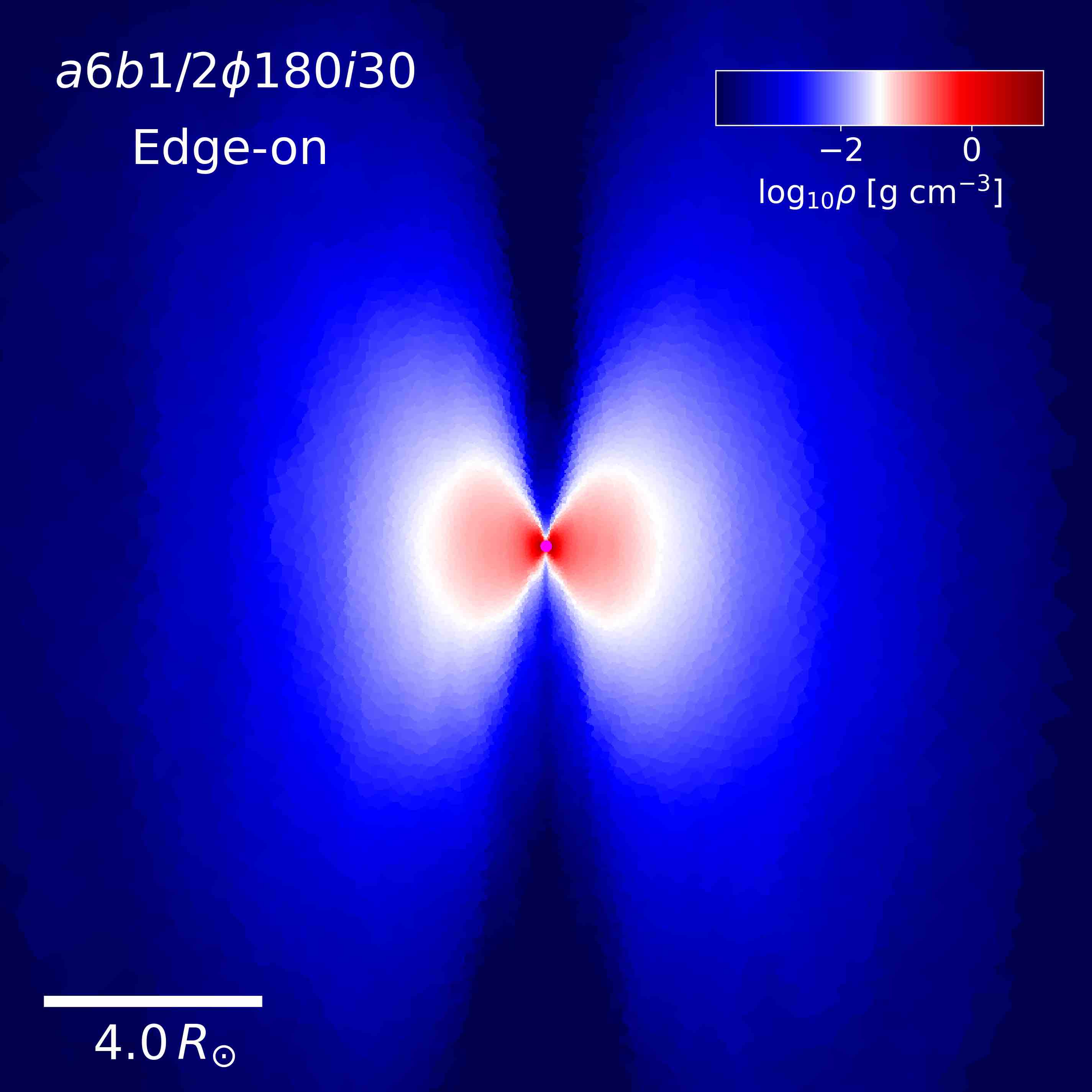}\\
\caption{Face-on (\textit{left}) and edge-on (\textit{right}) density distribution of disks around the BH that disrupts the star at the first closest approach in four selected models with $i=30^{\circ}$ or $150^{\circ}$, for Model 6. $a4b1\phi180i30$ ($1^{\rm st}$ row), Model 15. $a4b1/2\phi180i150$ ($2^{\rm nd}$ row), Model 18. $a2b1/2\phi180i30$ ($3^{\rm rd}$ row), and Model 22. $a6b1/2\phi180i30$ ($4^{\rm th}$ row), at the end of the simulations. The white horizontal bar at the bottom-left corner of each panel shows the spatial scale, $4\,{\rm R}_{\odot}$, except for the second row of panels where it is $2\,{\rm R}_{\odot}$. }
	\label{fig:diskdensity}
\end{figure*}

\begin{figure*}
	\centering
	\includegraphics[width=8.4cm]{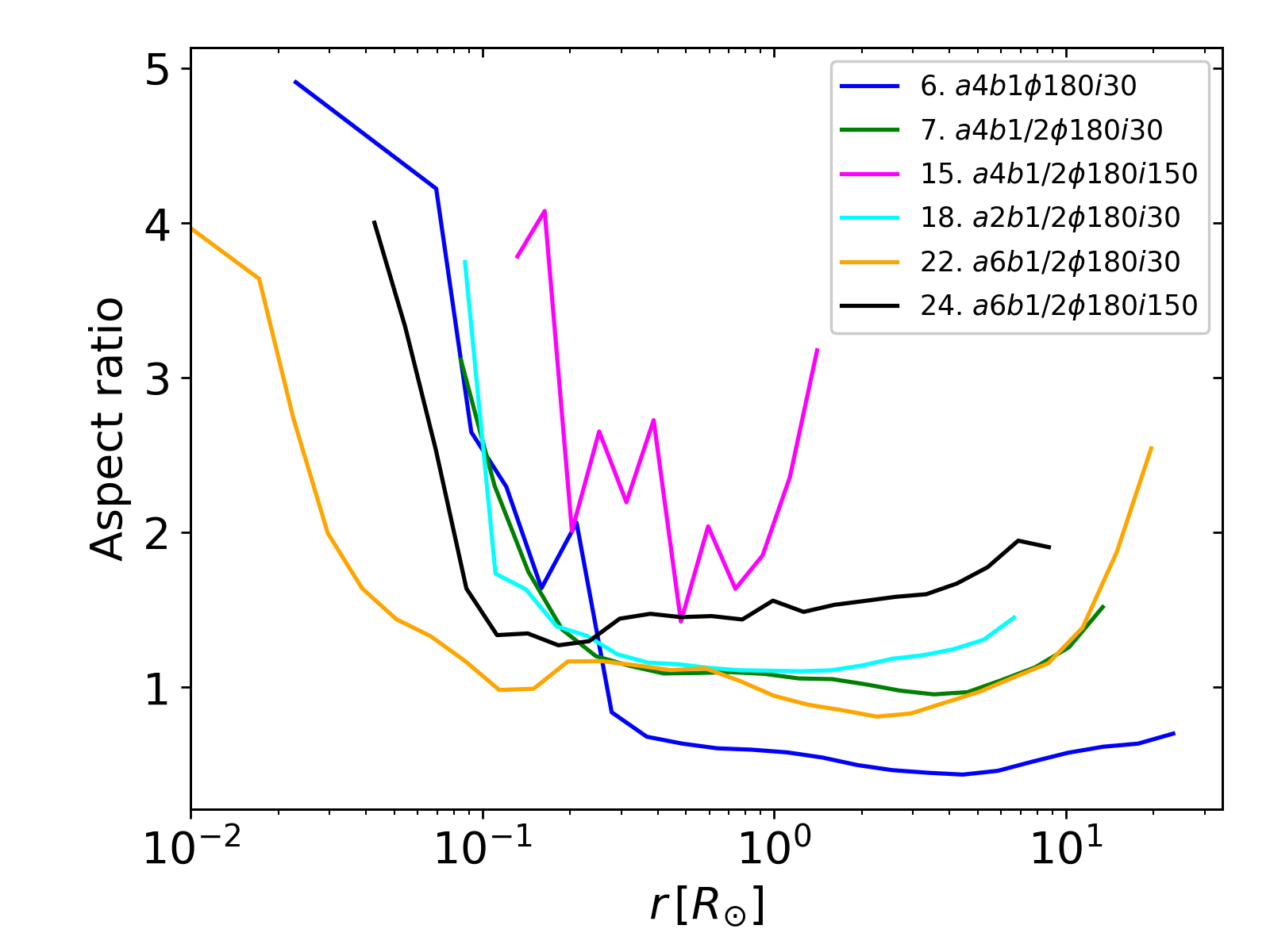}
	\includegraphics[width=8.4cm]{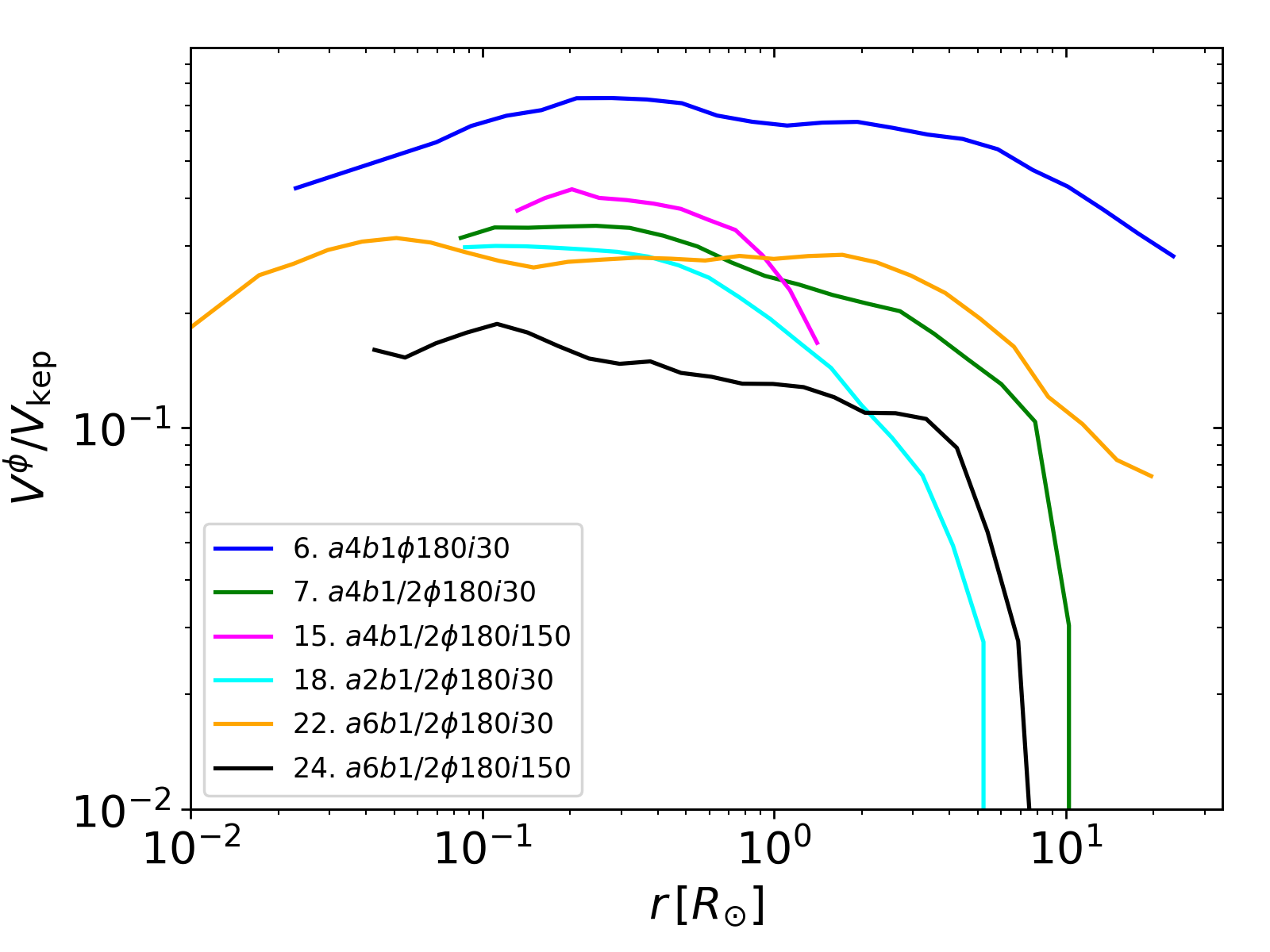}\\
	\includegraphics[width=8.4cm]{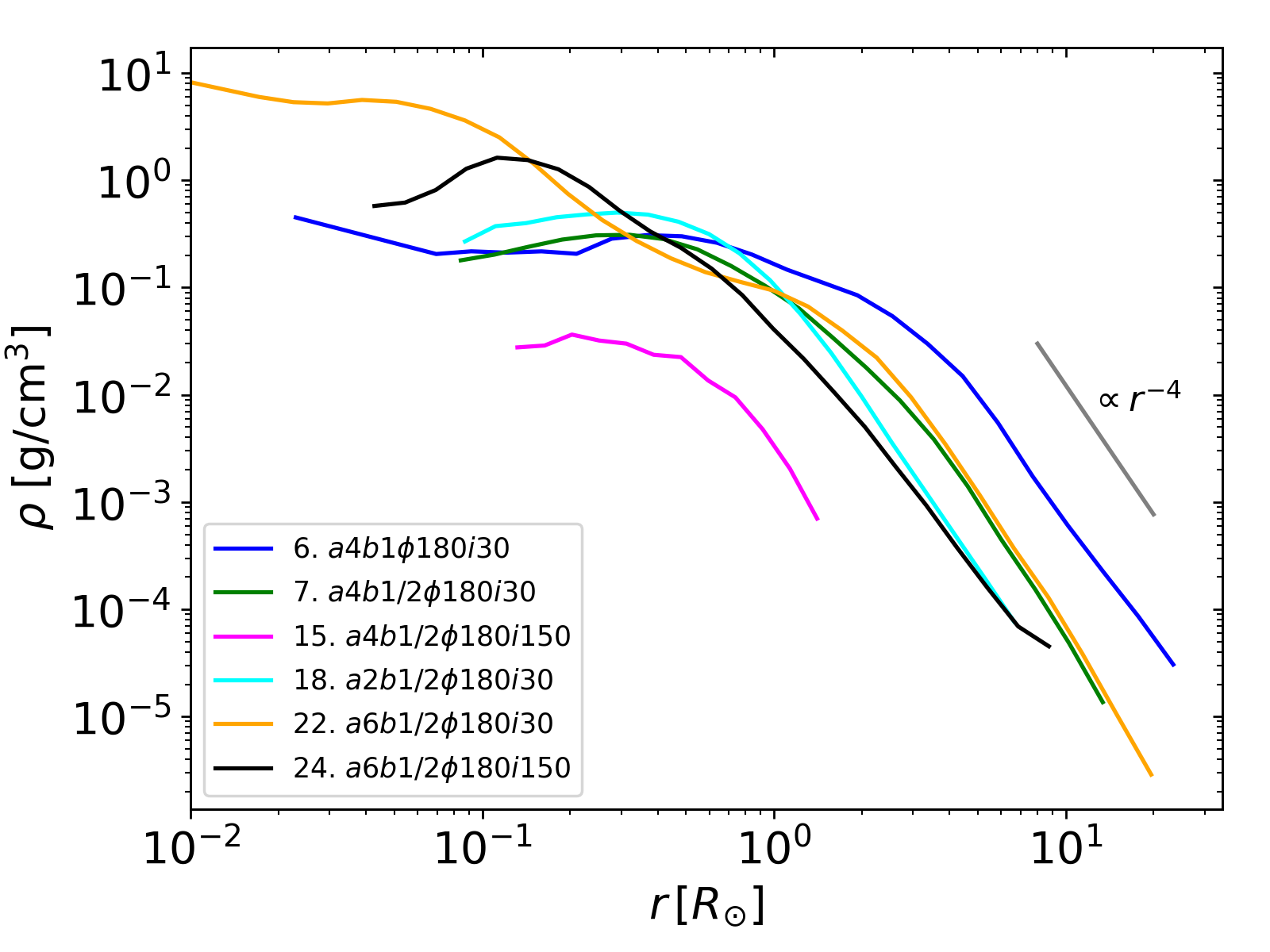}
	\includegraphics[width=8.4cm]{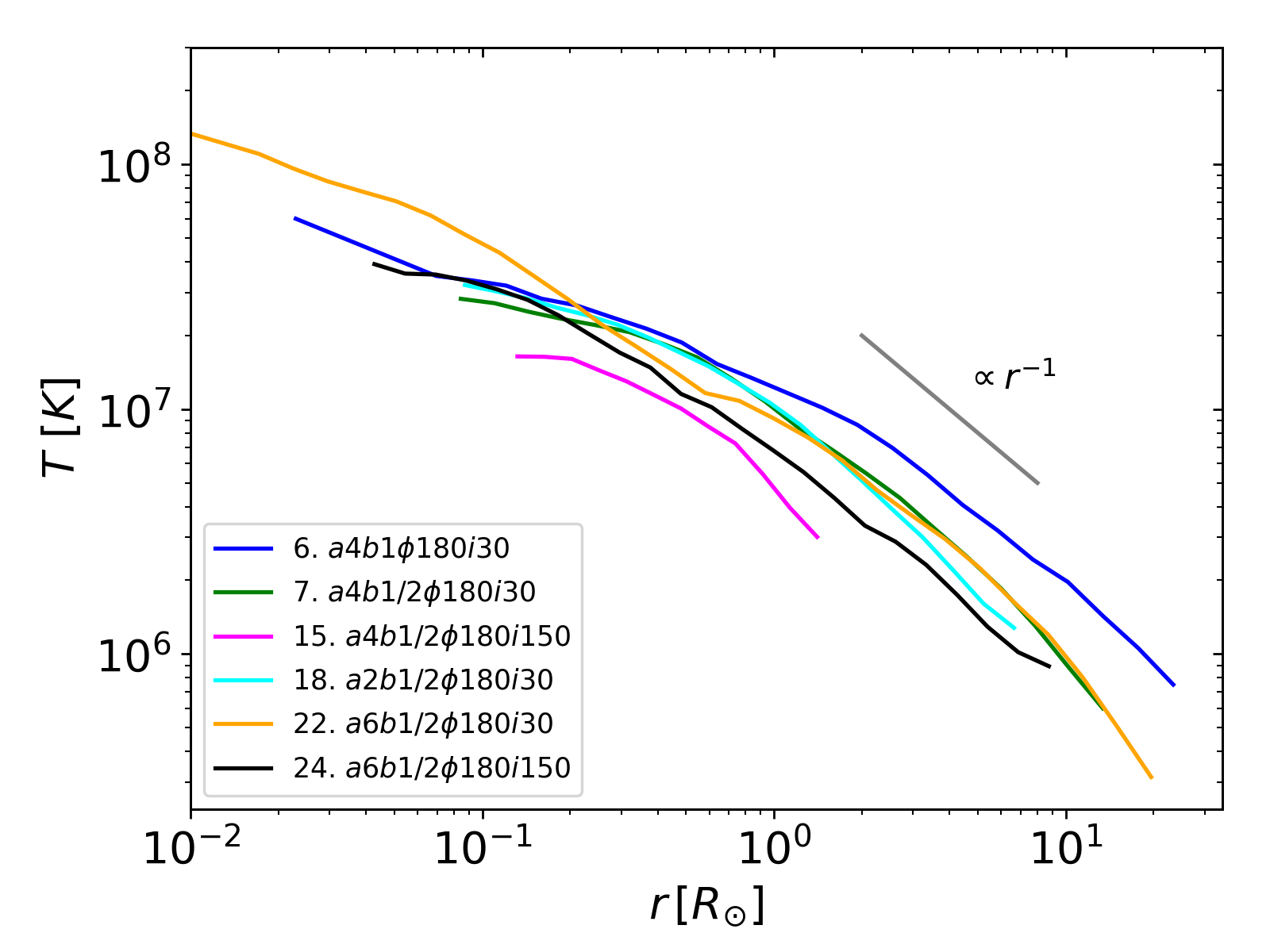}	
\caption{Profiles of the structure of the disks in simulations with $i=30^{\circ}$ or $150^{\circ}$ where a BBH forms, including the four models shown in Figure~\ref{fig:diskdensity}: The aspect ratio, defined as the ratio of the density scale height to the cylindrical radius $r$ (top-left), the ratio of the mass-weighted average of the azimuthal velocity along the midplane within the scale height to the Keplerian velocity $v_{\rm kep}$ (top-right), the average density along the midplane within the scale height (bottom-left), and the mass-weighted average of the temperature along the midplane within the scale height (bottom-right).  
All the reported quantities are measured at the end of the simulations.}
	\label{fig:diskprofile}
\end{figure*}

\begin{figure}
	\centering
	\includegraphics[width=8.6cm]{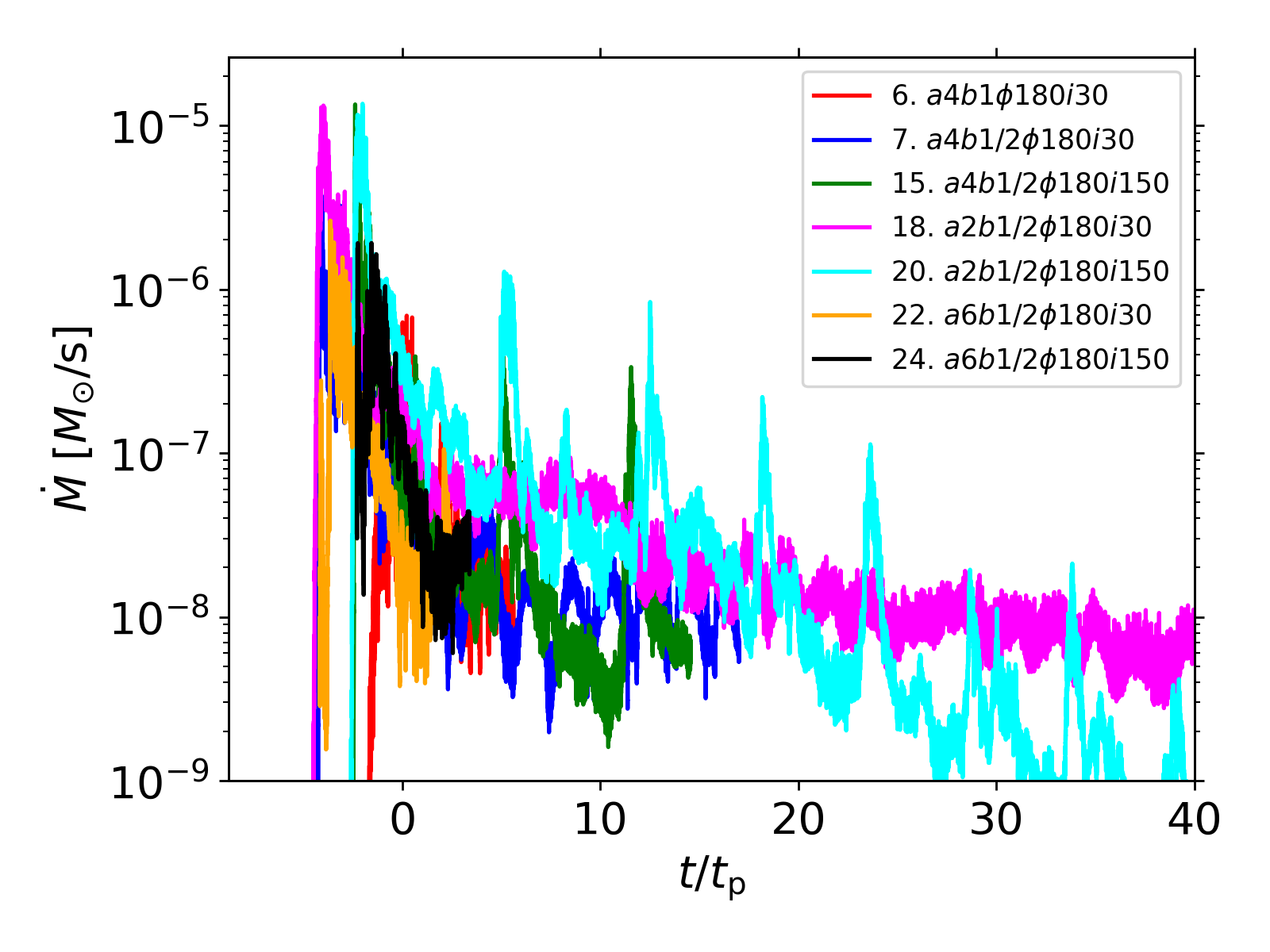}
\caption{The accretion rates of the initially single BHs that fully destroy the star in BBH-forming simulations with $\phi=180^{\circ}$, and $i=30^{\circ}$ or $150^{\circ}$.}
	\label{fig:accretion}
\end{figure}

\subsection{Accretion}

Our simulations show that stars can be disrupted in three-body interactions between BH-star binaries and single BHs via strong interactions with very small impact parameters, i.e., collisions. In such events, a merging BBH can subsequently form, and at least one of the BHs is surrounded by an accretion disk which can create EM transient phenomena. To zero-th order, the disk structure and features of the accretion rate can be imprinted onto light curves of such events. 

The refinement scheme adopted for the simulations allows us to resolve the gas structure down to $0.01\,{\rm R}_{\odot}\simeq 10^{3}\,r_{\rm g}$ from the BH. Although the regions that we can resolve are still too far from the BH to be directly related to the accretion process, we can provide an accurately resolved large scale structure of the disks formed in star-destroying events, which can be used as initial conditions for detailed disk simulations. Here, we define a disk as a group of gas cells tightly bound to the BH and coherently orbiting in the azimuthal direction. The outer edge of the disk is defined as the radius containing 99\% of the total bound mass orbiting at a velocity exceeding 1\% of the local Keplerian speed $v_{\rm kep}(r)=\sqrt{G[M(<r)+M_{\bullet}]/r}$, where $M(<r)$ is the mass enclosed within $r$. 

We show in Figure~\ref{fig:diskdensity} both the face-on (\textit{left} panels) and edge-on (\textit{right} panels) density distributions of the disks around the BH that destroys the star at the first encounter in four example models, and in Figure~\ref{fig:diskprofile} the radial profiles of the aspect ratio, the density, the temperature, and the rotational velocity for all models where an accretion disk forms. The aspect ratio $h/r$ is defined as the ratio of the first-moment density scale height, averaged over a given cylindrical radius, to the cylindrical radius. Here, we excluded Model 20. $a2b1/2\phi180i150$ in this analysis because the BH in that model is surrounded by a nearly spherical gas cloud, not by a disk. But we provide the accretion rate for that model also, shown in Figure~\ref{fig:accretion}.

We find that the disks are thick and pressure-supported, and mostly confined within $r\simeq30\,{\rm R}_{\odot}$. In general, the aspect ratio $h/r$ (\textit{top-left} panel of Figure~\ref{fig:diskprofile}) is comparable to or greater than order unity up to the outer edge of the disks. $h/r$ declines from $h/r\simeq3-5$ to $h/r\simeq 1$ outwards. The rotational velocity $v^{\phi}$ near the mid-plane is sub-Keplerian ($v^{\phi}/v_{\rm kep}\simeq 0.1 - 0.6$), indicating the disk is not rotationally supported. The velocity ratio remains the same  out to the outer disk edge. The density of the inner region stays flat at $\rho\simeq (0.1-5)\gram \cm^{-3}$ up to 0.1 - 0.2 of the disk size, then declines steeply following a $r^{-4}$ power-law. On the other hand,  the temperature does not show such flatness at $r\lesssim \Rsol$, but continuously decreases following a  $r^{-1}$ power-law. 

Finally, we present in Figure~\ref{fig:accretion} the accretion rate of the initially single BHs that fully destroy the star at the first closest encounter. The general trend is that, upon disruption or collision, the accretion rate dramatically increases up to $\dot{M}\simeq (10^{-6}-10^{-5})\Msol\,{\rm s}^{-1}$ and it takes around 80-100 hours until $\dot{M}$ declines by a factor of 100 from its peak. When the binary is eccentric and the pericenter distance is sufficiently close, a periodic perturbation from the other BH at periastron results in periodic bursts on a time scale $\simeq$ the orbital period (e.g., Model 15. $a4b1/2\phi180i150$, and Model 20. $a2b1/2\phi180i150$). Although the accretion rate is super-Eddington, the total accreted mass is at most $0.1\Msol$ ($\lesssim 0.4\%$) until the end of the simulation, and the magnitude of the BH spin driven by accretion can be as large as $0.01$.

We have to caution that such extremely high accretion rates for stellar-mass BHs would result in strong outflows \citep[e.g.,][]{Skadowski2014}, which would regulate the accretion rate. Although we have realized a significant improvement in resolving gas motions near the BHs compared to \citetalias{Ryu+2023} thanks to using refinement, since feedback from the BHs is not included in our simulations it is likely that our accretion rates are overestimated. Nonetheless, if the luminosity is mostly driven by accretion, the features revealed in the accretion rate (e.g., periodic bursts) could possibly be imprinted in the light curves.

\begin{figure}
	\centering
	\includegraphics[width=8.6cm]{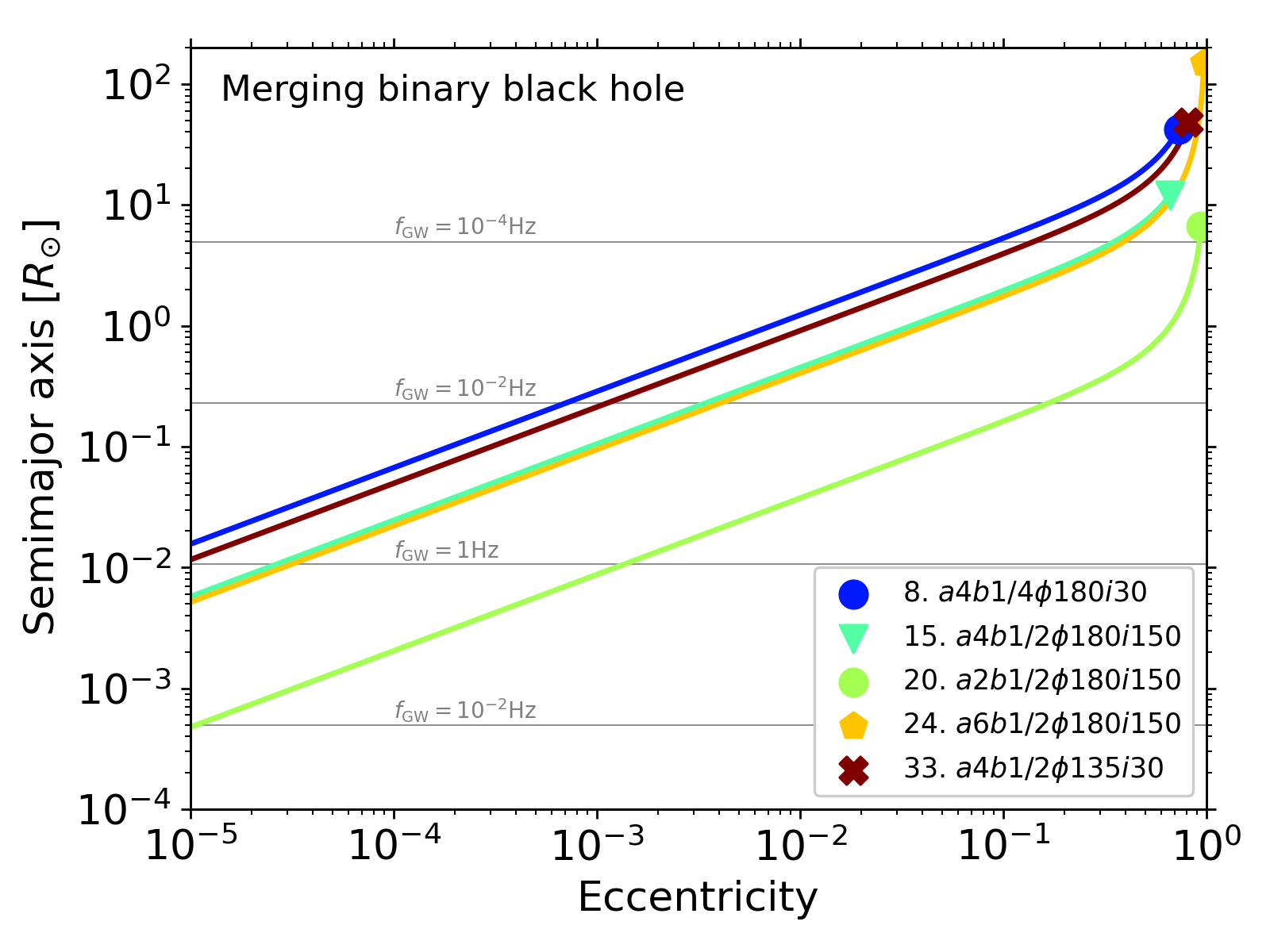}
\caption{The evolution of $a$ and $e$ of the five merging BBHs formed in three-body interactions due to GW emission. The markers depict $a$ and $e$ of the final merging binary black holes. The four grey horizontal lines indicate the semimajor axes at which the rest-frame GW frequency (twice the orbital frequency) is $f_{\rm GW} = 10^{-4}$, $10^{-2}$, $1$, and $10^{2}$ Hz, respectively. }
	\label{fig:mergingBBH}
\end{figure}

\section{Discussion}\label{sec:discussion}

\subsection{Formation of merging binary black holes}
Our simulations show that close three-body encounters between a BH-star binary and a single BH can create a merging BBH (see the \textit{top-left} panel of Figure~\ref{fig:binary_orbit}). One possibly dominant formation process we identified is the close interaction between the star and the incoming BH at the first closest approach, resulting in a stellar disruption, followed by the formation of a BBH. 5 out of 11 BBHs formed in our simulations would merge in a Hubble time via GW emission. The semimajor axes of the merging BBHs are $\lesssim 114\Rsol$ and their eccentricities are quite high, $0.66\lesssim e\lesssim0.97$. If the required conditions are met ($r_{\rm p}\simeq 0.5\, a$, encounters between the star and the incoming BH 
%with similar masses 
at the first closest approach), this type of encounters can form, albeit likely rarely, a very compact eccentric BBH: $t_{\rm GW}\simeq 10^{4}$ yr in Model 20. $a2b1/2\phi180i150$. 

To see whether the merging BBHs can have residual eccentricities when they enter the frequency band of LIGO (10 Hz to 10kHz), we evolve the five binaries assuming their orbits evolve purely via GW emission until $t_{\rm GW}= P$, where $P$ is the binary orbital period. We solve Equations 5.6 and 5.7 in \citet{Peters+1964} simultaneously using a 4th-order Runge-Kutta method with an adoptive step size of $10^{-3} t_{\rm GW}$. As a sanity check, we confirmed that our numerical solutions are consistent with the analytic solution (Equation 5.11 in \citealt{Peters+1964}) within fractional errors of $\lesssim10^{-8}$. Figure~\ref{fig:mergingBBH} shows the evolution of $a$ and $e$ of the five merging BBHs, starting from those found in our simulations (marked as scatters near the top-left corner). As shown in the figure, by the time the BBHs enter the LIGO frequency band, their residual eccentricities would be very small ($e< 10^{-5}$). 

Nonetheless, the circumbinary gas produced by the disruption of the star may affect the (at least early) orbital evolution, which may hence deviate from the purely GW-driven evolution considered above. The gas-binary interaction and resulting binary evolution remains an active topic of study. A growing number of numerical works have suggested that a binary surrounded by a circumbinary disk can expand \citep[e.g.,][]{Miranda+2017, Munoz2019,Duffell+2020} and can be driven into an eccentric orbit \citep[e.g.,][]{Zrake+2021,DorazioDuffell2021}, depending on the disk and binary parameters, as opposed to the predictions from the commonly held picture of surrounding gas driving binaries into shrinking circular orbits \citep[e.g.,][]{Armitage+2002}. However, given the limited parameter space explored in previous work, it is not straightforward to predict the evolution of our unequal-mass, very eccentric BBHs surrounded by a possibly misaligned disk, based on the results from the previous work. 

The remaining 6 BBHs with GW-driven merger timescales longer than a Hubble time are hard binaries in typical stellar cluster environments. This means that those binaries could become potential GW event candidates via weak interactions with other objects and a few strong interactions like the ones considered in this study. 

\subsection{Electromagnetic counterparts of binary black hole merger}

The close association of BBHs and stellar disruptions can have important implications for EM counterparts of BBH mergers. At the time the BBH forms, there would be a prompt EM transient phenomenon due to the stellar disruption. The very high accretion rate (Figure~\ref{fig:accretion}), along with the accretion-driven BH spin and magnetic field of debris inherited from the star, suggests that a jet can be launched. For such a case, the luminosity powered by the jet would track the accretion rate as $\propto\dot{M}c^{2}$ with an uncertain efficiency factor. We also found that both BHs can be surrounded by the stellar debris and accrete, possibly suggesting that both BHs may be able to launch jets simultaneously, potentially leading to a unique observational signature. 

In addition to the prompt EM emission, the existence of the surrounding gas when the BBH forms may result in a possible EM counterpart at the time of merger. This is a quite similar situation as found in \citet{Ryu+2022} where an initially hard BBH encounters with a single star and becomes surrounded by gas debris after disrupting the star. \citet{Perna2016} studied the evolution of an initially hyper-Eddington accretion disk which cools and shuts down the magnetorotational instability before the disk material is fully accreted. Under these conditions, the ``dead disk'' is expected to survive until the BBHs merge, and to heat up and re-ignite during the merger process, hence yielding a possible EM counterpart to the GW event. 

\subsection{Varieties of encounters}\label{subsec:variety}

Although we consider three-body encounters between a circular binary and a single object with similar masses (the largest mass ratio is 0.5), there could be a variety of these types of events involving, e.g., initially eccentric binaries and a wide range of masses of encountering objects.

Encounters involving massive stars (i.e., $10\Msol$)  are likely to occur during the early evolutionary stages of star clusters unless there is another episode of star formation, since stars with mass $>10\Msol$ would collapse to compact objects in tens of Myrs. Therefore, over the full cluster lifetime, the overall rate would indeed be higher for encounters involving less massive MS stars because such binaries would survive longer. Using Monte Carlo simulations of globular clusters, \citet{Kremer+2018} showed that up to 10 detached BH-MS binaries can exist in clusters at an age of $10-12$ Gyr, and the typical mass of the companion MS stars is $\lesssim 1-2\Msol$, depending on the cluster properties. Even for this case, strong interactions between a low-mass MS star and the incoming BH at the first closest approach would have higher chances of forming BBHs than for the other cases where two BHs meet first. 

At later times when all massive stars collapse to compact objects, interactions between stars and BHs with significantly different masses would be more probable. If the star is significantly less massive than both BHs, the interactions would be effectively two-body with small perturbations of the BH orbits by the star. However, if the star was disrupted by the incoming BH as in the \textit{BBH-forming} encounters, the stellar disruption would generate bright EM flares. Furthermore, resulting momentum kicks and gas dynamical friction would facilitate the formation of BBHs, unless the momentum kick is given to increase the relative kinetic energy of the BHs. This process would be most efficient when the star and the incoming BH have comparable masses. 

If the encountering binary is eccentric, the binary members would spend most of their orbital time near apocenter, indicating that the cross-section would be enhanced by a factor of $1+e$. Unlike the increase in $e$ in our models when initially circular binaries are considered, the final binaries can be circularized depending on the direction of the momentum kick associated with close interactions between the two objects at the first closest approach. We already demonstrated in Figures~\ref{fig:example1} and \ref{fig:example2} that the momentum kick acts to add to or remove the momentum of the BH in the binary, depending on whether the orbit is initially in a prograde or retrograde direction. 

\section{Summary and Conclusions}\label{sec:conclusion}

In this work we have investigated the outcomes of three-body encounters between a $20\Msol$ BH -- $10\Msol$ star circular binary and a $10\Msol$ stellar-mass BH, using a suite of hydrodynamical simulations with the moving-mesh code {\small AREPO}. We have focused on the formation of BBHs, the conditions required for their formation, and the EM emission from those systems. We have considered a wide range of encounter parameters, i.e., varying the binary size ($a\simeq 34$, 68, 101$\Rsol$), the impact parameter ($a/4 - a$), the inclination angle ($i=0^{\circ}$, 30$^{\circ}$, 120$^{\circ}$, 150$^{\circ}$, and 180$^{\circ}$), and the phase angle ($\phi=0^{\circ}$, 45$^{\circ}$, 90$^{\circ}$ and 135$^{\circ}$, 180$^{\circ}$, 225$^{\circ}$, 270$^{\circ}$, 315$^{\circ}$), while we have kept fixed the masses of the star and of the BHs.  

We have categorized the encounters into two classes depending on their outcomes. This classification is primarily determined by which types of objects meet at the first closest approach. When the star and the incoming single BH encounter first, their close interaction imparts a momentum kick to the BH, resulting in a dramatic decrease in the BH's speed. The BH is subsequently captured by the other bystander BH, forming a BBH. In this case, the star is frequently destroyed due to its close encounter with the BH. On the other hand, when two BHs encounter first, either the original binary's orbit is simply perturbed (prograde encounters), or the originally single BH captures the star, forming a new binary (member exchange, retrograde encounters). Although the most frequent outcomes are BH-star binaries, a disruption of the star and BBH formation are still possible. 

The most important factors that determine the outcomes are the phase angle and the impact parameter. As explained above, the phase angle primarily demarcates the boundary between ``BBH-forming'' encounters and ``non-BBH-forming'' encounters. The impact parameter on the other hand affects the strength of interactions: for $r_{\rm p}>a$, the incoming BH interacts weakly with the binary. As a result, the binary orbit is perturbed, or the binary members are exchanged. For $ r_{\rm p}\lesssim a$, interactions can become significant, possibly resulting in a disruption of the star when the star and the BH meet at the first closest approach. Although our simulations do not cover the entire parameter space for this type of encounters, the key dynamical processes can be extrapolated within this class of encounters to other initial parameters, and possibly also to other astrophysical systems (e.g., three-body encounters involving a massive black hole having a stellar companion and an isolated BH, forming extreme mass ratio inspirals). 
 
The close correlation between BBH formation and stellar disruption in our systems has interesting implications for the formation channel of BBHs and EM counterparts of their merger. We confirm that three-body encounters between a BH-star binary and a BH can produce merging BBHs. In addition, we find that the BH that disrupts the star in the \textit{BBH-forming} encounters is promptly surrounded by an optically and geometrically thick disk with accretion flows towards the BH exceeding the Eddington limit. If a jet is launched from the system, the jet luminosity would likely track the accretion rate. If the disk remains long-lived and revives at merger, EM counterparts can be produced at the time of the BBH merger. 

Our order-of-magnitude estimate for the encounter rate suggests that this type of encounters may be rarer than other types of three-body encounters considered in \citetalias{Ryu+2022} (i.e., between binary BHs and single stars) and \citetalias{Ryu+2023} (i.e., between stellar-binaries and single BHs). However, given the simplified assumptions made here, 
more detailed estimates should be made for these encounters, taking their specific astrophysical environments accurately into account.

\section*{Acknowledgements}
The authors are grateful to the referee for constructive comments and suggestions. This research project was conducted using computational resources (and/or scientific computing services) at the Max-Planck Computing \& Data Facility. The simulations were performed on the national supercomputer Hawk at the High Performance Computing Center Stuttgart (HLRS) under the grant number 44232. 
The authors gratefully acknowledge the scientific support and HPC resources provided by the Erlangen National High Performance Computing Center (NHR@FAU) of the Friedrich-Alexander-Universität Erlangen-Nürnberg (FAU) under the NHR project b166ea10. NHR funding is provided by federal and Bavarian state authorities. NHR@FAU hardware is partially funded by the German Research Foundation (DFG) – 440719683. R.~Perna acknowledges support by NSF award AST-2006839.

\section*{Data Availability}
Any data used in this analysis are available on reasonable request from the first author.

\bibliographystyle{mnras}
%\bibliography{biblio.bib} 

\end{document}